\documentclass{aa}  

\usepackage{graphicx}
\usepackage{txfonts}
\usepackage{comment}
\usepackage{hyperref}      
\usepackage{csquotes}

\bibpunct{(}{)}{;}{a}{}{,} 
\hypersetup{colorlinks=true,linkcolor=blue,citecolor=blue,filecolor=blue,urlcolor=blue,}

\newcommand{\pc}    {\,{\rm pc}}

\newcommand{\Msun}  {\,{\rm M_{\odot}}}
\newcommand{\Gyr}   {\,{\rm Gyr}}
\newcommand{\Myr}   {\,{\rm Myr}}

\newcommand{\K}     {\,{\rm K}}

\newcommand{\gcm}  {\,{\rm g\,cm^{-3}}}

\begin{document}

   \title{Type Ia supernova feedback effects on globular clusters of different masses}
    \titlerunning{Type Ia SNe effects on GCs of different masses}
     \authorrunning{E. Lacchin}

   \author{E. Lacchin,
          \inst{1,} \inst{2,} \inst{3,} \inst{4,} \inst{5},
          M. Donati \inst{2,} \inst{6},
          F. Calura\inst{1},
         C. Nipoti\inst{2},
         A. Lupi\inst{6,7},
         A. Yaghoobi \inst{8,9}.
}
   \institute{INAF - OAS, Osservatorio di Astrofisica e Scienza dello Spazio di Bologna, via Gobetti 93/3, I-40129 Bologna, Italy\\
    \email{elena.lacchin@inaf.it}
   \and
 Dipartimento di Fisica e Astronomia "Augusto Righi", Alma Mater Studiorum - Unversità di Bologna, via Gobetti 93/3, I-40129 Bologna, Italy
    \and
Dipartimento di Fisica e Astronomia “Galileo Galilei”, Univerisità di Padova, Vicolo dell’Osservatorio 3, I-35122, Padova, Italy 
    \and
   Institut f{\"u}r Theoretische Astrophysik, ZAH, Universit{\"a}t Heidelberg, Albert-Ueberle-Stra{\ss}e 2, D-69120, Heidelberg, Germany
   \and
 INFN–Padova, Via Marzolo 8, I–35131, Padova, Italy
 \and 
Dipartimento di Scienza e Alta Tecnologia, Università degli Studi dell’Insubria, via Valleggio 11, I-22100, Como, Italy
\and
INFN, Sezione di Milano-Bicocca, Piazza della Scienza 3, I-20126 Milano, Italy
 \and
 Institute for Research in Fundamental Sciences (IPM), Larak Garden, 19395-5531 Tehran, Iran
 \and
 Univ Lyon, Univ Lyon1, Ens de Lyon, CNRS, Centre de Recherche Astrophysique de Lyon UMR5574, F-69230, Saint-Genis-Laval, France
}

   \date{Received September 15, 1996; accepted March 16, 1997}

 
  \abstract
    {Through 3D hydrodynamical simulations, we explore the impact of Type Ia supernova (SN) explosions on the star formation history and chemical properties of second-generation (SG) stars in young globular clusters with masses of $10^5-10^6\Msun$. We assume that the SG is formed out of the asymptotic giant branch (AGB) ejecta of first-generation stars plus pristine interstellar medium gas which is modelled as a uniform gas moving at a constant velocity towards the cluster. We tested two values for the infalling gas density of $10^{-24}$ and $10^{-23} \gcm$. Type Ia SNe start to explode together with the release of gas from the most massive AGB stars. Three simulated models are analyzed. In the low-mass and low-density scenario, we find that SNe Ia quench star formation which however restarts when the gas cools down again in between two explosions. SG stars are dominated by a He-rich population ($Y>0.33$), which is poorly diluted by pristine gas. In the high-mass models, star formation is mildly affected, while the He composition is significantly altered as exploding SNe prevent the accretion of pristine gas and therefore extremely helium-rich stars form. In the high-density model, such weak gas accretion leads to a maximum enhancement in helium mass fraction much larger than the observed one and not correlating with the initial cluster mass as found in models without Type Ia SNe. As for the iron content, small spreads have been found in all models, but the SG is less homogeneous than the FG, at variance with current observations.   }


   \maketitle 
%

\section{Introduction}

Over the past decades, numerous photometric and spectroscopic studies have revealed that Galactic globular clusters (GCs) host multiple stellar populations, challenging the long-standing view of GCs as single stellar populations (SSPs). The first indications of multiple populations within GCs emerged from studies on the chemical composition of their stars. Some stars displayed chemical abundances rarely observed in field stars, with anomalies primarily involving light elements such as C, N, Na, O (see, for example, \citealt{carretta2009c,carretta2009b,Carretta2009a,gratton2019} and references therein).

Subsequent photometric investigations further supported these findings by identifying splits in sequences within the colour–magnitude diagram of GCs, providing additional evidence for multiple populations in the same cluster (\citealt{lee1999,pancino2000,piotto2005,marino2008,milone2017,Milone2012}). Stars with chemical compositions akin to those in the field are thought to represent the first generation (FG) of stars, while \enquote{anomalous} stars -- those enriched in He, N, Na, and depleted in O and C -- are associated with subsequent generations.
Such chemical variations have been observed in globular clusters in nearby galaxies such as the Magellanic Clouds \citep{mucciarelli2009}, Fornax \citep{larsen2014}, and M31 \citep{nardiello2019}, making multiple stellar populations ubiquitous.

Until some years ago, both generations were thought to share the same metallicity content apart from a small group of clusters, labelled Type II GCs, where a spread in metallicity was observed \citep{johnson2015,milone2017}. Recently, both photometric \citep {legnardi2022,lardo2022} and spectroscopic \citep{marino2019} studies of Galactic GCs 
(such as NGC 3201 and M 2) have revealed a metallicity spread as large as $0.3\ {\rm dex}$ in FG stars \citep{lardo2023}, 
suggesting that these stars were formed out of chemically inhomogeneous gas. 
Surprisingly, the second generation instead displays a much narrower spread, which could provide important information on how \enquote{anomalous} stars were formed.

The precise mechanisms and timeline for the formation of chemically peculiar stars, and in general globular clusters, remain still unclear despite the number of scenarios proposed in the literature. None of the proposed models fully accounts for all the observed features and trends without facing some challenges (see \citealt{renzini2015,bastian2018,gratton2019}). Among the various theories, the asymptotic giant branch (AGB) scenario has been the most extensively explored in the past years (\citealt{dantona2004,dercole2008,bekki2011}). In this scenario, the first generation of stars forms all at once with the same chemical composition. Following this, feedback from massive stars, both in the pre-supernova (SN) and SN phases, clears the system of any gas, preventing the enrichment of the remaining material by massive star ejecta \citep{calura2015}. Afterwards, a second generation of stars can form from the ejecta of FG AGB stars mixed with gas that has the same composition as FG stars (see \citealt{dercole2010,calura2019,yaghoobi2022}).
Other models propose that second-generation (SG) stars arise from gas expelled by fast-rotating massive stars \citep{decressin2007}, massive interacting binaries \citep{demink2009}, very massive stars \citep{vink2018}, supermassive ($m \sim  10^4 \Msun$) stars \citep{denissenkov&hartwick2014,gieles2018}, supergiant stars \citep{szecsi2018}, black hole accretion disks \citep{breen2018}, stellar mergers \citep{wang2020}.   

We focus here on the AGB scenario, which has been significantly explored in the last years \citep{calura2019,lacchin2021,lacchin2022,mckenzie2021,yaghoobi2022,yaghoobi2022ion,yaghoobi2024}. In particular, our aim is to extend the work of \citet{lacchin2021} studying the effects of Type Ia SNe on the star formation of lower mass clusters, such as those simulated by \citet{yaghoobi2022}. Type Ia SNe have long been assumed to be responsible for the quenching of the star formation of the second generation based on the pioneering results of \citet{dercole2008}, who performed 1D hydrodynamic simulations of GCs with initial masses of $10^{6-7}\Msun$ with and without Type Ia SN feedback. \citet{dercole2008} found that in the models with SN Ia feedback, once the first SNe were going off, the gas was rapidly wiped out of the system leading to an almost immediate quenching of the star formation. However, their SNe were located in the centre of the cluster, and this could significantly increase the energy deposited in the inner regions, where stars are forming. In \citet{lacchin2021}, we performed 3D hydrodynamic simulations of a $10^{7}\Msun$ cluster and obtained that, with a more realistic spatial distribution of the SNe, star formation was not halted immediately after the first explosions. In the extreme case where a high interstellar medium density is assumed, the star formation rate is not affected by SN Ia explosions. On top of that, we have shown that when star formation is reduced, the helium mass fraction of the newborn stars is much higher than expected. The pristine gas is not able to continuously provide helium-poor gas to dilute the helium-rich AGB ejecta, and therefore most of the stars are formed from undiluted material. Later, \citet{yaghoobi2022} performed similar simulations for lower-mass clusters without including Type Ia SNe. Combining their results with the ones of \citet{calura2019}, they were able to obtain positive correlations between the SG-to-total number ratio and maximum He enhancement in SG stars as a function of the initial cluster mass in agreement with observations.

In this work, we include Type Ia SNe in the low-mass cluster models of \citet{yaghoobi2022} to explore the role of this source of feedback both on the star formation and on the chemical composition of SG stars and determine whether the scaling relations derived by \citet{yaghoobi2022} are affected or not by the presence of Type Ia SNe.

The paper is organized as follows: in Section \ref{sec:simuset} we describe the setup of the simulations and the main physical ingredients included in the simulations. In Section \ref{sec:results} we present the results of our set of simulations which we discuss in Section \ref{sec:discussion}. Finally in Section \ref{sec:conclusions} we draw the conclusions of our work.

\section{SIMULATION SETUP}

\begin{table}
\centering
\caption{Parameters of the simulations.}
\hspace{-0.5cm}
\resizebox{1.05\columnwidth}{!}{
\renewcommand{\arraystretch}{1.40}
\begin{tabular}{llc} 
\hline
\hline
Parameter & Description & Adopted values \\
\hline
$M_{\rm FG}$      & Stellar mass of the FG                           & $10^5-10^6 {\rm M_{\odot}}$\\
$a$    & Plummer radius of FG stellar distribution        & 3 pc\\
$\rho_{\rm pg} $  & Density of the pristine gas                      & ${\rm 10^{-24}; 10^{-23} g\ cm ^{-3}} $ \\
$v_{\rm pg}  $    & Pristine gas velocity relative to the cluster    & ${\rm 2 \times 10^6 cm\ s^{-1}}$   \\
$Z_{\rm pg}  $    & Metallicity of the pristine gas                  & 0.001  \\
$X_{\rm He} $     & Helium mass fraction of the pristine gas         & 0.246  \\
$X_{\rm Fe} $     & Iron mass fraction of the pristine gas           & $3.774 \times 10^{-5}$   $^{(a)}$    \\
$T_{\rm pg} $     & Temperature of the pristine gas                  & $10^4 {\rm  K}$ \\
$T_{\rm floor}$   & Minimum temperature                              & $10^3{\rm K}$ \\
$t_{\star} $      & Star formation time-scale                        & 0.1 Gyr \\
\hline
\hline
\end{tabular}
}
\label{tab:param}
$^{(a)} $ taken from \citet{dercole2012} 
\end{table}

\label{sec:simuset}

In this work, we extend the study of \citet{lacchin2021} to lower masses. In particular, we model clusters with a mass of $10^5-10^6 {\rm M_{\odot}}$ and a half-mass radius of $4 \pc$ as done by \citet{yaghoobi2022}, but we include the feedback from Type Ia supernovae
The initial radius has been assumed to be equal for both cluster masses, as observations have revealed a flat-radius cluster mass relation below $M_{\rm cl}=10^6\Msun$ \citep{Krumholz2019}.
We adopted a customized version of the RAMSES hydrocode \citep{teyssier2002} to study how Type Ia SNe affects the formation and the chemical composition (in terms of helium and iron) of SG stars. The code uses a second-order Godunov scheme to
solve the Euler equations and a particle-mesh solver to compute
the dynamical evolution of particles. For all simulations, we fix the box size to $L^3={\rm (50 \ pc)^3}$ which is refined exploiting the adaptive mesh refinement (AMR) technique. The maximum resolution reached in our simulations is ${\rm 0.1\ pc}$ as in \citet{yaghoobi2022}, and much smaller than in \citet{lacchin2021} since the clusters we are modelling here are much more compact and, therefore, higher resolutions are required to characterize them. In \citet{lacchin2021}, we have not exploited the AMR as the whole box was refined at the maximum level shortly after the beginning of the simulations due to the AGB ejecta. In these simulations, instead, the large dimension of the box compared to the size of the cluster makes AMR very effective.

\subsection{Initial setup}

\label{sec:initinput}
We briefly describe the setup of our simulations which is the same as in \citet{lacchin2021} and \citet{yaghoobi2022}: we therefore advise the reader to consult these works for further details. 
The starting point of all our runs $t_0$ lies $31.3$ Myr after the formation of the FG as in \citet{yaghoobi2022}, when core-collapse SNe have already cleared out the system from gas \citep{calura2015} and have carved a hole in the interstellar medium (ISM). The cluster is then composed of low- and intermediate-mass stars belonging to the FG, which is modeled as a static \citet{plummer1911} profile with Plummer radius $a=3 \pc$. Following \citet{dercole2016}, we assume that the cluster is orbiting in the disk of a high-$z$ star-forming galaxy. This implies that the cluster meets the ISM gas asymmetrically. We thus fix the cluster at the centre of our computational box and, at time $t_{\rm inf}$, gas is imposed to flow inside the box from one of the boundaries.
 The value of $t_{\rm inf}$ depends on the stagnation point of the core-collapse supernovae (CC-SN) driven bubble, i.e. where the ram pressure of the SN-driven wind balances the ISM pressure. The location of this point depends on the density of the ISM, $\rho_{\rm pg}$, and the mass of the FG, $M_{FG}$, as described in \citet{dercole2016}. In all our simulations we calcuate $t_{\rm inf}$ through equation 11 of \citet{dercole2016}, as done in \citet{lacchin2021} and \citet{yaghoobi2022}.

In addition, 39 Myr after the FG formation, the FG starts releasing the ejecta of the most massive AGB stars and of Type Ia SNe. 

For all the six sides of our computational box we adopt outflow boundary conditions. At $t_{\rm inf}$, the gas is allowed to enter the box from the $yz$ plane at negative $x$ with a velocity $v_{\rm pg}$. 

In Table \ref{tab:param} we have listed the main parameters adopted in the present work, where we tested two values both for the density of the pristine gas $\rho_{\rm pg}$, to represent a typical dwarf galaxy \citep{marcolini2003} and star-forming regions in galaxies at high redshifts \citep{wardlow2017}, and for the mass of the FG $M_{\rm FG}$. The corresponding $t_{\rm inf}$, as well as a more detailed description of each model, are summarized in Table \ref{tab:simu}. Note that, in Table \ref{tab:simu}, as well as from now on, we report the time assuming $t_0=31.3$ Myr as the time zero, so the beginning of AGB pollution, $t_{\rm AGB}$ will start at 7.7 Myr, together with Type Ia SN explosions. The infall starts earlier in models M5I24 and M6I23, while in model M6I24 it enters the box when both AGBs and SNe Ia feedback have already started. At variance with \citet{yaghoobi2022} we do not present here the model with $10^5 \Msun$ and ISM density $\rho_{\rm pg} =10^{-23} {\rm g\ cm^{-3}}$, as for this model the convergence is very poor as shown by \citet{yaghoobi2022}, especially when comparing the results with and without AMR.

\subsection{Star formation} 

The star formation is modeled in a subgrid fashion described in \citet{rasera&teyssier2006}. A star particle is allowed to form only in cells where: i) the gas is neutral, which means that the temperature is $T<2\times 10^4$ K, ii) the velocity field is convergent $\nabla\cdot {\bf v}<0$.

Once a cell satisfies both these requirements, the gas is converted into star particles following the \citet{schmidt1959} law:

\begin{equation}
\dot{\rho}_{\rm \star, SG} =\frac{\rho}{t_{\star}}
\end{equation}

where $t_{\star}$ corresponds to the star formation time scale, proportional to the free-fall time and set to $t_{\star}=0.1\Gyr$.  The mass associated with every stellar particle is $M_{\rm p}=N m_0$ where $m_0$ is the minimum mass a star can have and $N$ is sampled from a Poisson distribution. The particle is located in the centre of the cell and its velocity and chemical composition are the same as the parental gas cell.

\subsection{AGB feedback}
The mass and energy return of AGB stars is modeled by introducing a source term in both the mass and energy conservation equation. The mass injection by AGB stars, per units time and volume is:

\begin{equation}
\dot{\rho}_{\rm \star,AGB}=\alpha \rho_{\star},
\end{equation}
where $\alpha$ is the the specific mass return rate of the FG component, while $\rho_{\star}$ is the local density of FG stars. 

The injected energy by AGB stars per unit time and volume is: 

\begin{equation}
S=0.5\alpha \rho_{\star} (3 \sigma^2+ v^2+v_{\rm wind}^2),
\end{equation}

where $\sigma$ is the 1D velocity dispersion of FG stars, while $v$ and $v_{\rm wind}$ are, respectively, the gas velocity and the wind velocity of AGB stars. We assume $ v_{\rm wind} \sim 2\times 10^6 {\rm cm\ s^{-1}}$  \citep{dercole2008}. 

In addition, we trace the helium abundance of both the gas and the SG stars adopting the same yields as in \citet{calura2019} and \citet{lacchin2021} for AGB stars. No iron is assumed to be produced in low- and intermediate-mass stars; therefore, the iron composition of AGB ejecta is the same as that of the pristine gas (see Table \ref{tab:simu}).

\subsection{Type Ia Supernova feedback}
\label{sec_ia}
Together with the AGB feedback, here we also model Type Ia SNe as in \citet{lacchin2021}. SNe Ia are spatially distributed following the \citet{plummer1911} profile, computed for a cluster of mass $M_{\rm FG}$ and Plummer radius 3 pc. The maximum radius at which SNe are located corresponds to half of the computational box size, which means that all of them are exploding inside it. Concerning the temporal distribution, we distribute them following the  \citet[their equation 14]{greggio2005} delay time distribution (DTD) for the single degenerate scenario. The rate of SNe Ia explosions depends on the mass of the system $M_{\rm FG}$ and on the assumed IMF. For a \citet{kroupa2001} IMF and in the time span we are interested in (of about 0.1 Gyr), the number of SNe Ia will be $\sim 10$ for a cluster with a mass $M_{\rm FG}=10^5 {\rm M_{\odot}} $ and $\sim 100$ for $M_{\rm FG}=10^6 {\rm M_{\odot}}$. In order to have a good sampling both of the DTD and of the
Plummer profile, we have created 1000 random realizations both for their spatial and temporal distributions and then chosen the one that deviates less from the mean. 
In addition to helium, we trace the evolution of iron, in order to study the effects of SNe Ia on the chemical composition of SG stars and compare the iron content with observations. Once a SN explodes, we inject a Chandrasekar mass of metals, out of which $0.5{\rm M_{\odot}}$ of iron \citep{scalzo2014}. We also assume that for every explosion  $10^{51}$ erg of thermal energy \citep{blondin2024} are injected into the system. 
As done by \citet{lacchin2021}, we have checked that our supernova remnants are well resolved in all our simulations. With the high resolution used here, none of the supernovae fall in the overcooling regime derived using the criterion provided by \citet{kim&ostriker2015}.

\subsection{Cooling}
We adopt the built-in RAMSES implementation for radiative gas cooling where the cooling rates of hydrogen, helium and metals are taken into account \citep{few2014}. We adopt a temperature floor of $T=10^3$ K (see \citealt{calura2019}), while the temperature of the inflowing gas is assumed to be $T=10^4$ K, a reasonable value for a warm photoionised ISM in a star-forming galaxy \citep{haffner2009}.

\begin{table}
\caption{Models description. {\it Columns}: (1) name of the model, (2) FG initial mass, (3) pristine gas density, (4) time of pristine gas reaccretion, (5) starting time of Type Ia SN explosions. Times listed in (4) and (5) are expressed assuming $t_0=31.3$ Myr as the time zero.} 
\begin{tabular}{ccccc} 
\hline
\hline
 Model &${\rm M_{FG} [M_\odot]}$& ${\rm \rho_{pg}[g\ cm^{-3}]}$  & $t_{\rm inf}\mathrm{[Myr]} $&${t_{\rm Ia}\mathrm{[Myr]}} $ \\
\hline
${\rm M5I24}$                  & $10^5  $                                       & $  10^{-24}$            & 3  & 7.7   \\
${\rm M6I23}$                  & $10^6 $                                         & $  10^{-23}$            & 3  & 7.7    \\
${\rm M6I24}$                  & $10^6$                                           & $  10^{-24}$            & 12.2  & 7.7    \\

\hline
\hline
\end{tabular}

\label{tab:simu}

\end{table}

\begin{figure*}
        \centering

        \includegraphics[width=0.346\textwidth,trim={0 1cm 0 0.0cm},clip]{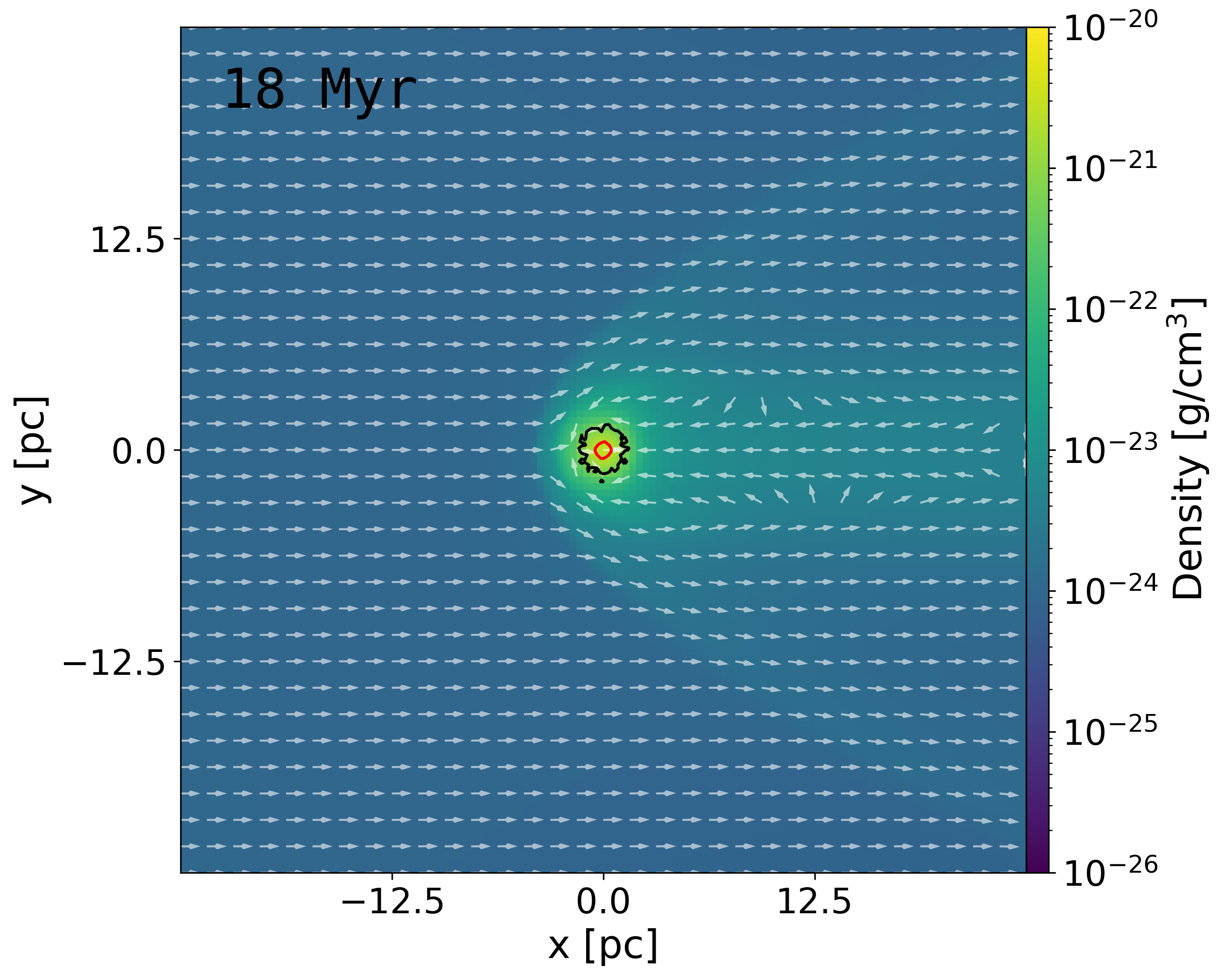}    
        \includegraphics[width=0.31\textwidth,trim={1.6cm 1cm 0 0.0cm},clip]{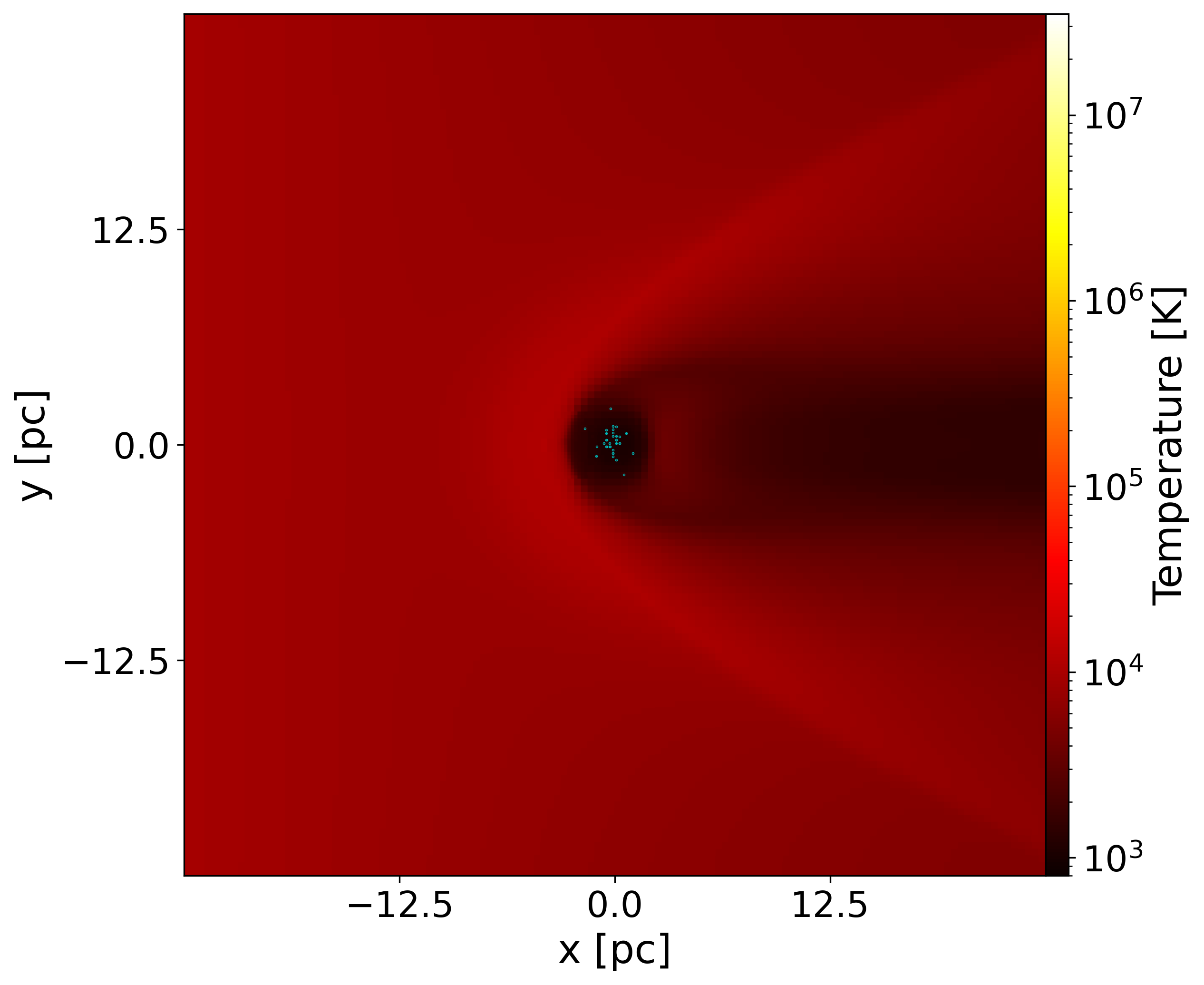}   
         \includegraphics[width=0.31\textwidth,trim={1.6cm 1cm 0 0.0cm},clip]{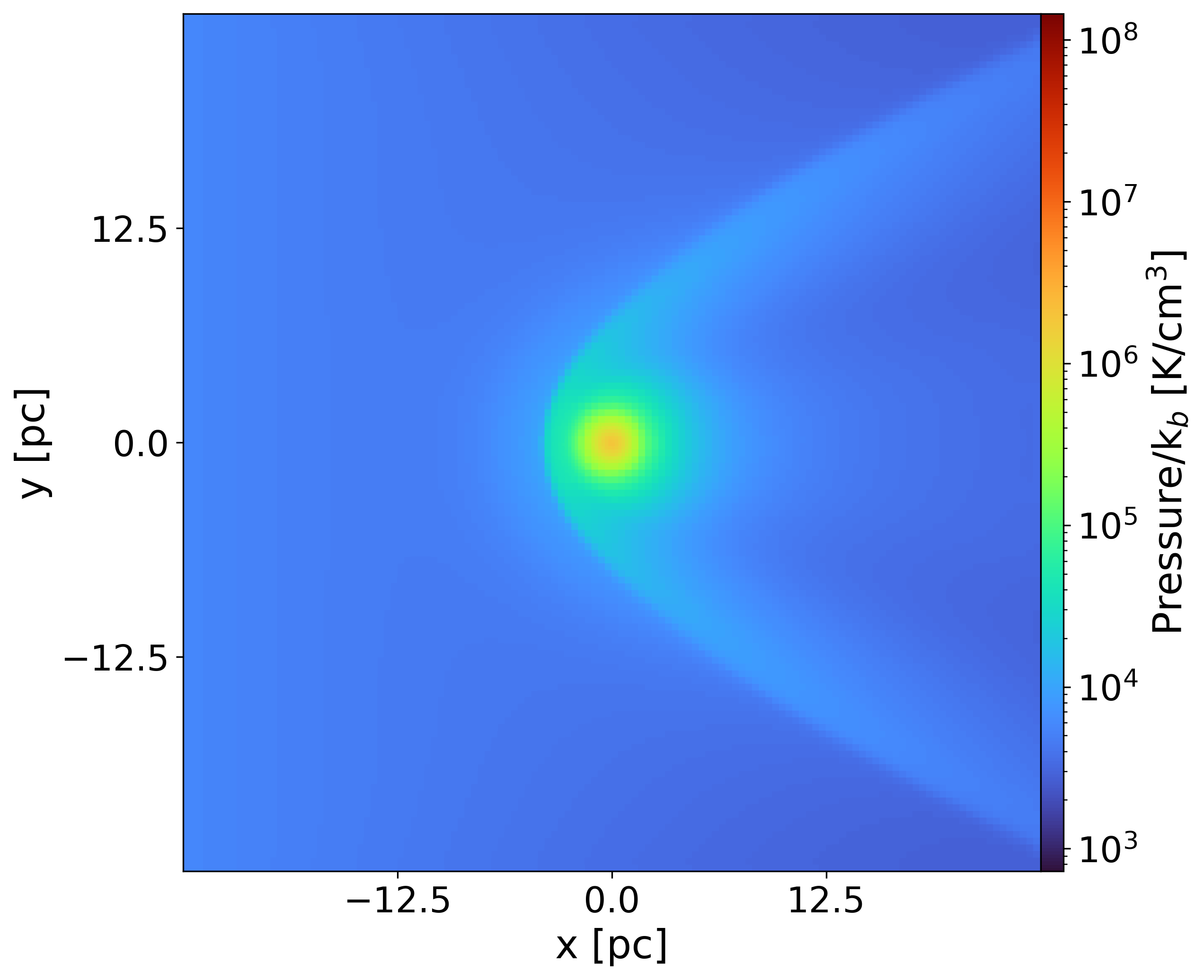}    
         \\
        \includegraphics[width=0.346\textwidth,trim={0 1cm 0 0.0cm},clip]{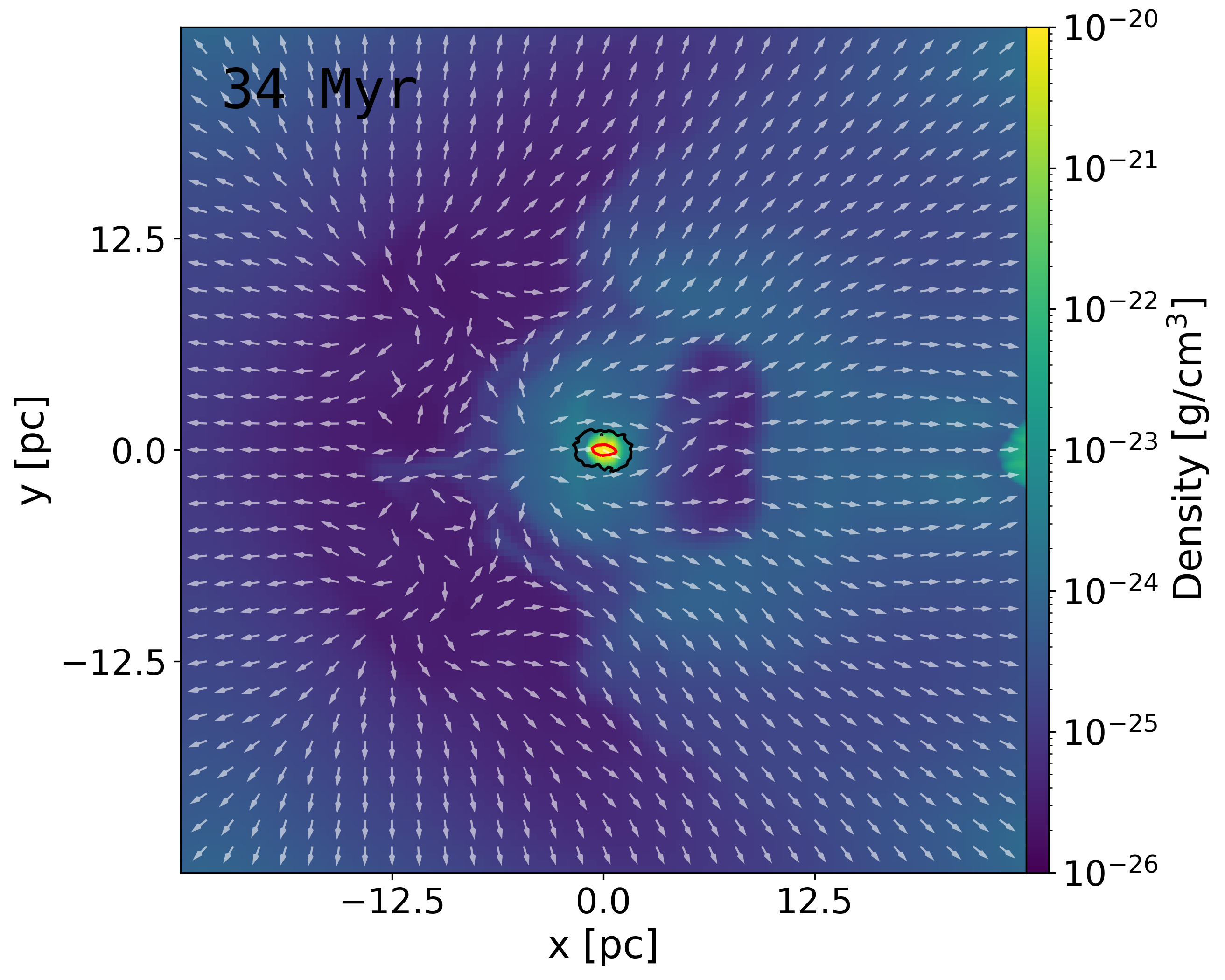}    
        \includegraphics[width=0.31\textwidth,trim={1.6cm 1cm 0 0.0cm},clip]{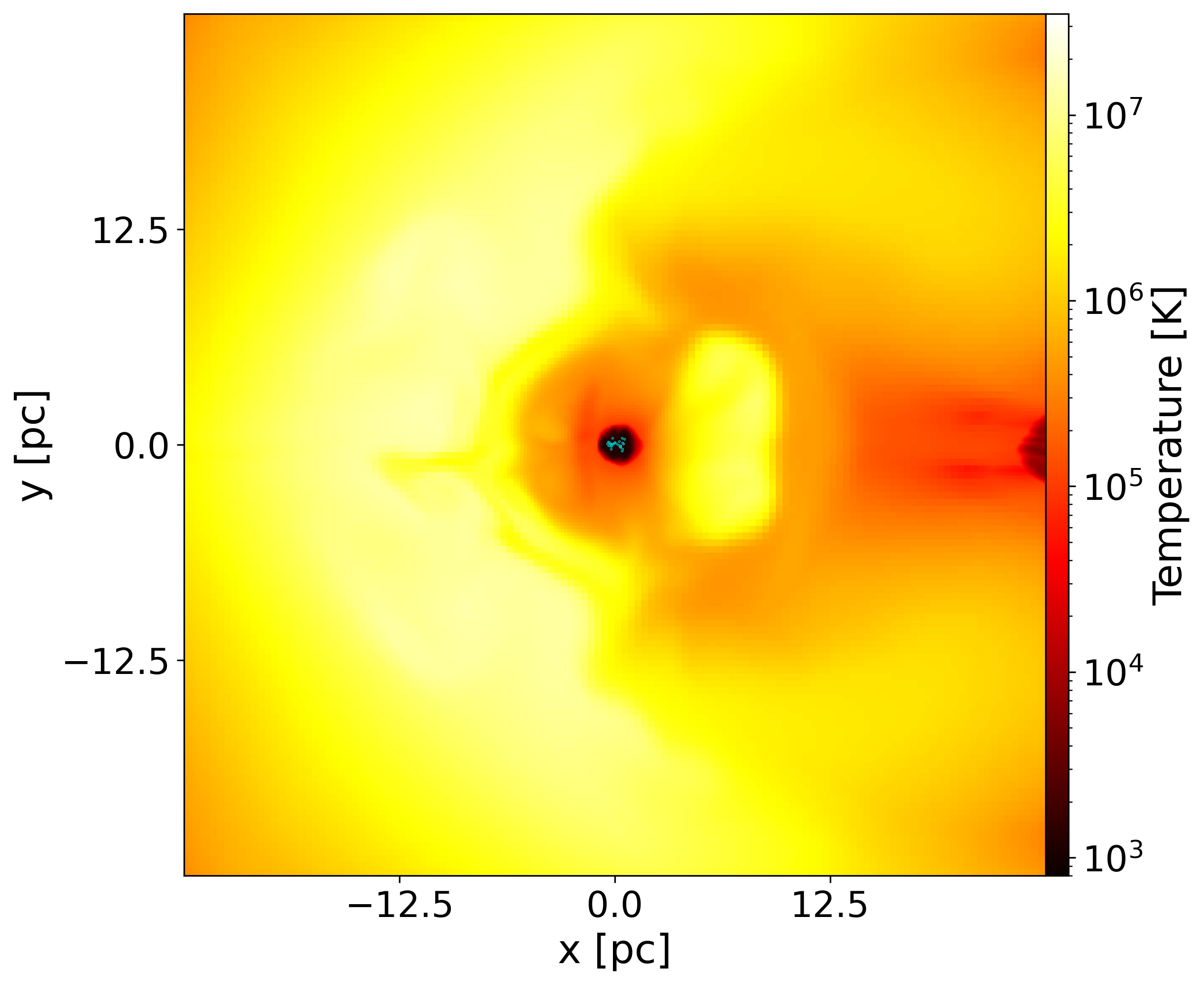}   
         \includegraphics[width=0.31\textwidth,trim={1.6cm 1cm 0 0.0cm},clip]{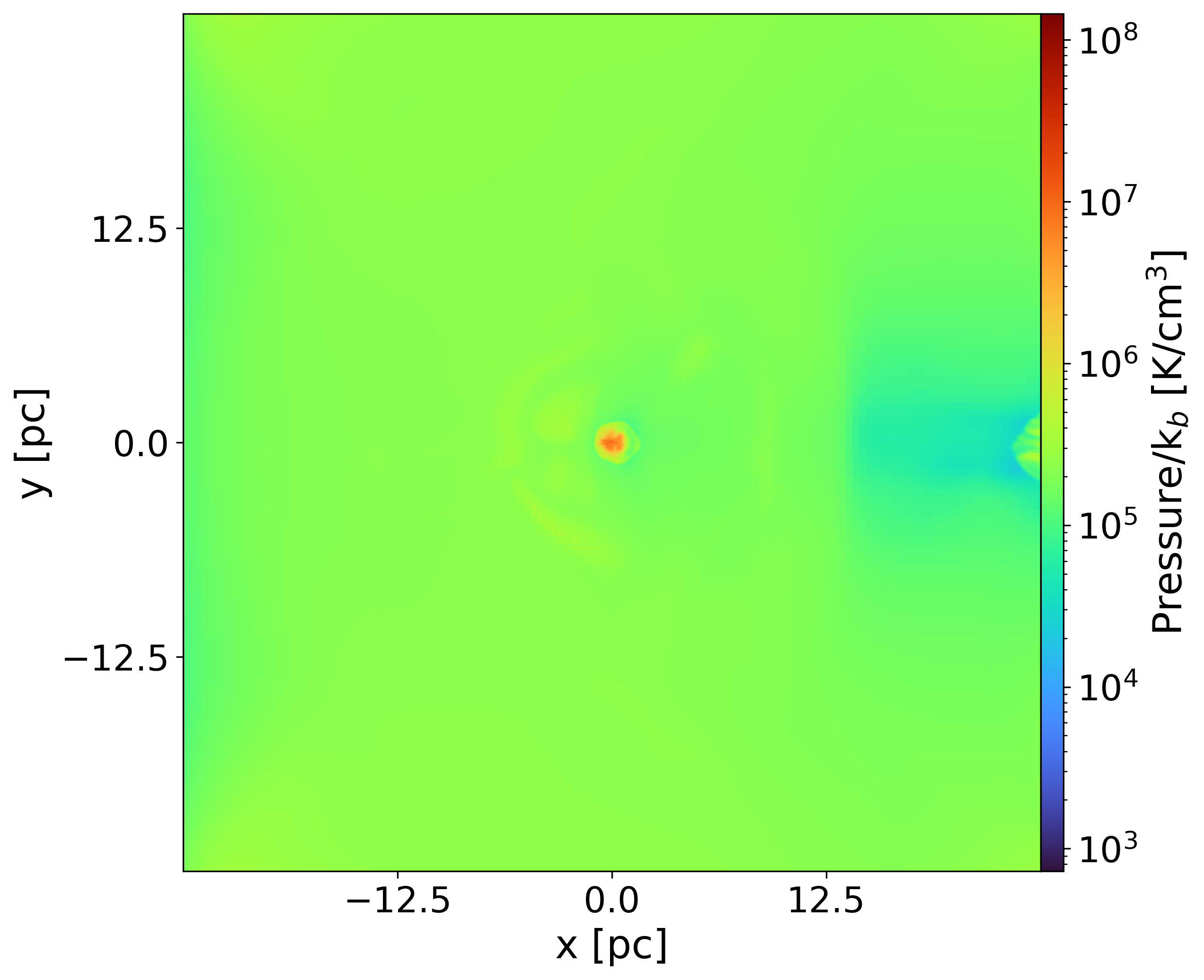}  
         \\
        \includegraphics[width=0.346\textwidth,trim={0 1cm 0 0.0cm},clip]{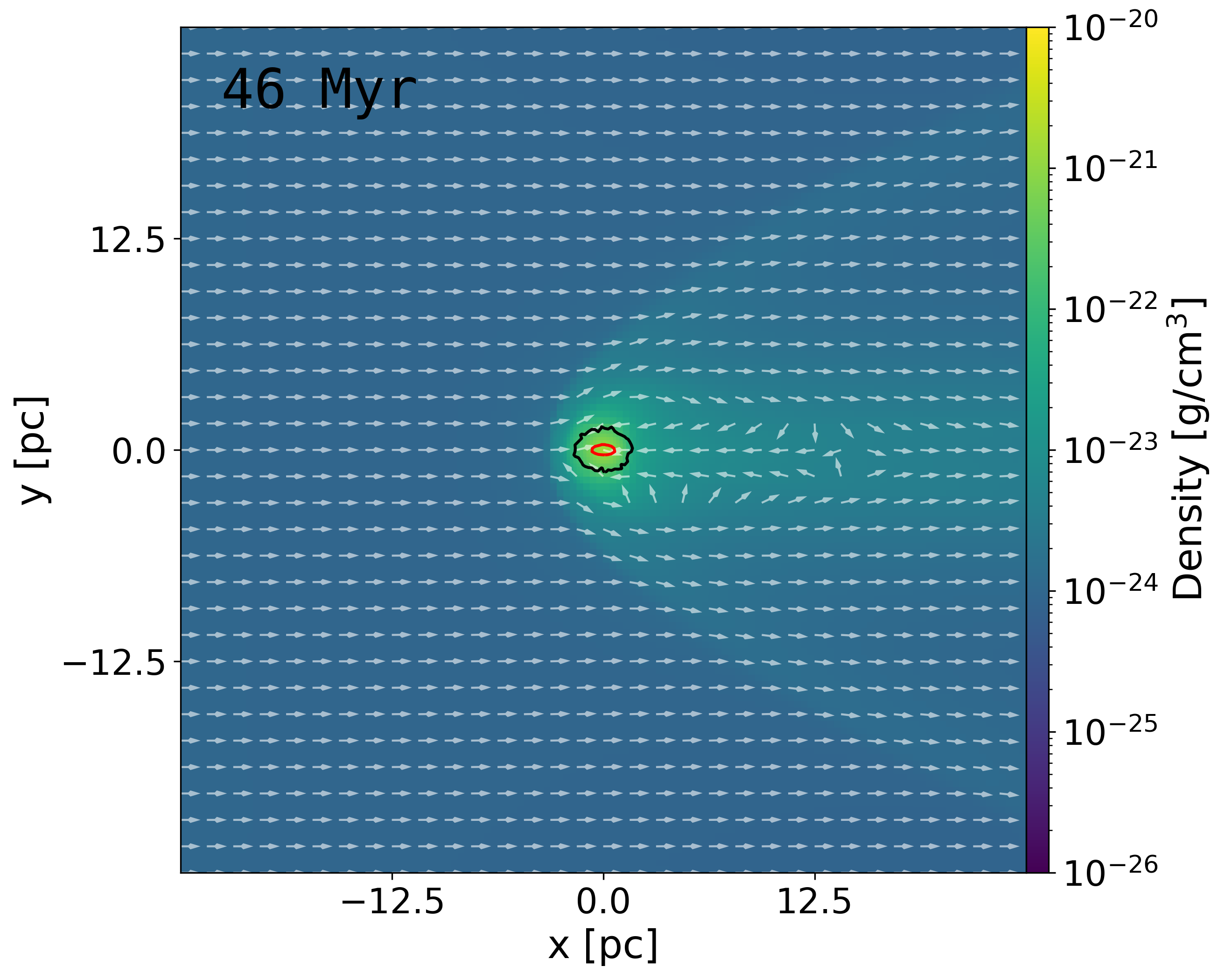}    
        \includegraphics[width=0.31\textwidth,trim={1.6cm 1cm 0 0.0cm},clip]{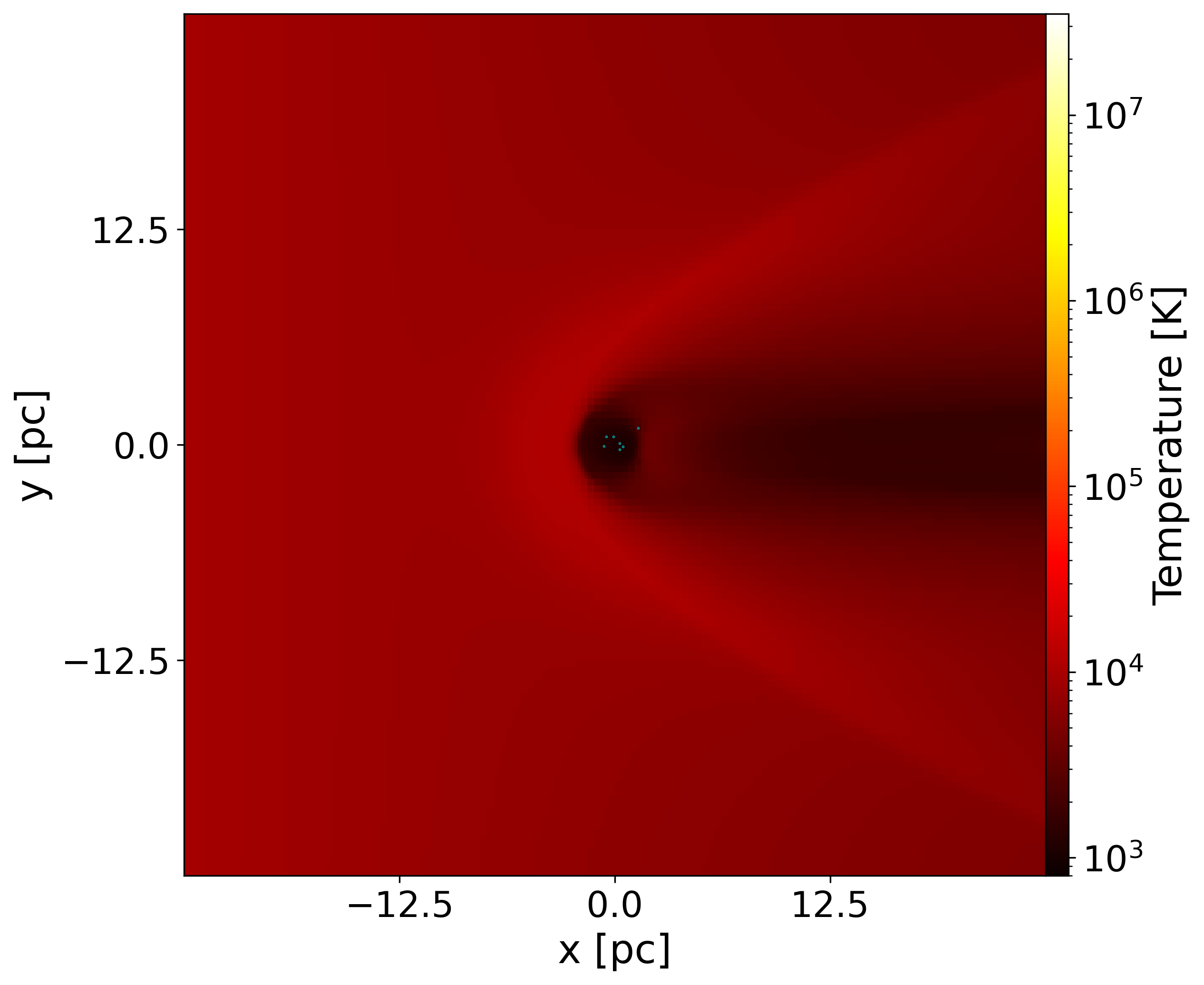}   
         \includegraphics[width=0.31\textwidth,trim={1.6cm 1cm 0 0.0cm},clip]{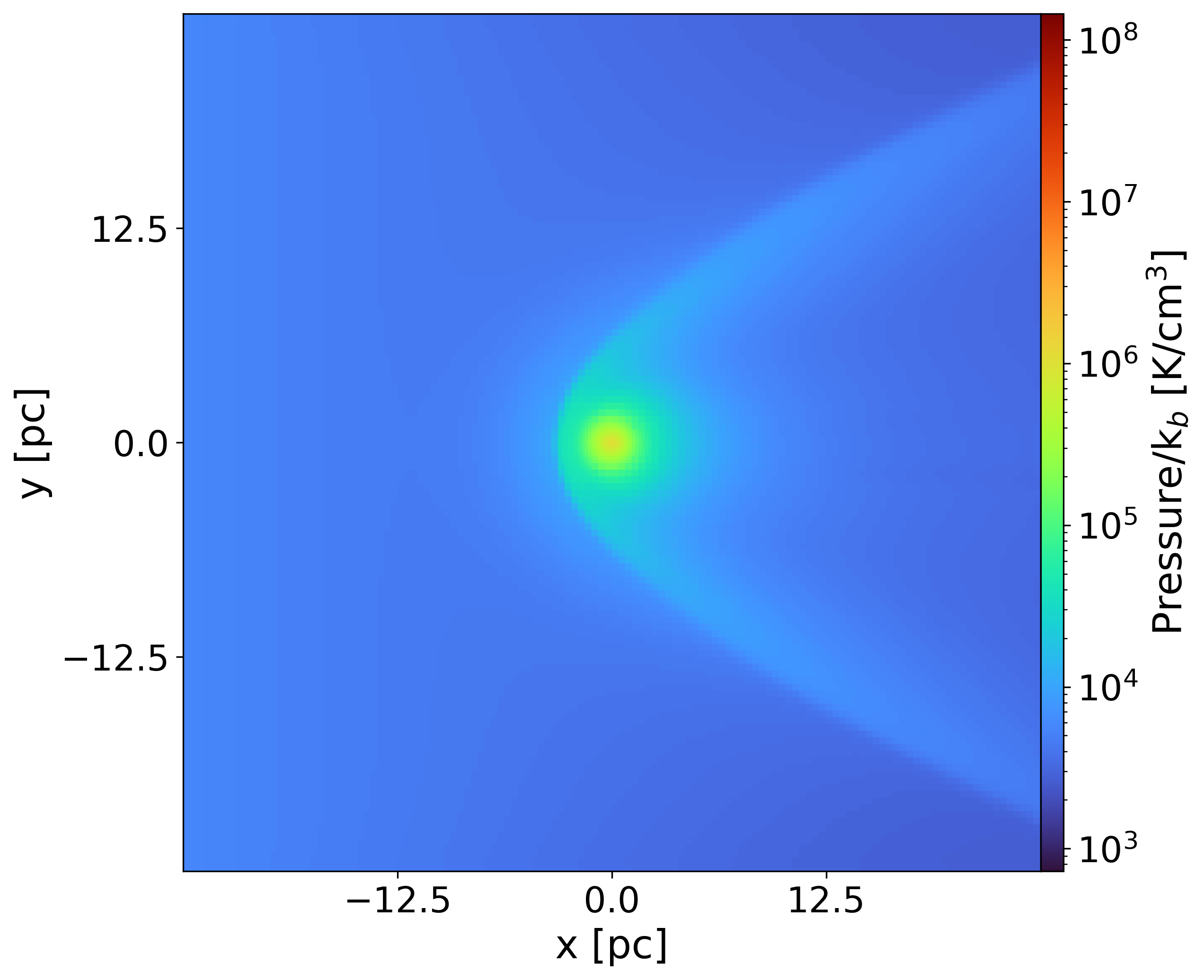}  
         \\
        \includegraphics[width=0.346\textwidth,trim={0 0cm 0 0.0cm},clip]{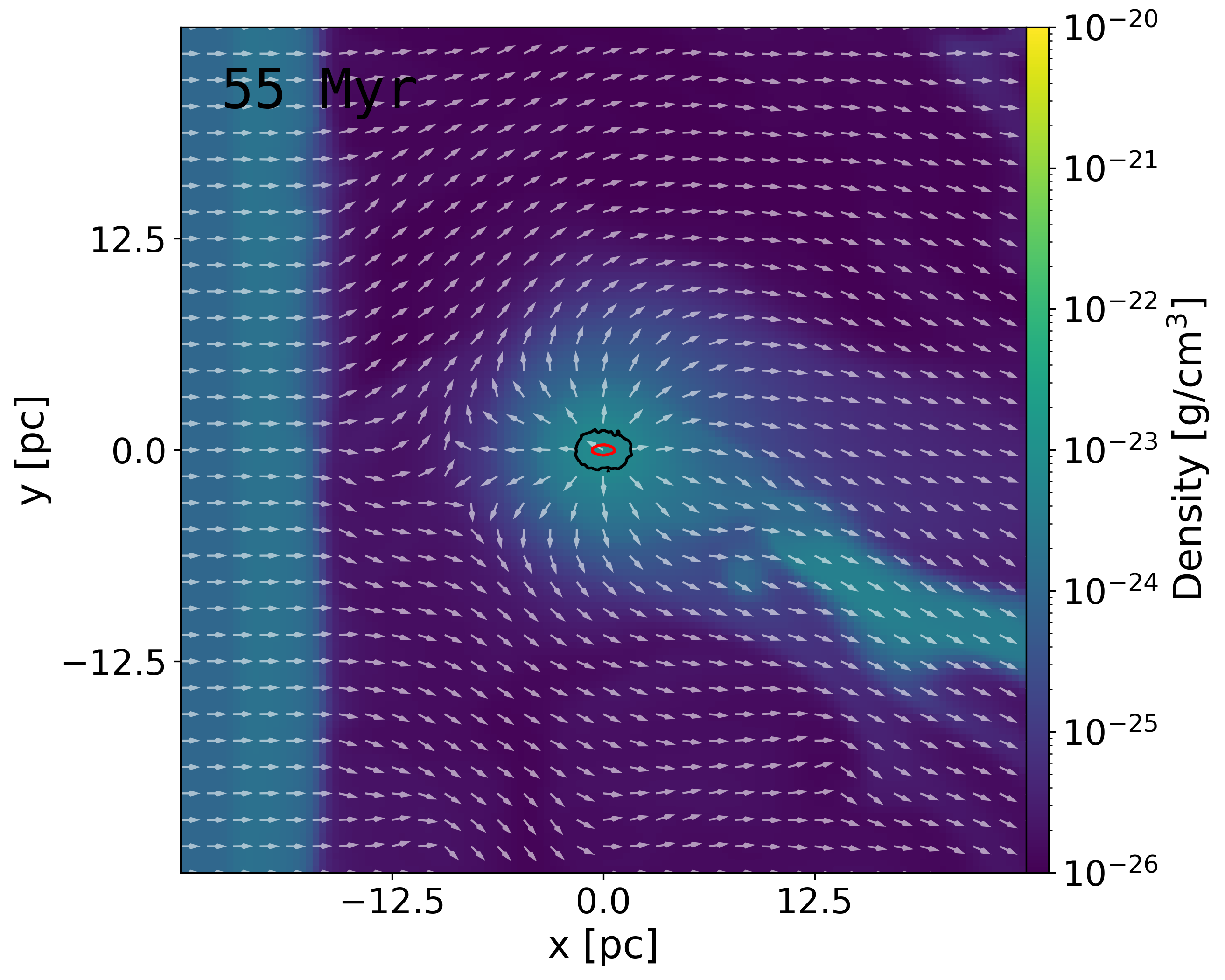}    
        \includegraphics[width=0.31\textwidth,trim={1.6cm 0cm 0 0.0cm},clip]{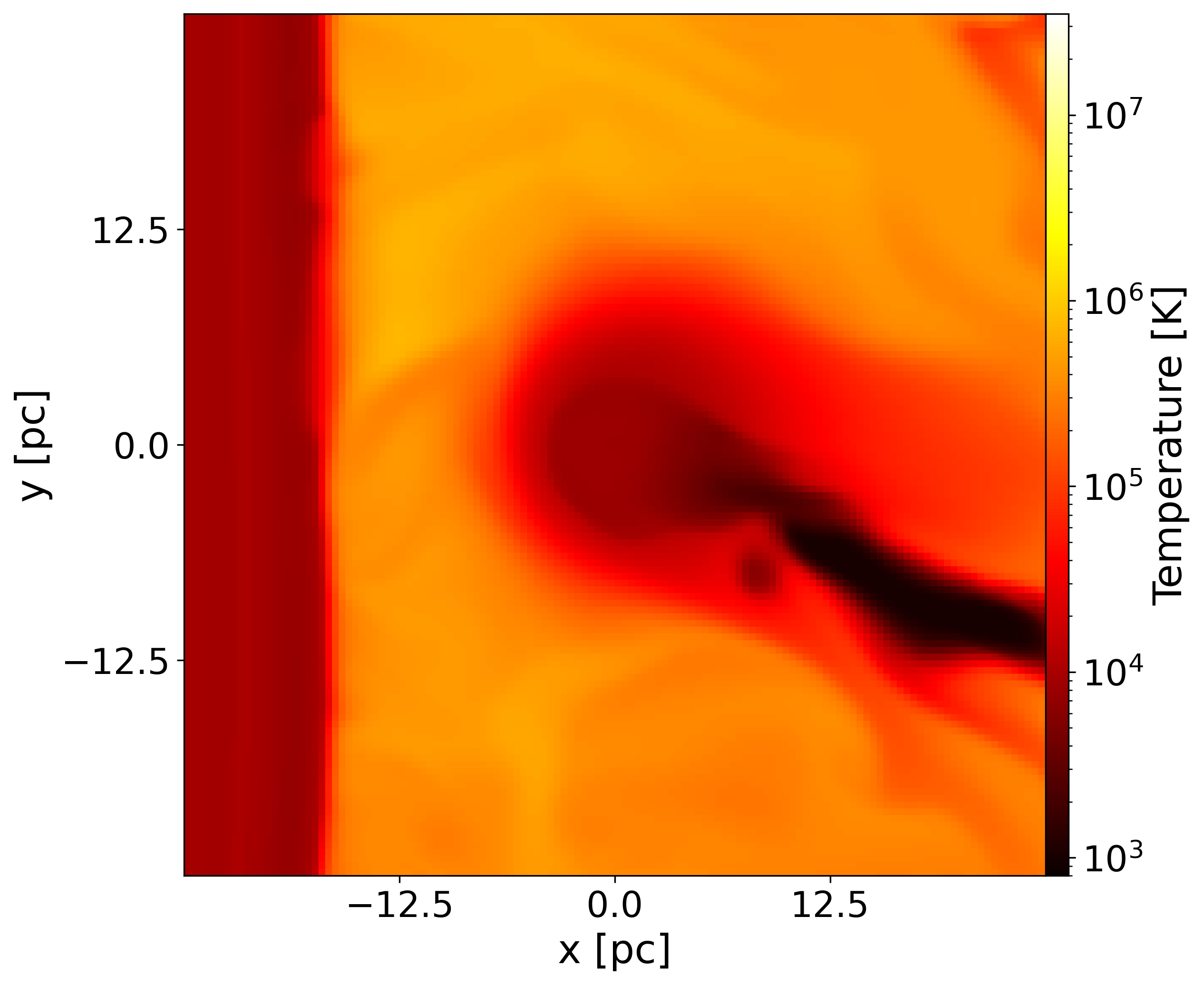}   
         \includegraphics[width=0.31\textwidth,trim={1.6cm 0cm 0 0.0cm},clip]{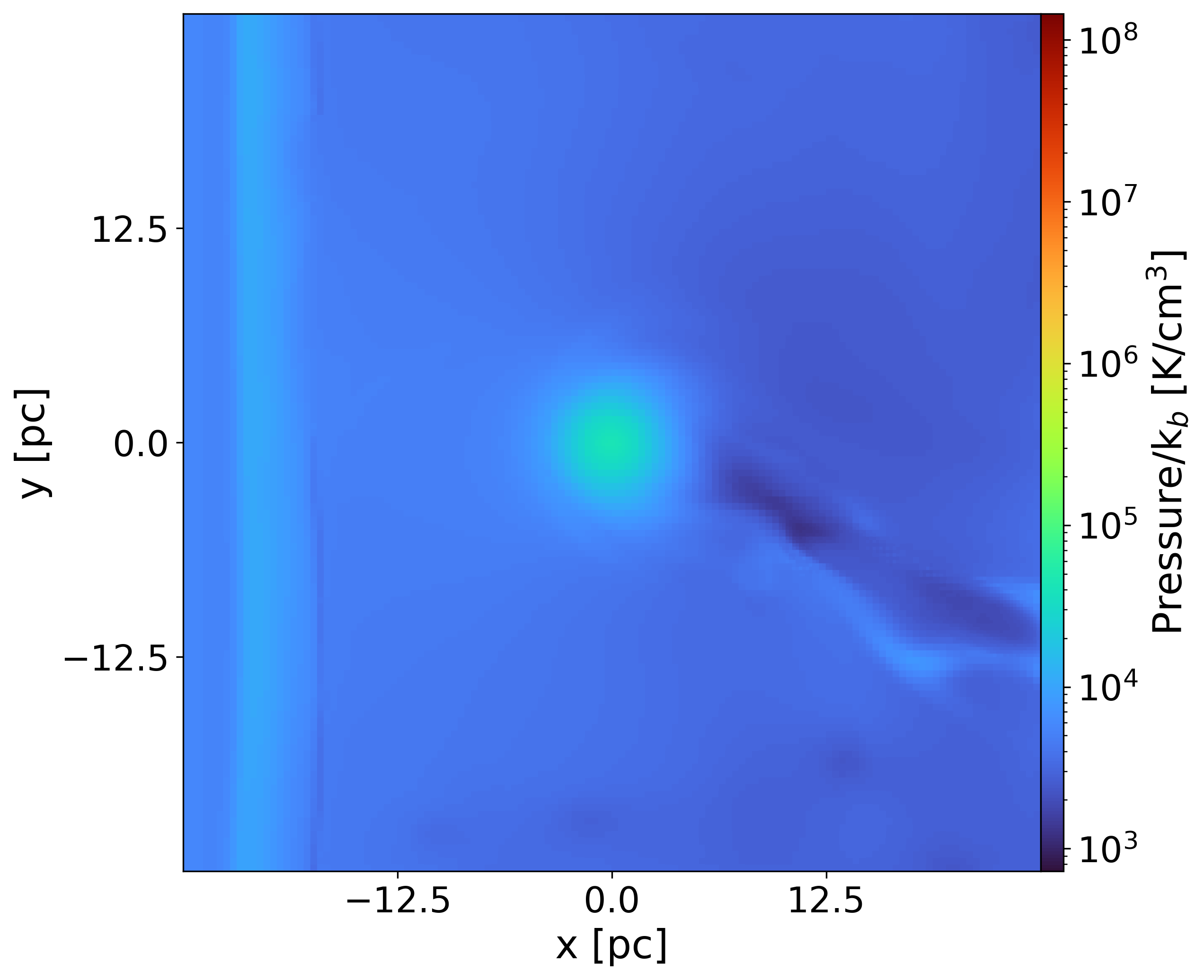}  


        \caption{Two-dimensional maps of the gas density (left-hand panels) and of the temperature (central panels) and the pressure (right-hand panels) on the x-y plane for the {\tt M5I24} simulation. The corresponding evolutionary time of each set of panels is reported in the density map, together with the velocity field denoted by the white arrows (all arrows have the same length, so no information on the intensity of the field is reported), and the contours describing the region enclosing the  75\% and 95\% of the SG mass in red and black, respectively. The cyan points on the temperature map represent newborn stars (age <0.05\Myr).}
  \label{fig:gas_map_M5I24}
\end{figure*}  

\begin{figure*}
        \centering

        \includegraphics[width=0.268\textwidth,trim={0 0cm 2.935cm 0.0cm},clip]{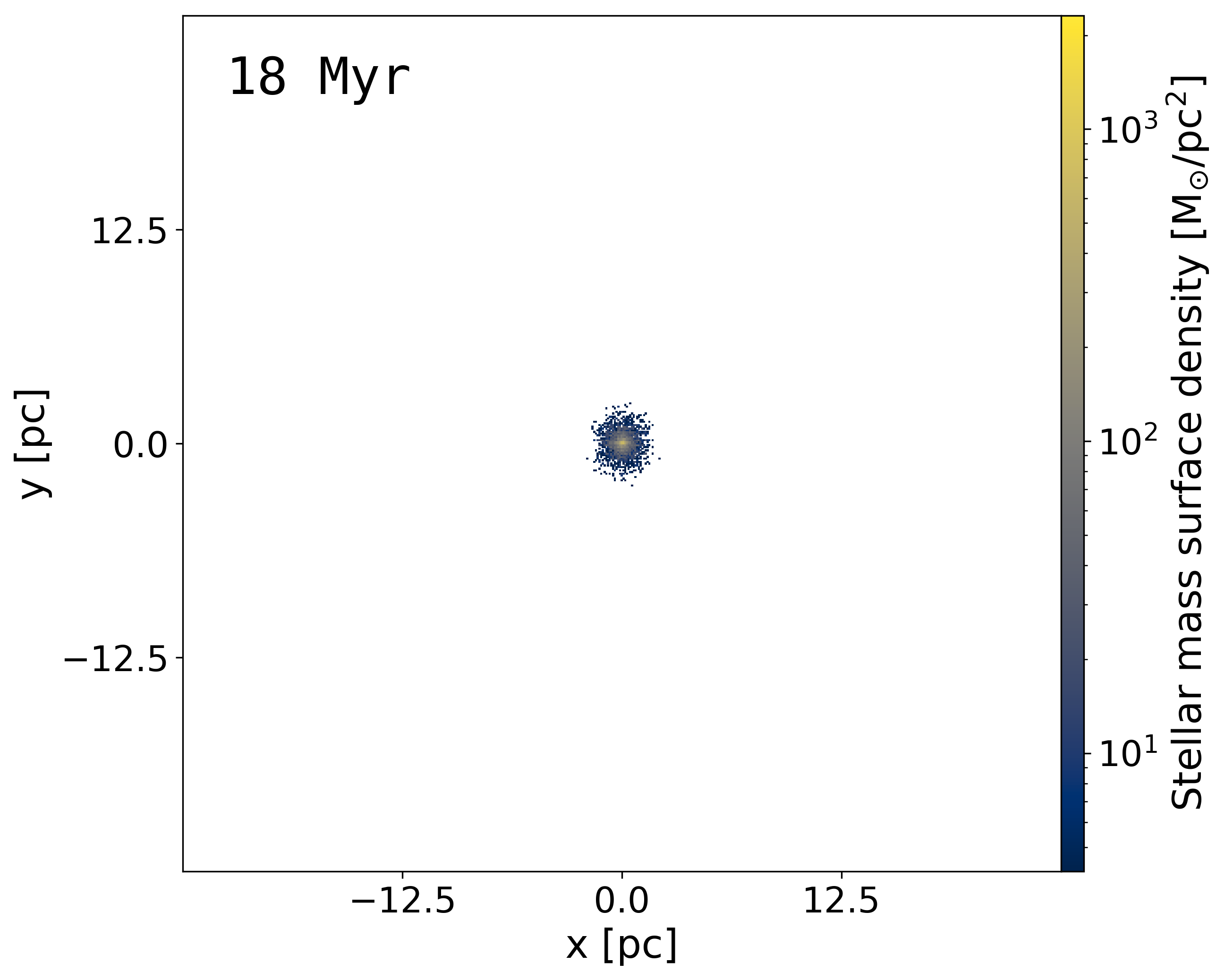}    
        \includegraphics[width=0.225\textwidth,trim={3.1cm 0cm 2.935cm 0.0cm},clip]{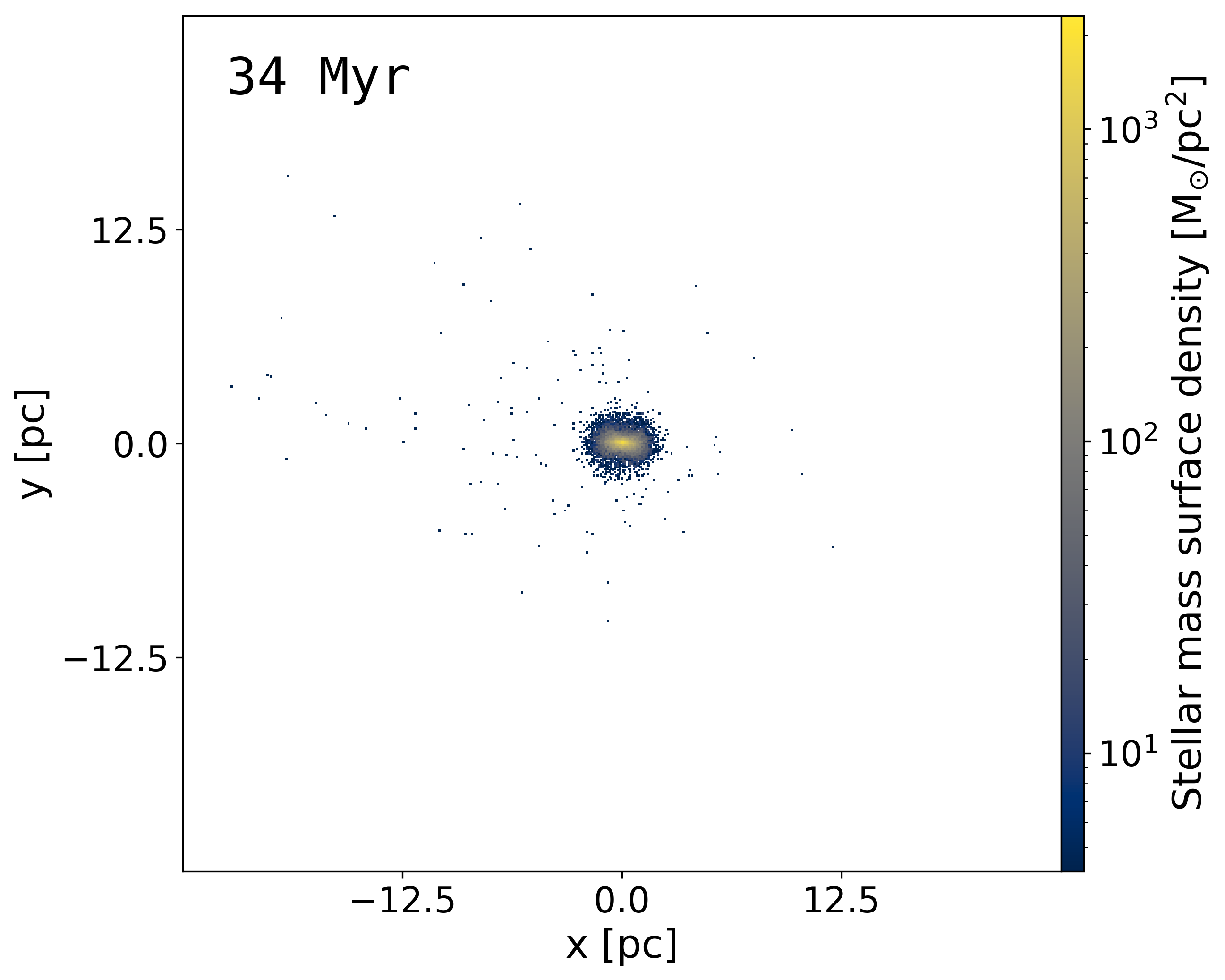}  
          \includegraphics[width=0.225\textwidth,trim={3.1cm 0cm 2.935cm 0.0cm},clip]{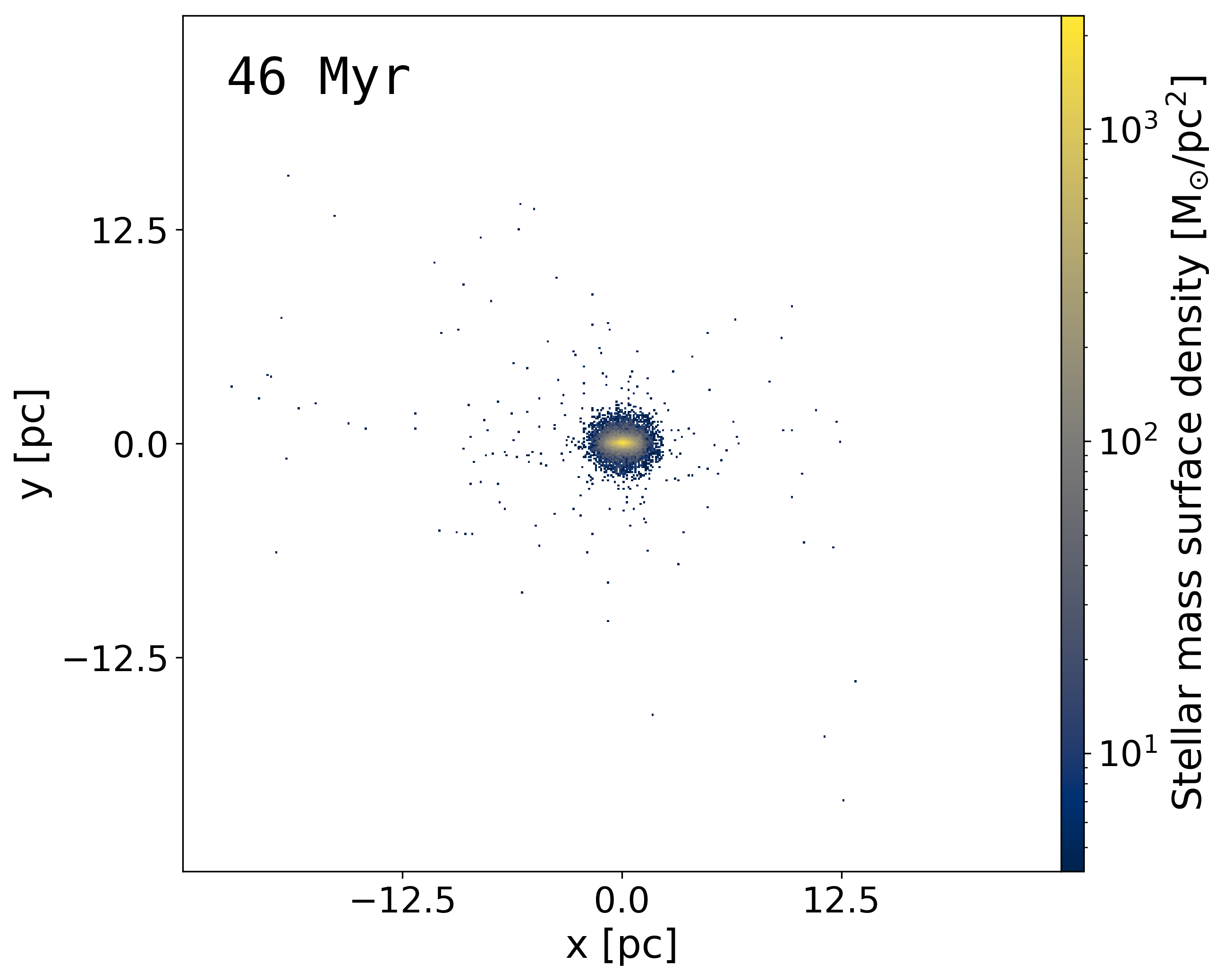}    
        \includegraphics[width=0.267\textwidth,trim={3.1cm 0cm 0cm 0.0cm},clip]{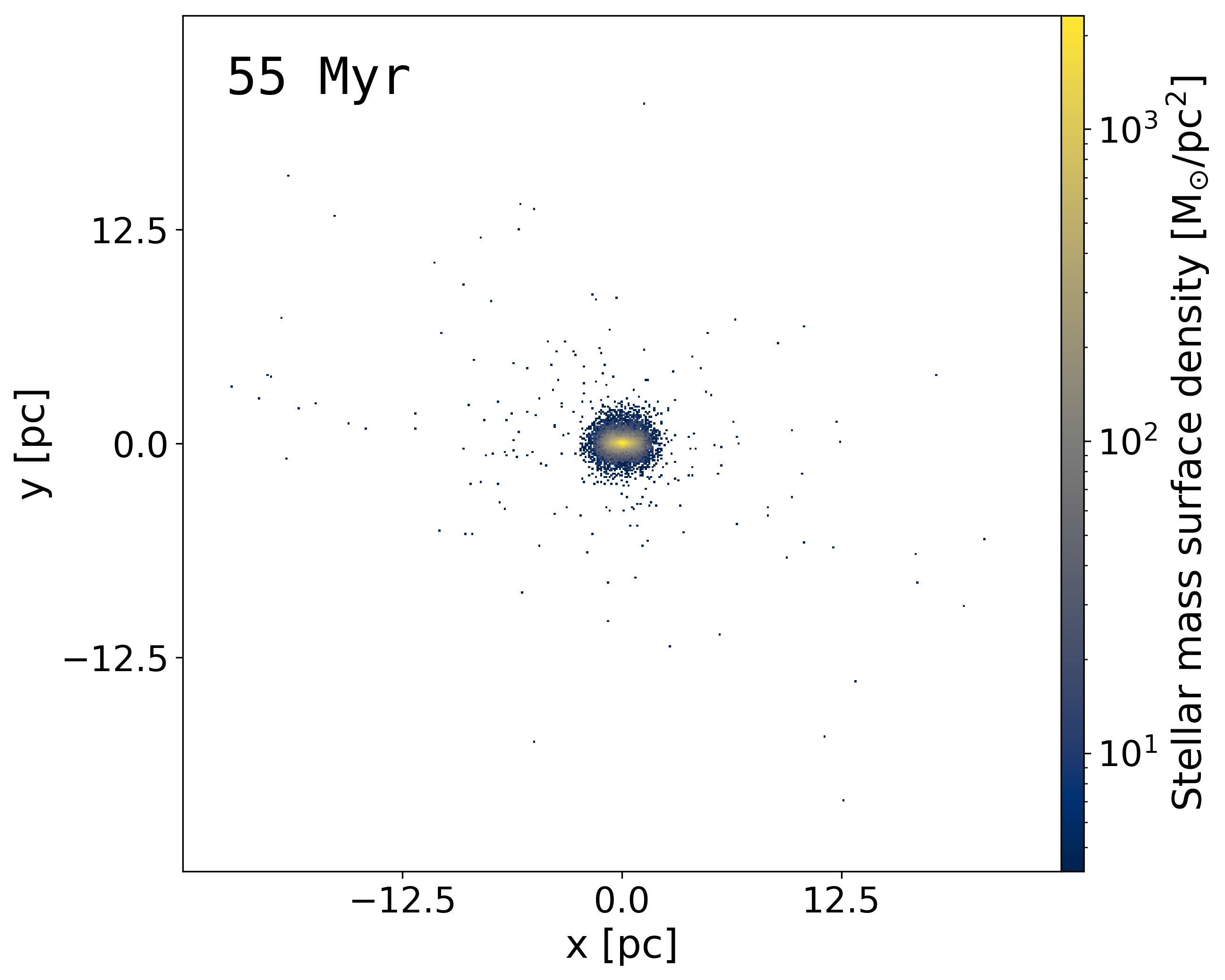}
        
        \caption{Two-dimensional mass surface density maps of the stellar component at four evolutionary times for the {\tt M5I24} simulation on the x-y plane.}
  \label{fig:part_map_M5I24}
\end{figure*}

\begin{figure*}
\centering
        \includegraphics[width=0.44\textwidth,trim={0cm 0cm 0cm 0.0cm},clip]{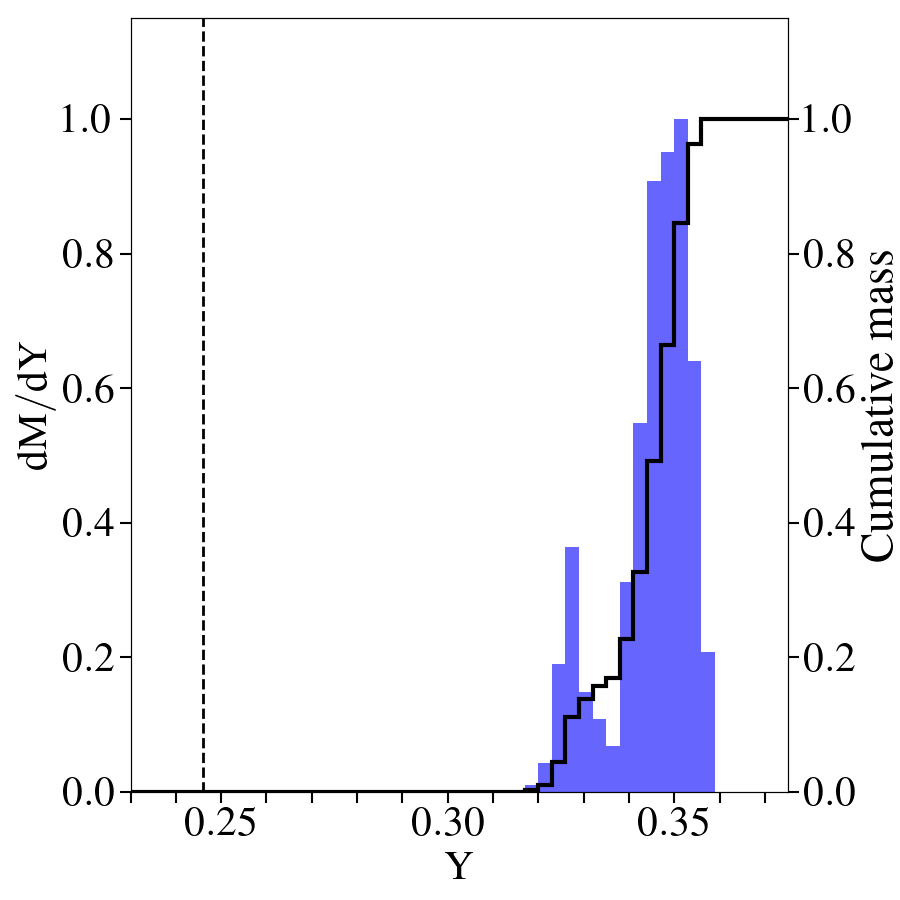}   
        \includegraphics[width=0.44\textwidth,trim={0cm 0cm 0 0.0cm},clip]{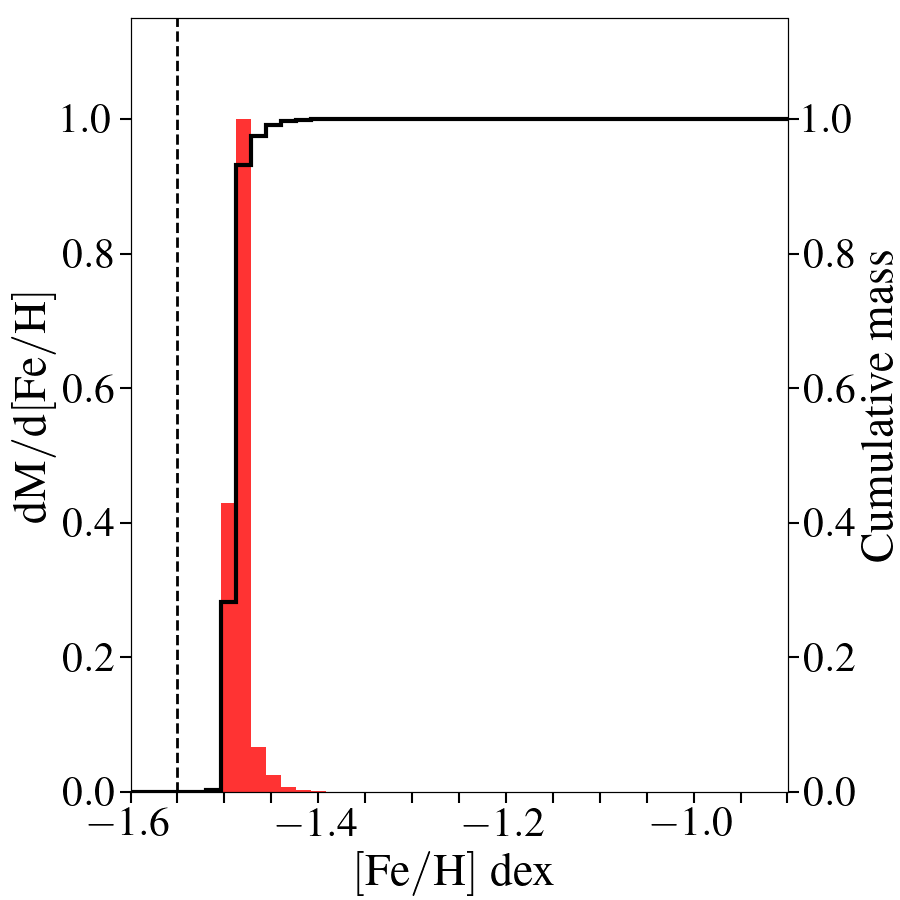} 
        \caption{The mass distribution of Y (left) and [Fe/H] ratio (right) of the SG stars at the same time of the profiles. The distributions have been derived by summing the stellar masses in each bin and then normalizing every distribution to its maximum value. The black dashed lines represent the pristine gas composition both for Y and [Fe/H] ratio while the solid black line represents the normalized cumulative mass.}
  \label{fig:prof_M5I24}
\end{figure*}

\section{Results}
\label{sec:results}
In this paper, we present the results for three models, all of them taking into account the feedback from Type Ia SNe.
We have performed two runs, the M6I23 and the M6I24, varying only
the density of the pristine gas (see Table \ref{tab:simu} for the details of all the
models) for the cluster with an initial mass of $10^6\Msun$, while for the cluster with mass $10^5\Msun$, we present here only the model M5I24, with a low-density ISM gas.

\subsection{Low-density models}

\begin{figure*}
  \includegraphics[width=0.95\linewidth]{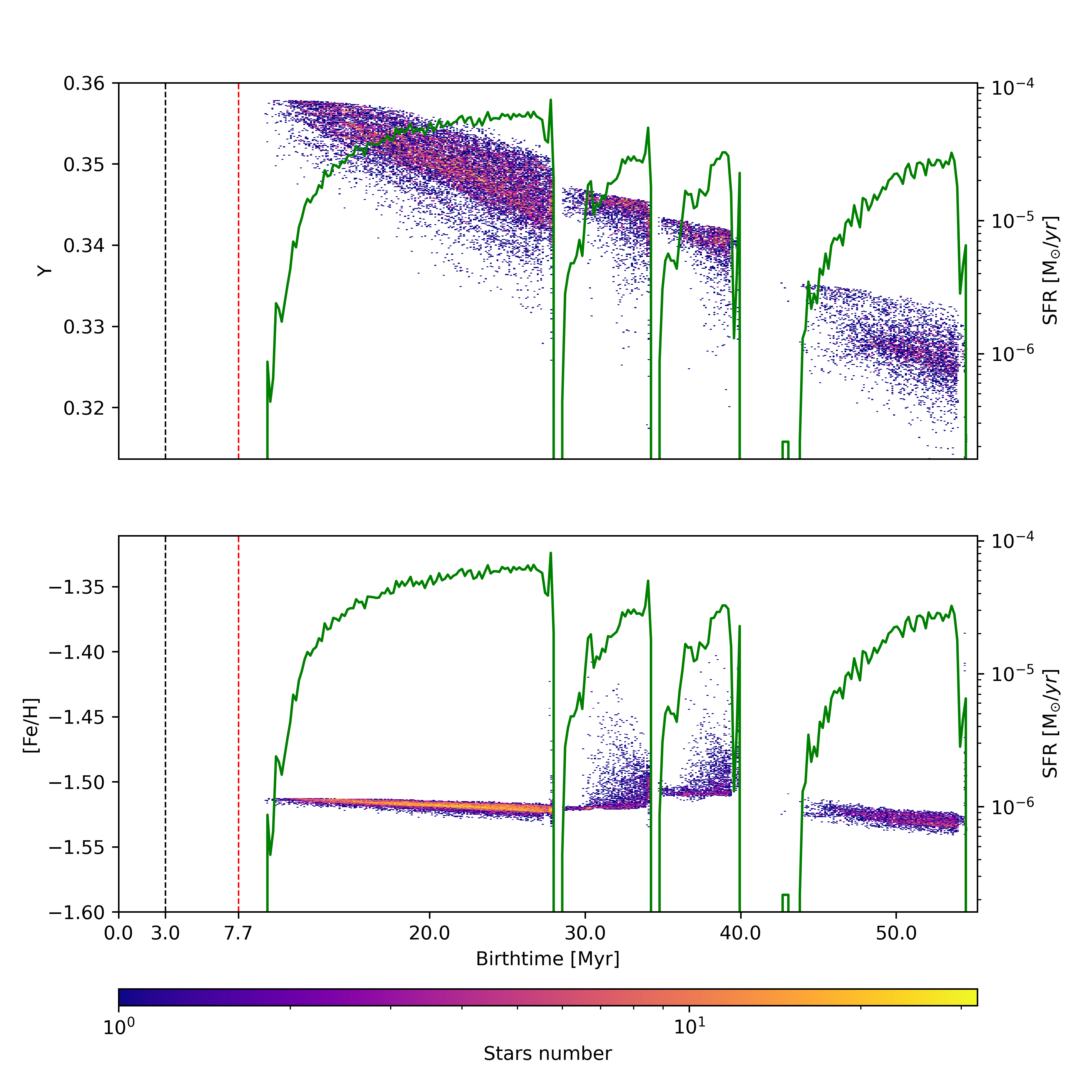}
  \caption{Stellar helium mass fraction (\textit{Y}) as a function of the stellar birthtime for the {\tt M5I24} model. The color scale indicates the number of SG stars. The black and red dashed vertical lines represent the time at which the infall and the AGB ejecta start to pollute the system, respectively (see \autoref{tab:simu}). The green solid line represents the star formation rate.}
  \label{fig:Y_birth_M5I24}
\end{figure*}

\begin{figure*}
        \centering

        \includegraphics[width=0.346\textwidth,trim={0 1cm 0 0.0cm},clip]{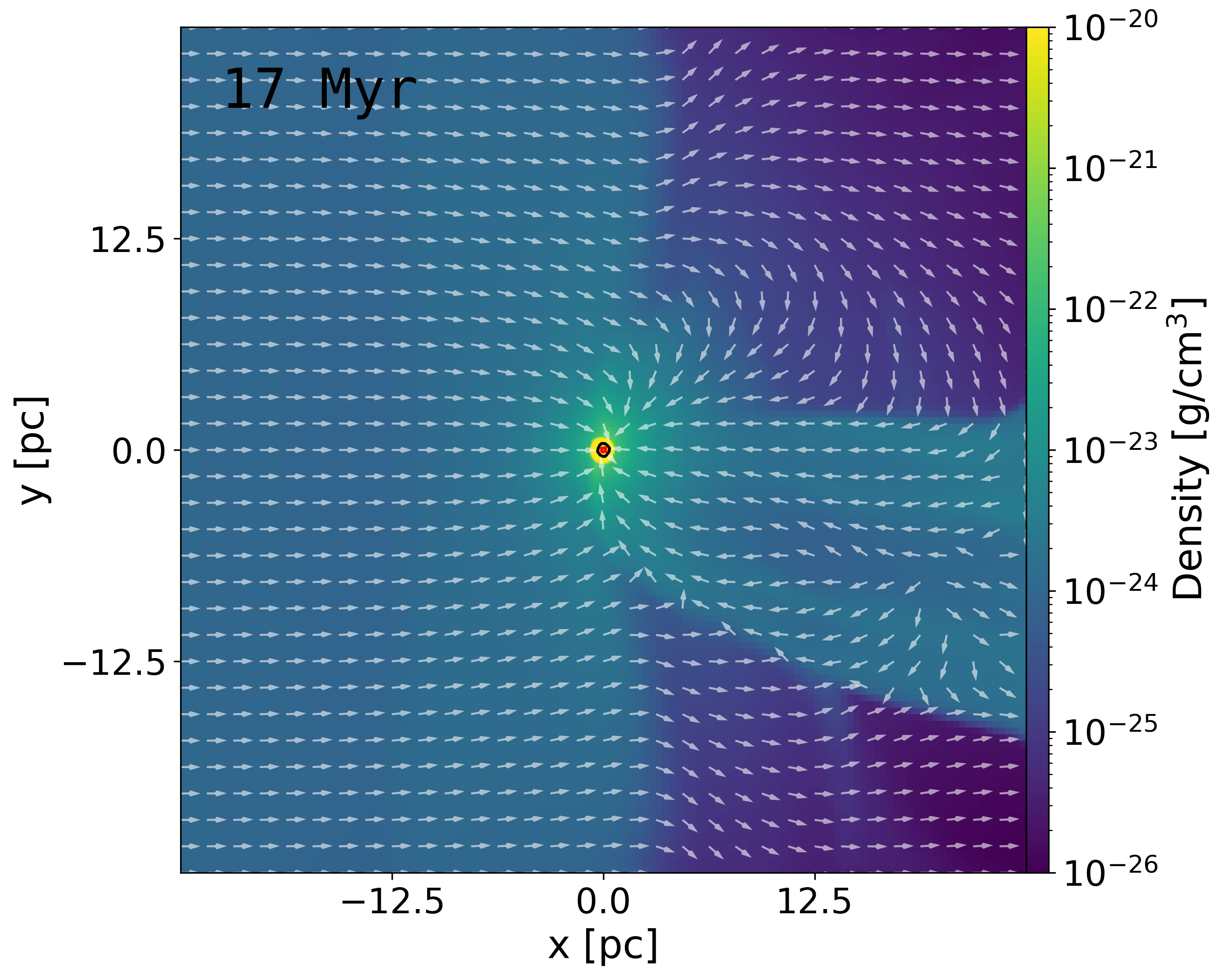}    
        \includegraphics[width=0.31\textwidth,trim={1.6cm 1cm 0 0.0cm},clip]{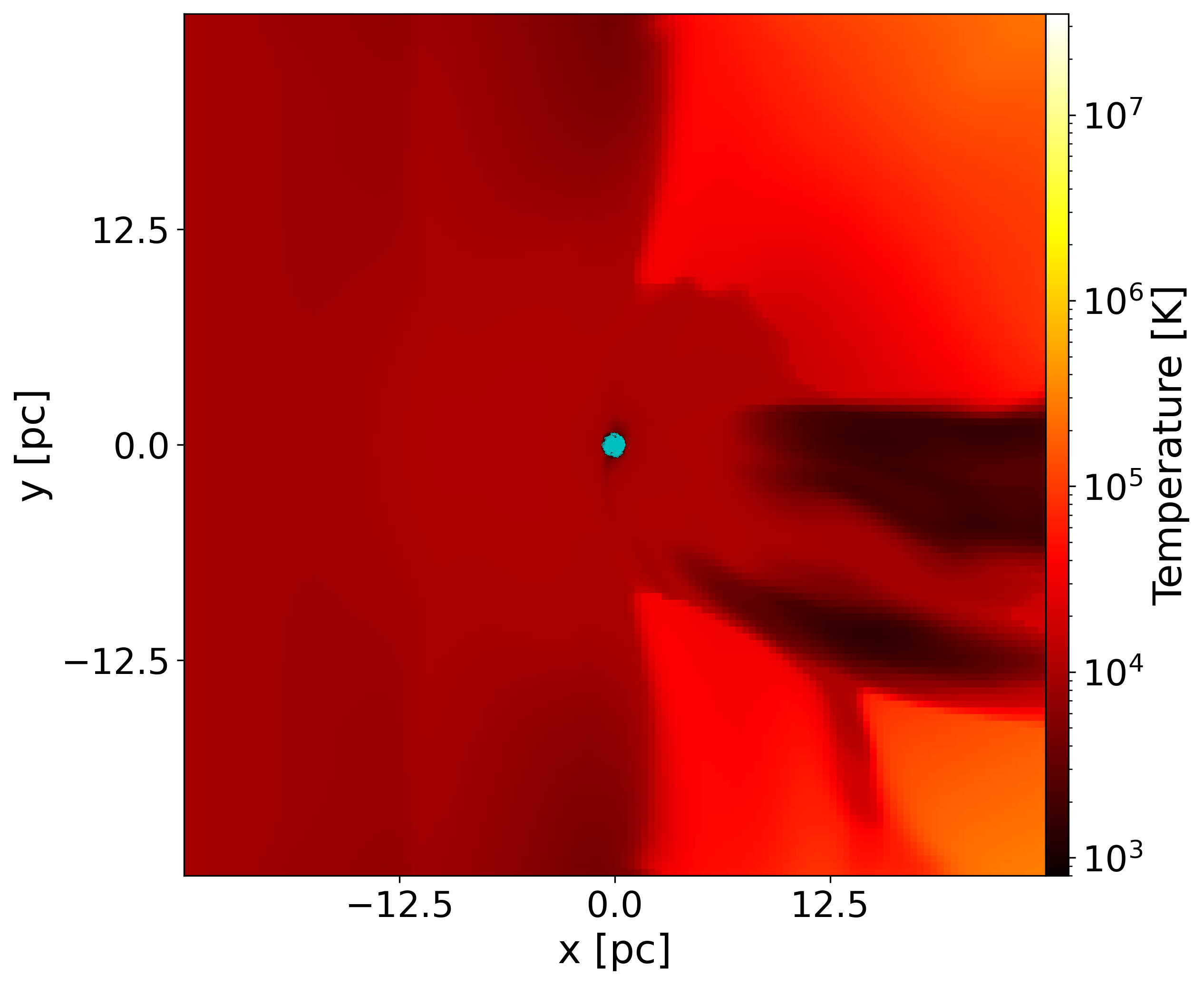}   
         \includegraphics[width=0.31\textwidth,trim={1.6cm 1cm 0 0.0cm},clip]{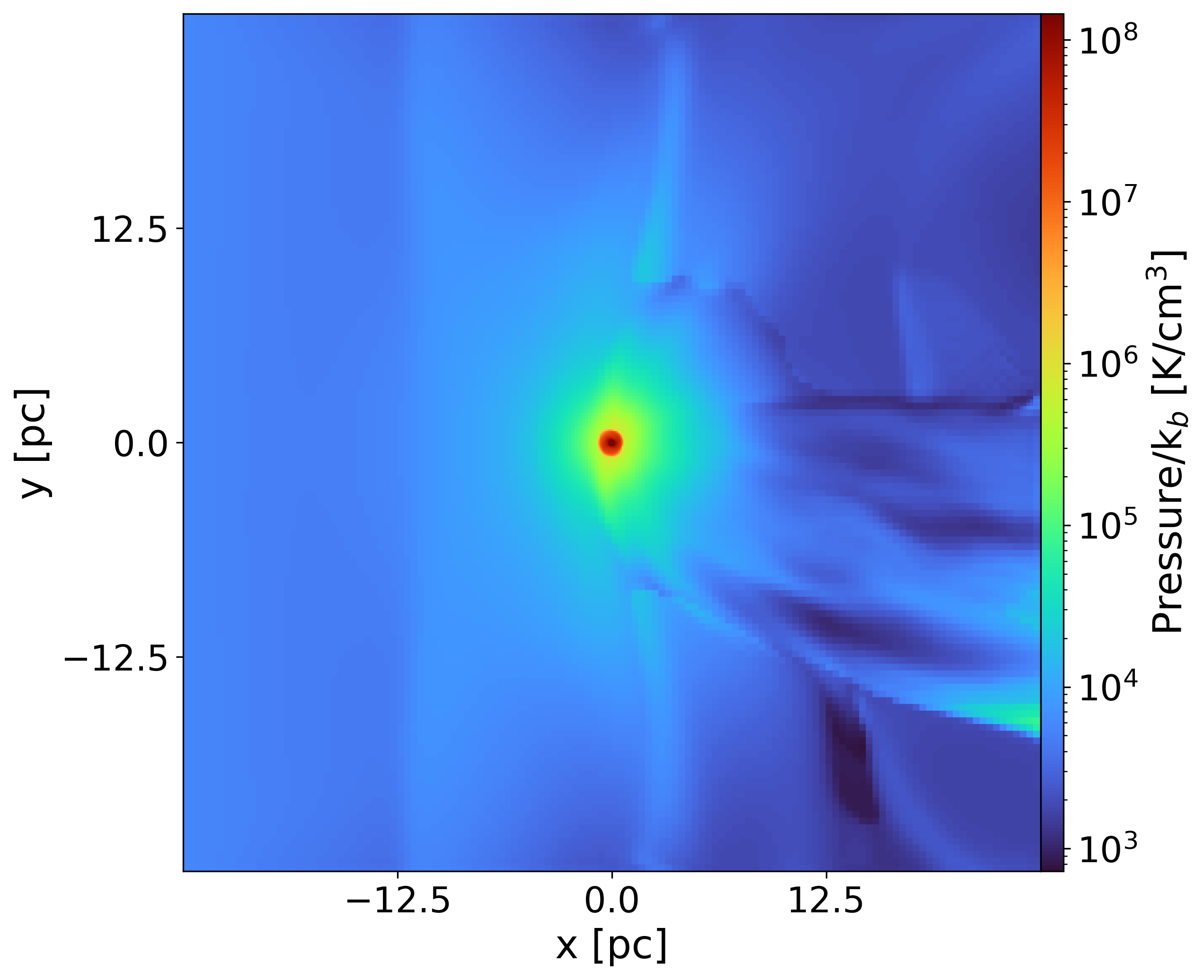}    
         \\
         \includegraphics[width=0.346\textwidth,trim={0 1cm 0 0.0cm},clip]{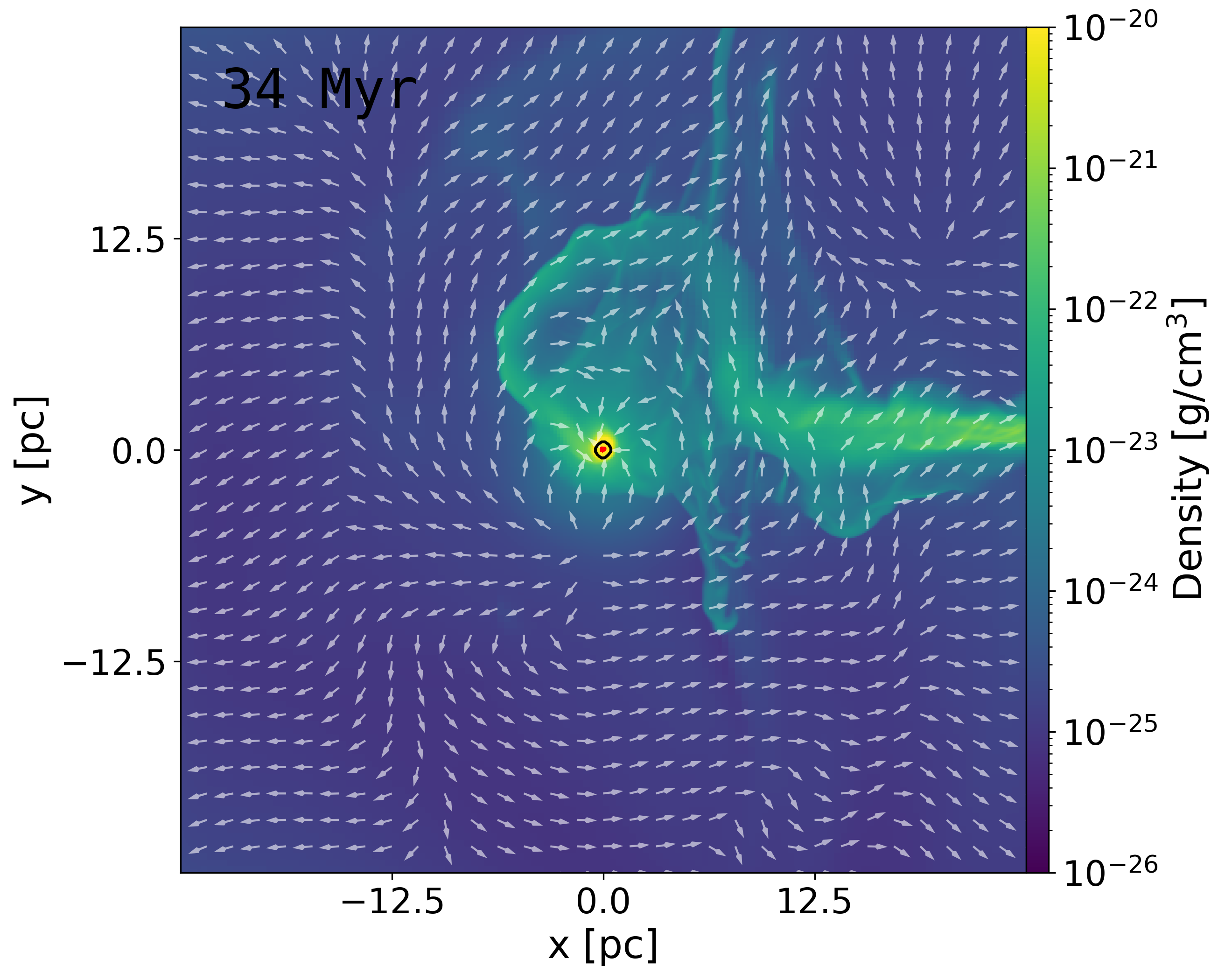}    
        \includegraphics[width=0.31\textwidth,trim={1.6cm 1cm 0 0.0cm},clip]{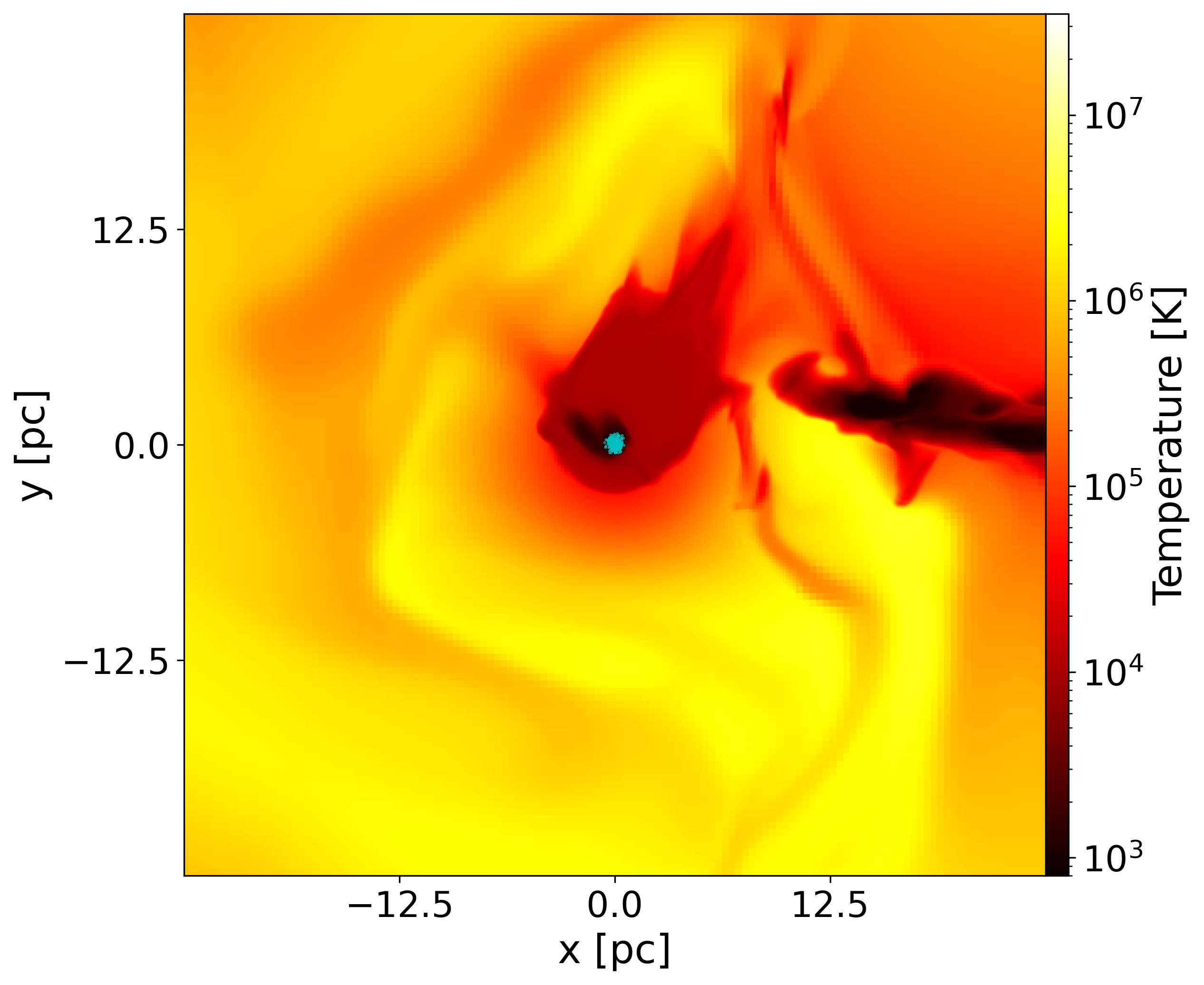}   
         \includegraphics[width=0.31\textwidth,trim={1.6cm 1cm 0 0.0cm},clip]{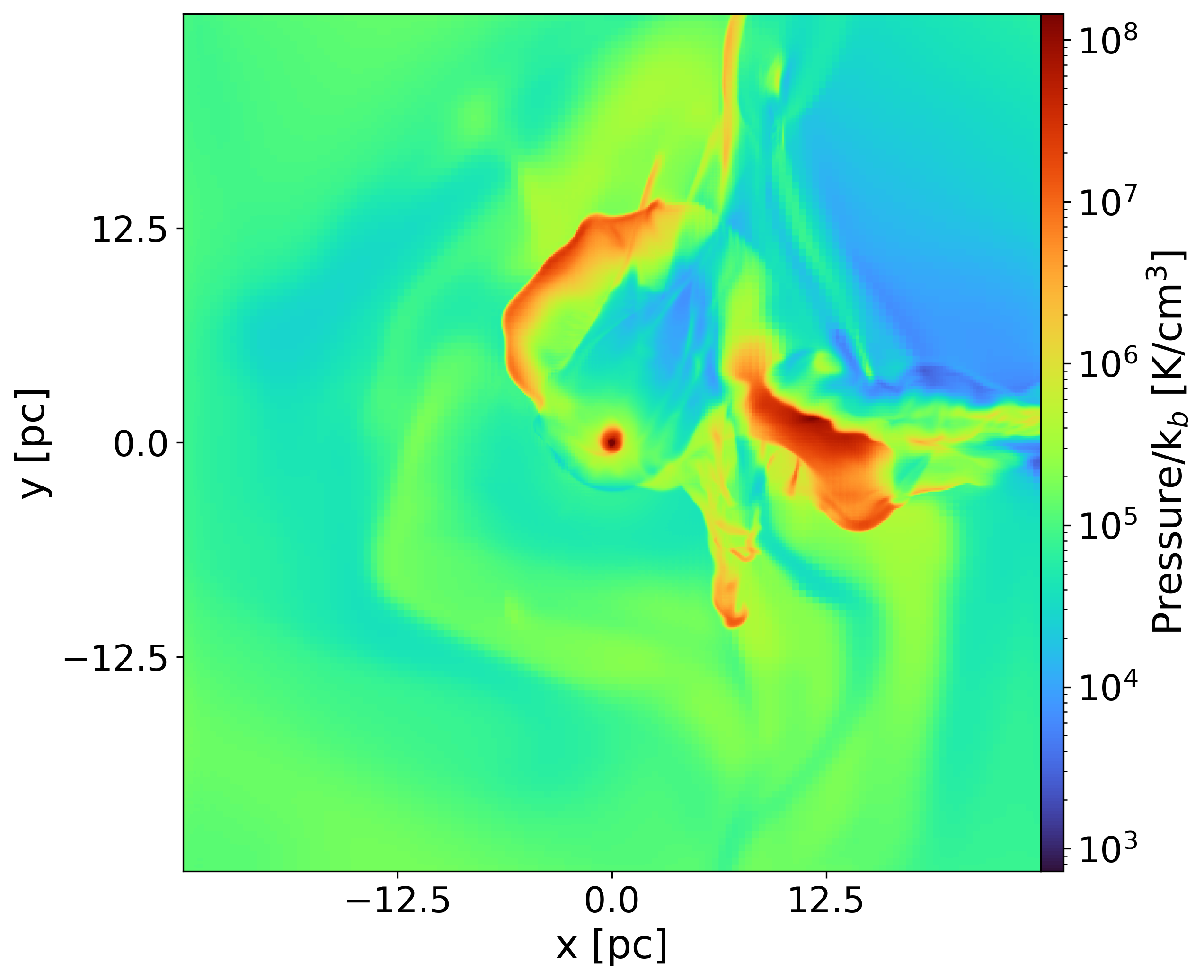}   
         \\
       \includegraphics[width=0.346\textwidth,trim={0 1cm 0 0.0cm},clip]{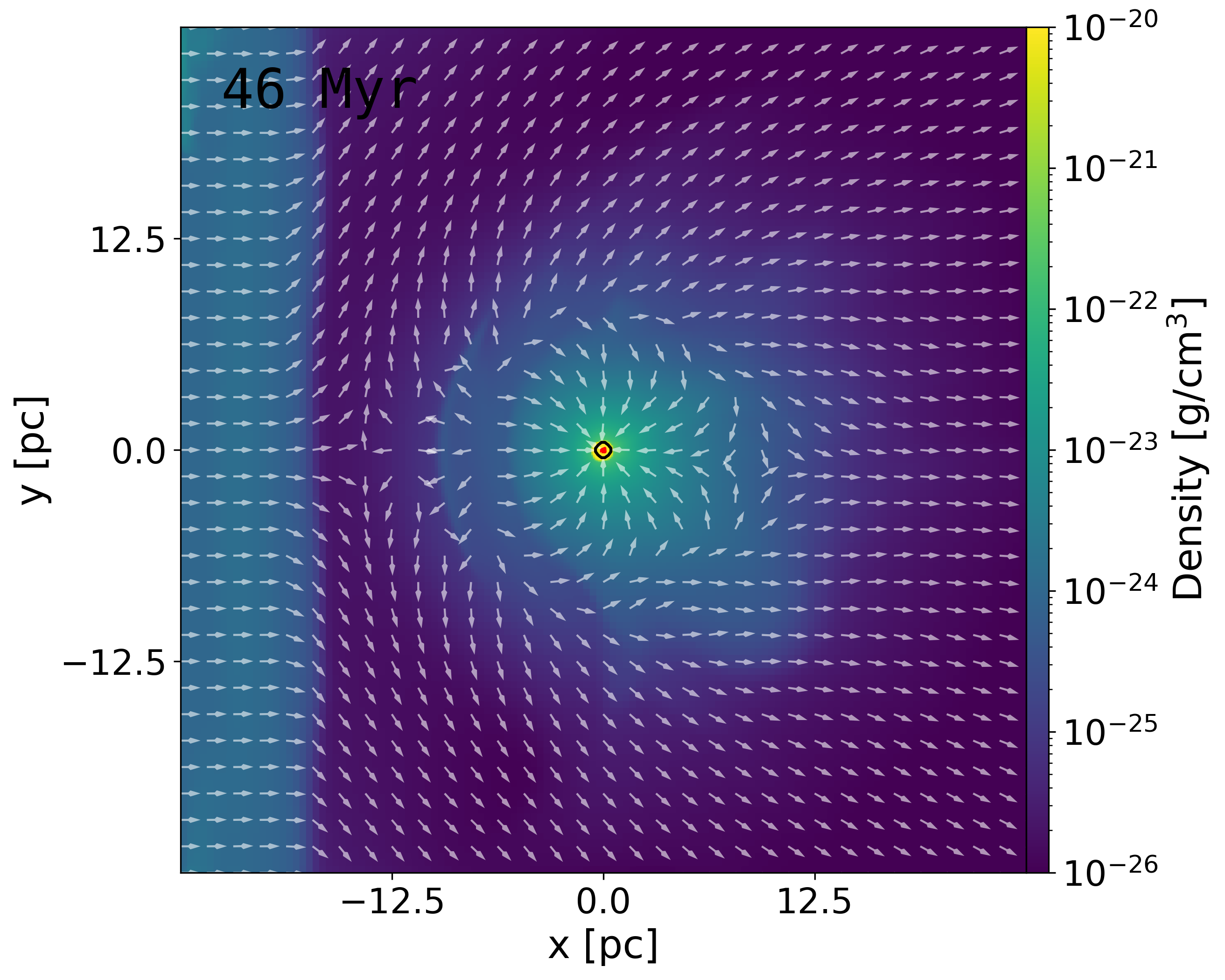}    
        \includegraphics[width=0.31\textwidth,trim={1.6cm 1cm 0 0.0cm},clip]{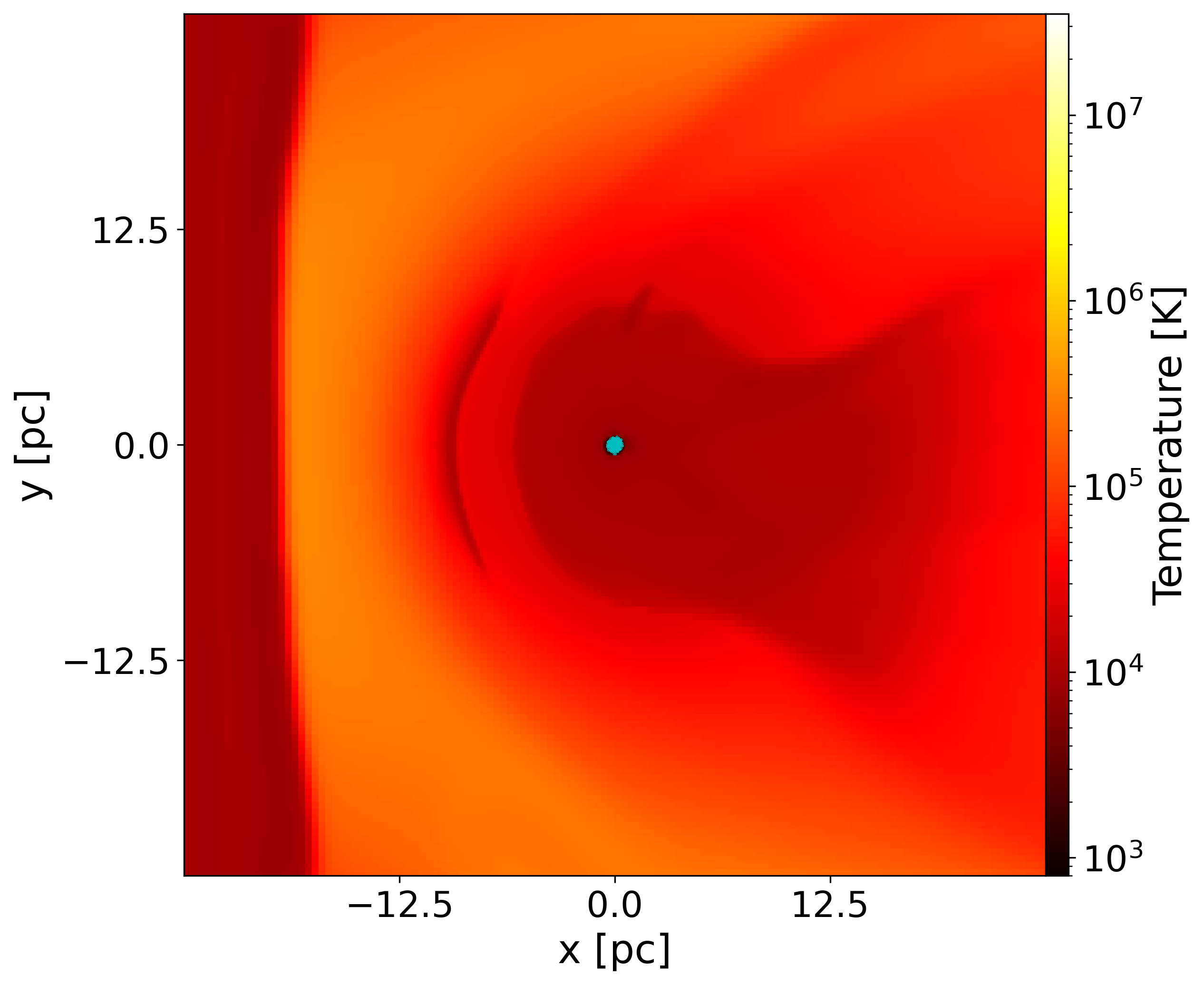}   
         \includegraphics[width=0.31\textwidth,trim={1.6cm 1cm 0 0.0cm},clip]{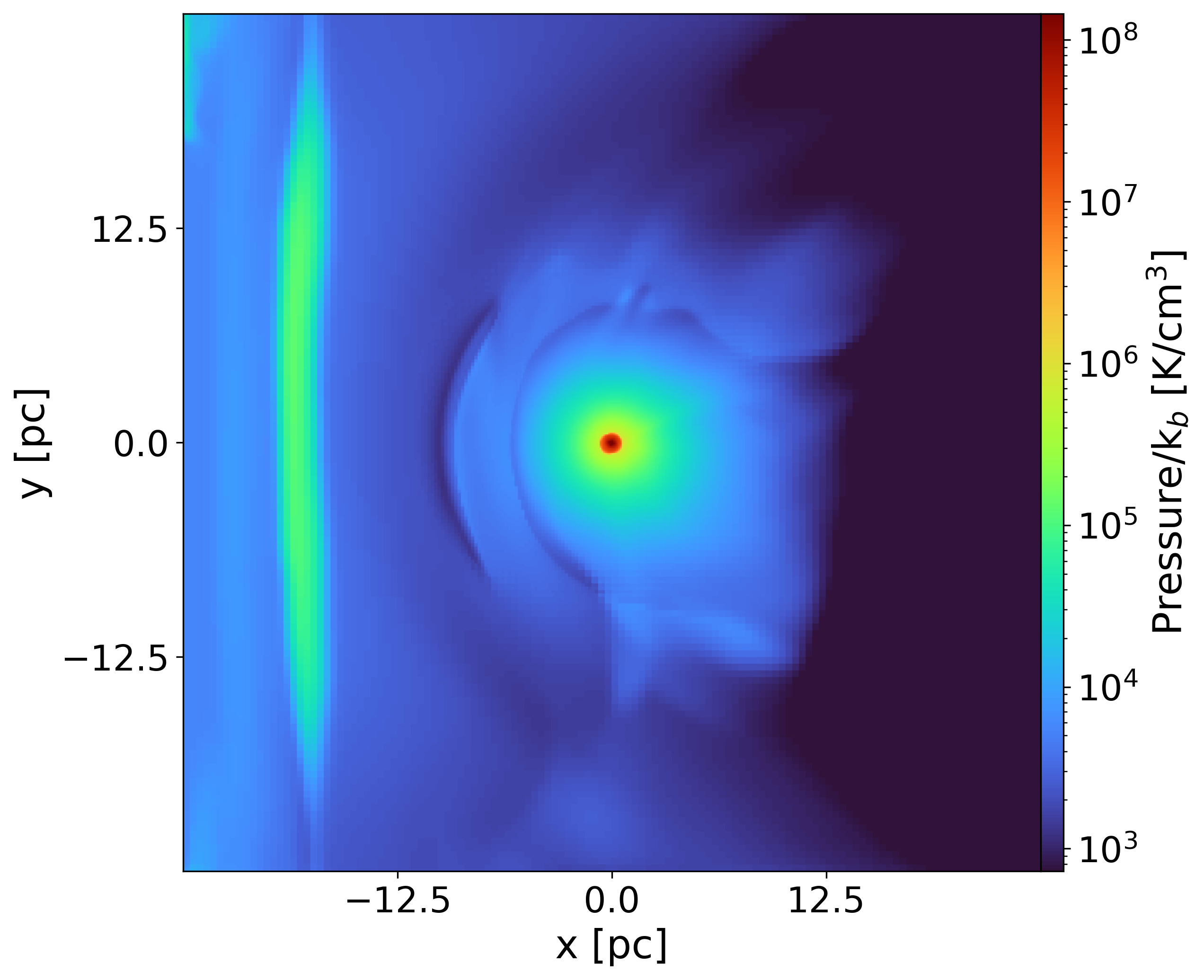}   
         \\
        \includegraphics[width=0.346\textwidth,trim={0 0 0 0.0cm},clip]{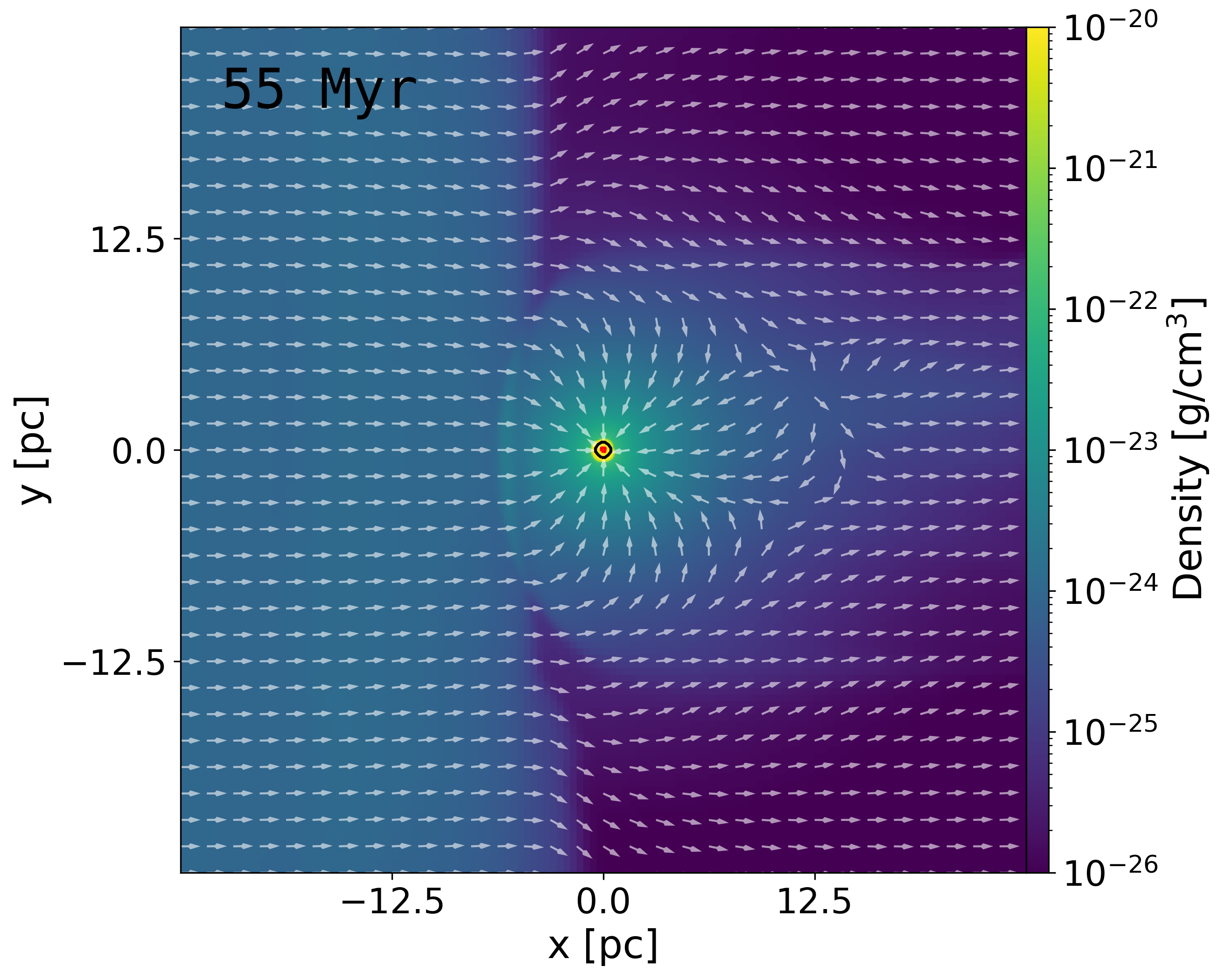}   
        \includegraphics[width=0.31\textwidth,trim={1.6cm 0 0 0.0cm},clip]{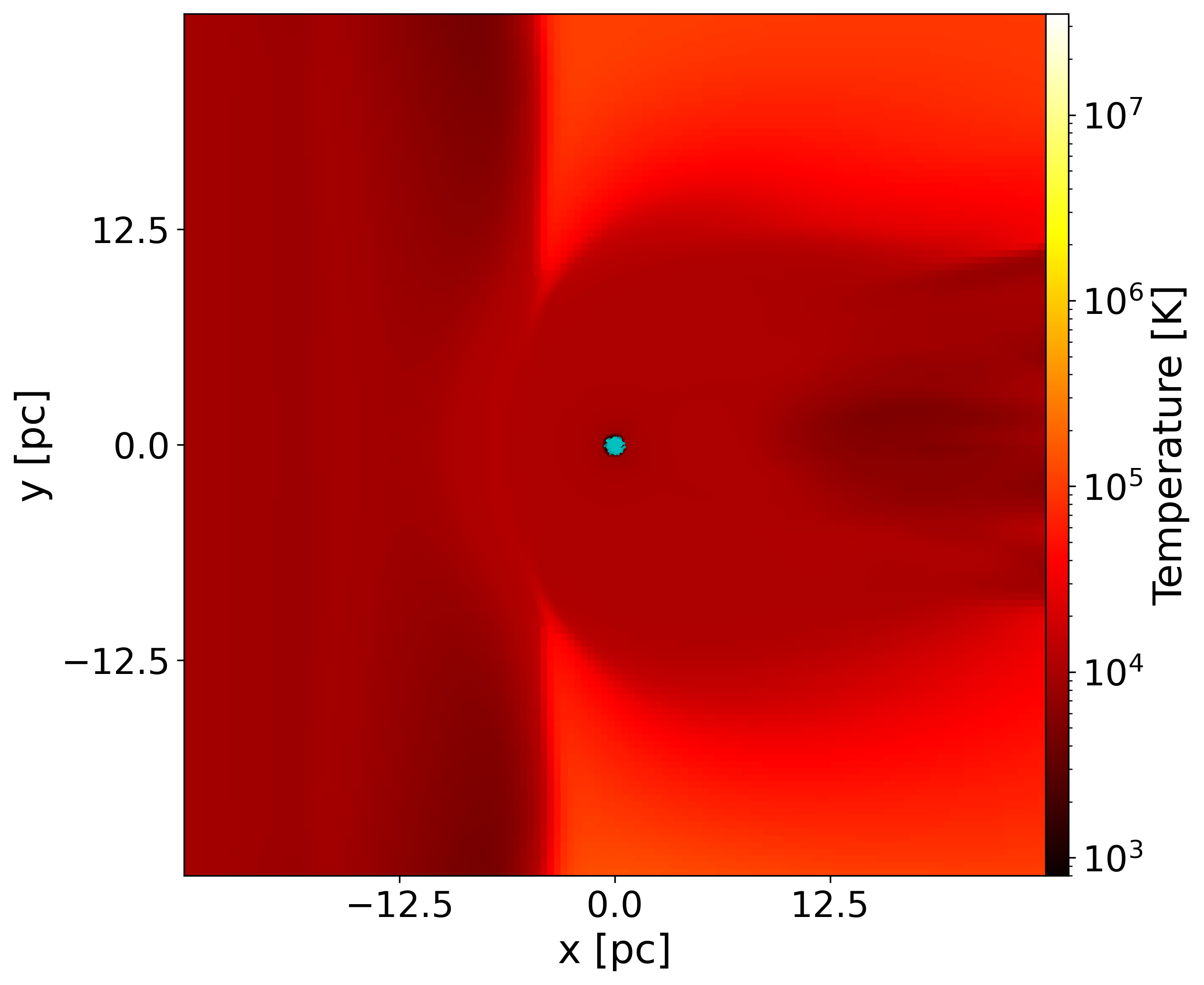}   
         \includegraphics[width=0.31\textwidth,trim={1.6cm 0 0 0.0cm},clip]{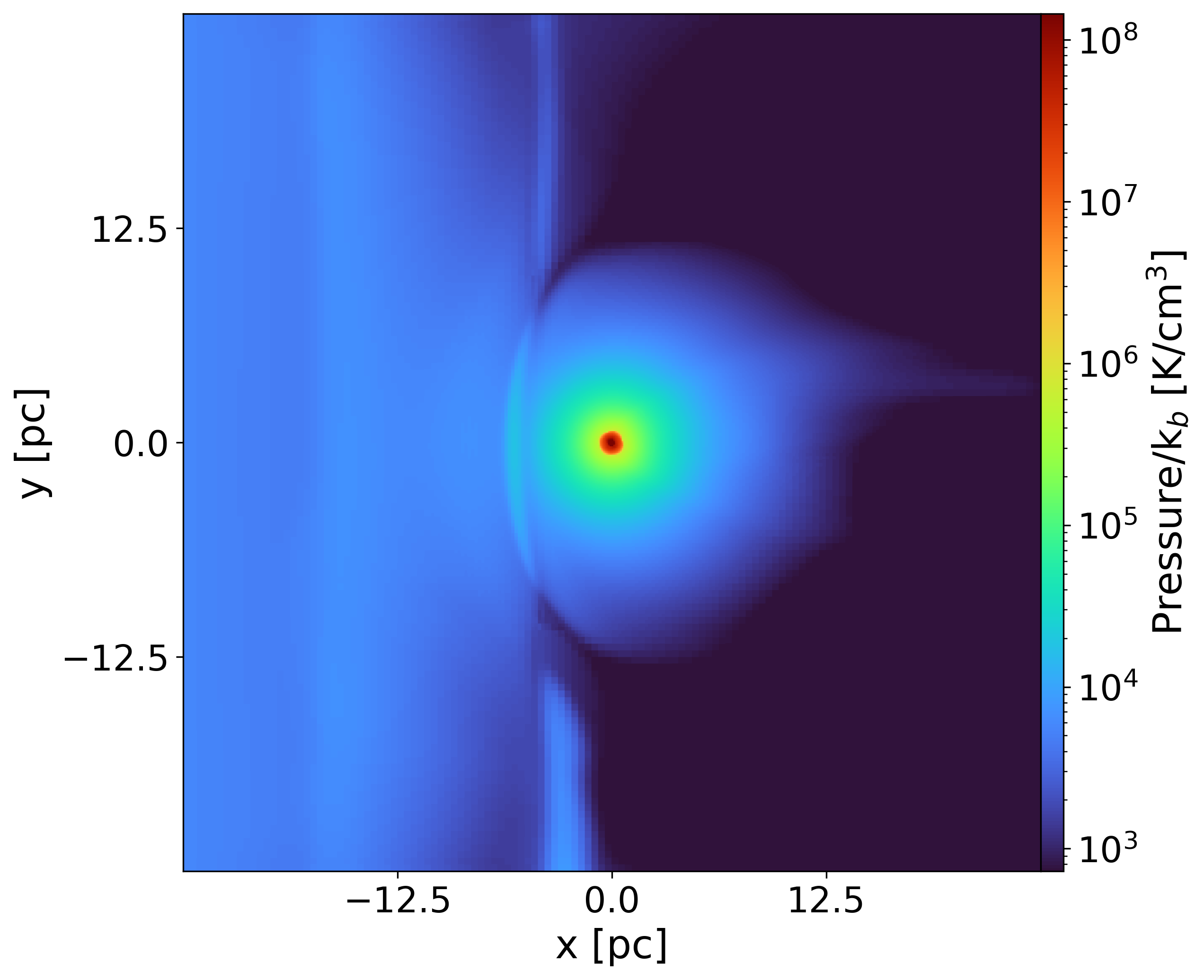}  


        \caption{The same as Figure \ref{fig:gas_map_M5I24}, but for model {\tt M6I24}.}
  \label{fig:gas_map_M6I24}
\end{figure*}

\begin{figure*}
        \centering

        \includegraphics[width=0.268\textwidth,trim={0 0cm 2.935cm 0.0cm},clip]{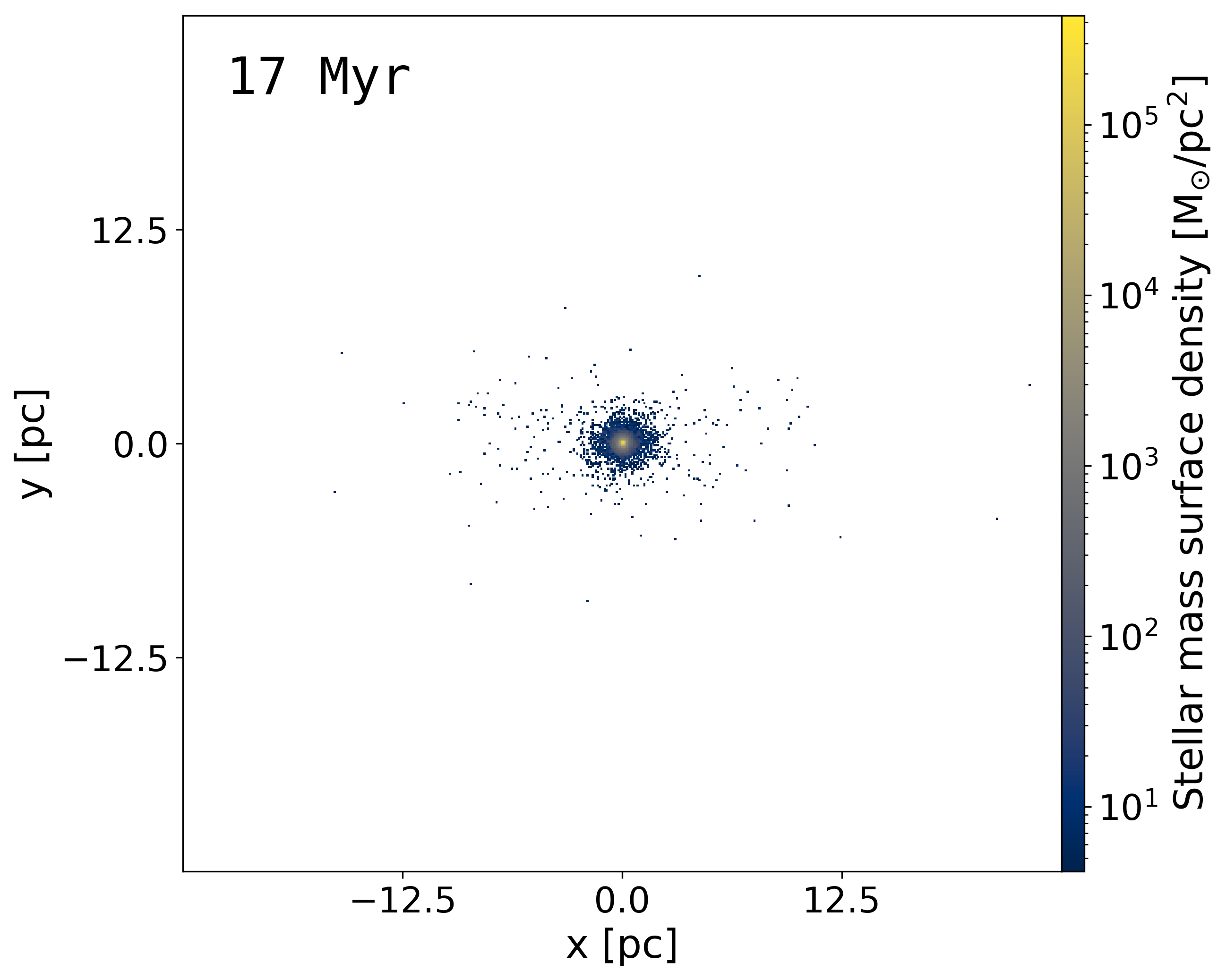}    
        \includegraphics[width=0.225\textwidth,trim={3.1cm 0cm 2.935cm 0.0cm},clip]{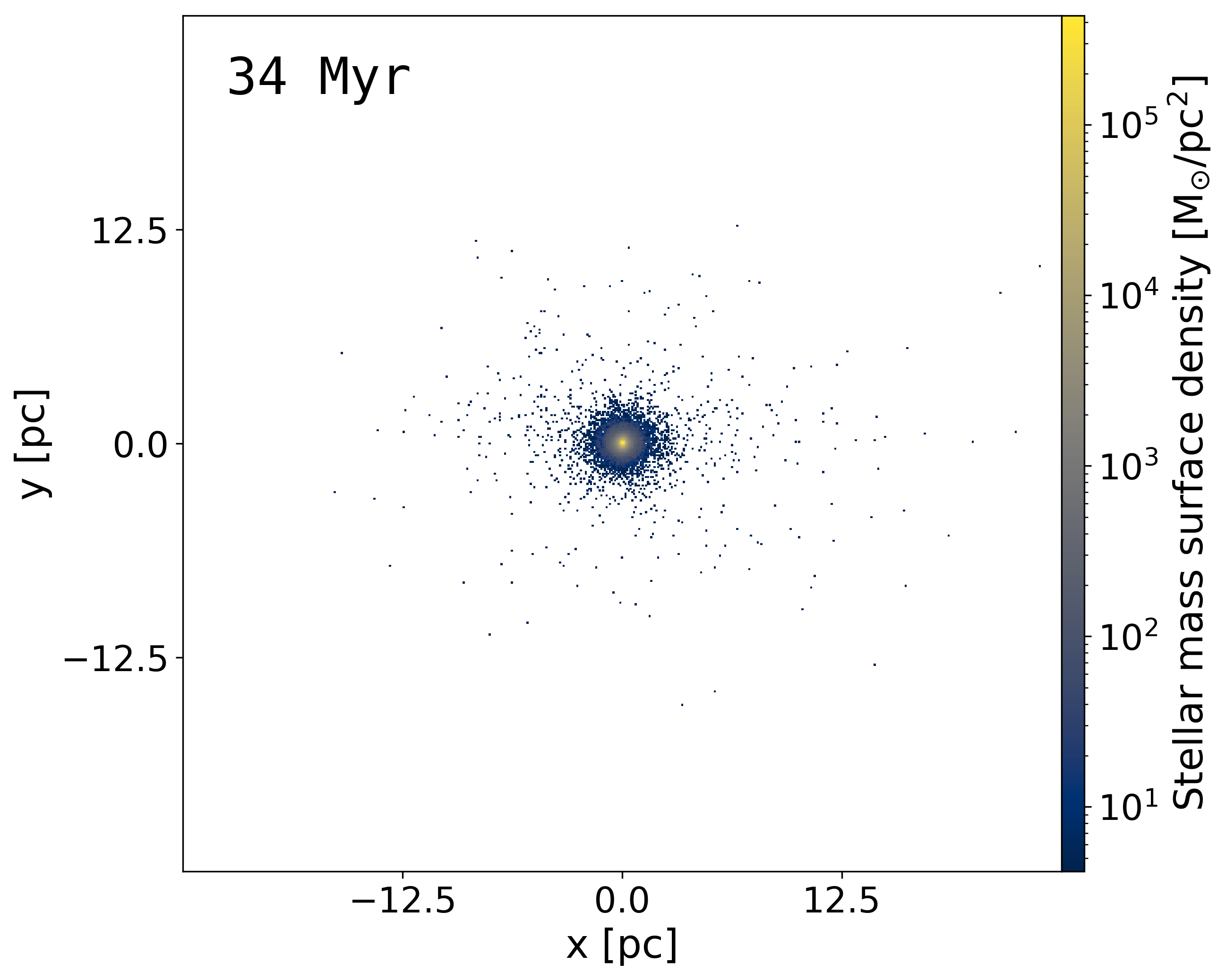}  
          \includegraphics[width=0.225\textwidth,trim={3.1cm 0cm 2.935cm 0.0cm},clip]{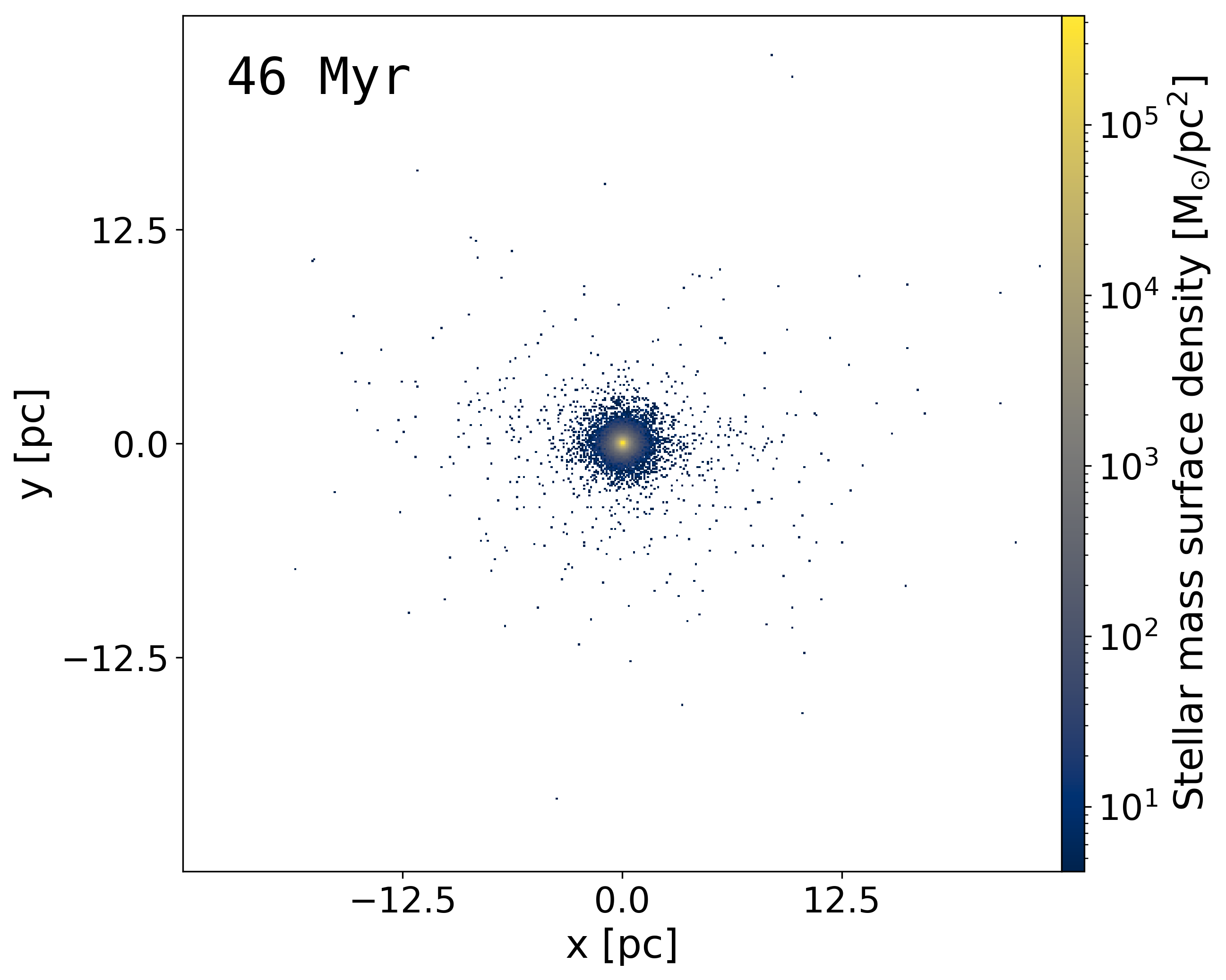}    
        \includegraphics[width=0.267\textwidth,trim={3.1cm 0cm 0cm 0.0cm},clip]{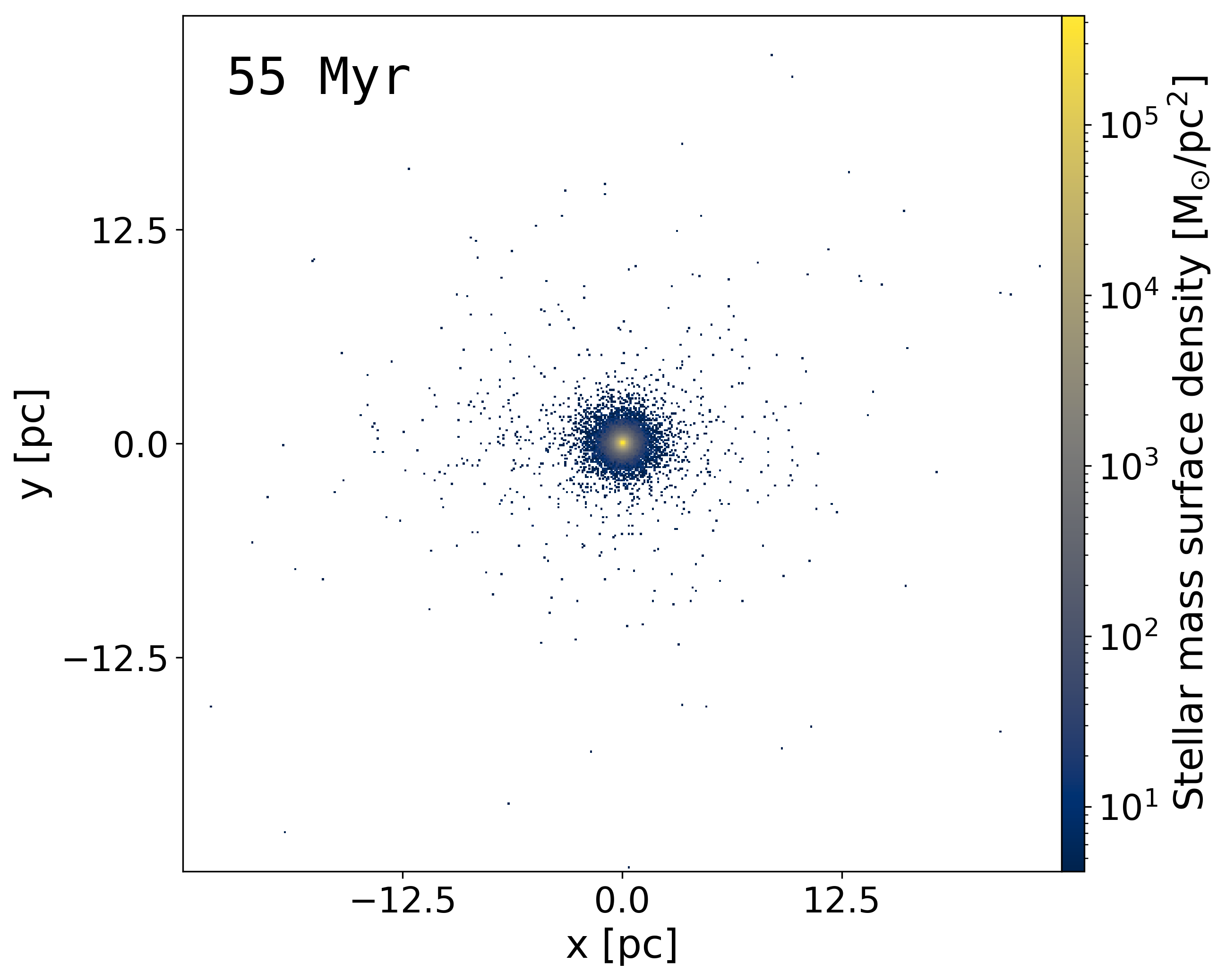}

        \caption{The same as Figure \ref{fig:part_map_M5I24}, but for model {\tt M6I24}.}
  \label{fig:part_map_M6I24}
\end{figure*} 

\begin{figure*}
        \centering
        \includegraphics[width=0.44\textwidth,trim={0cm 0cm 0cm 0.0cm},clip]{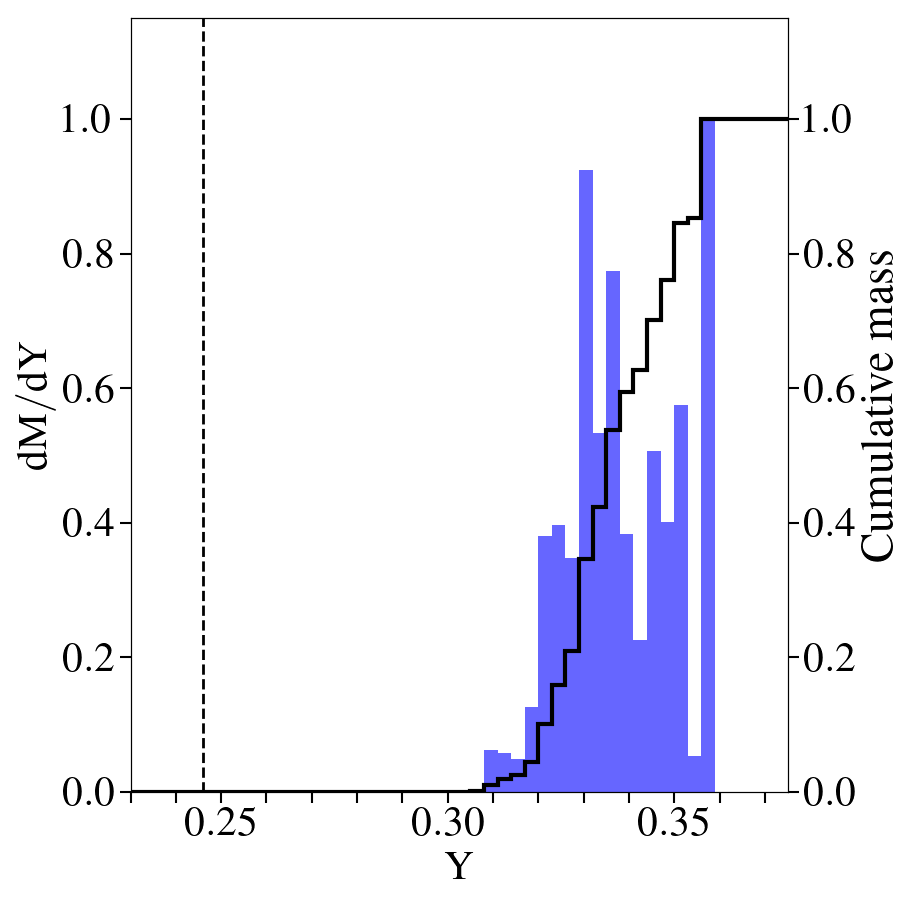}  
        \includegraphics[width=0.44\textwidth,trim={0cm 0cm 0 0.0cm},clip]{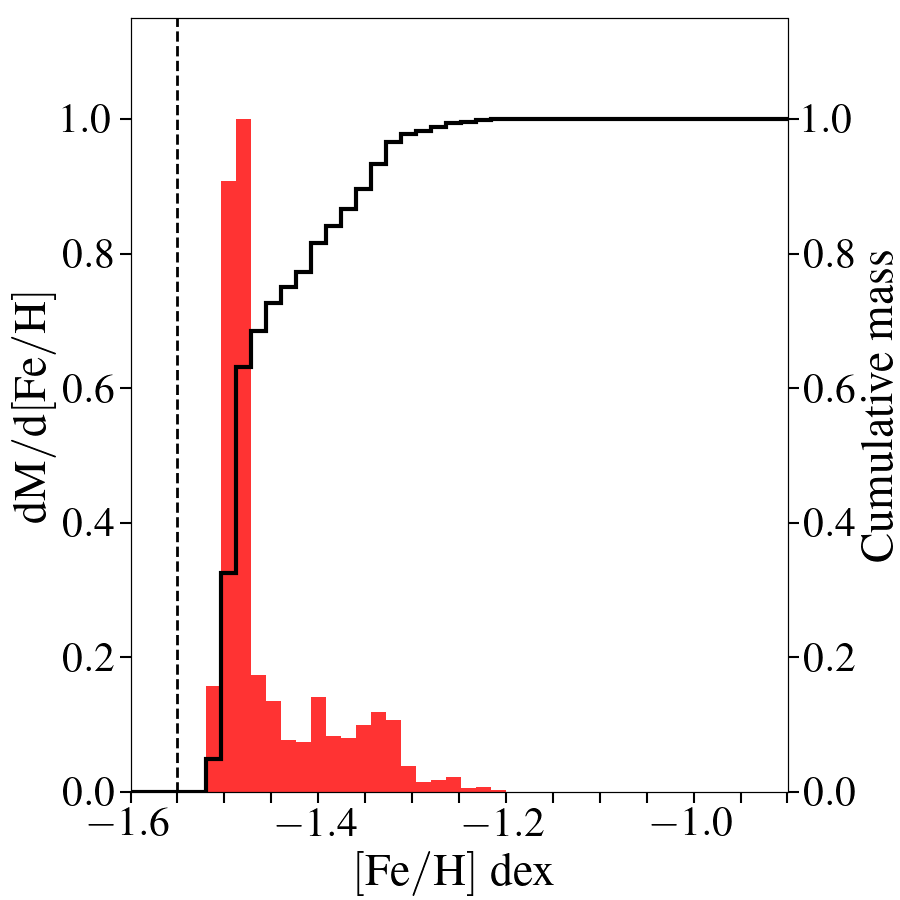} 
        \caption{The same as Figure \ref{fig:prof_M5I24}, but for model {\tt M6I24} at 55 Myr.}
  \label{fig:prof_M6I24}
\end{figure*} 
\begin{figure*}
  \includegraphics[width=0.95\linewidth]{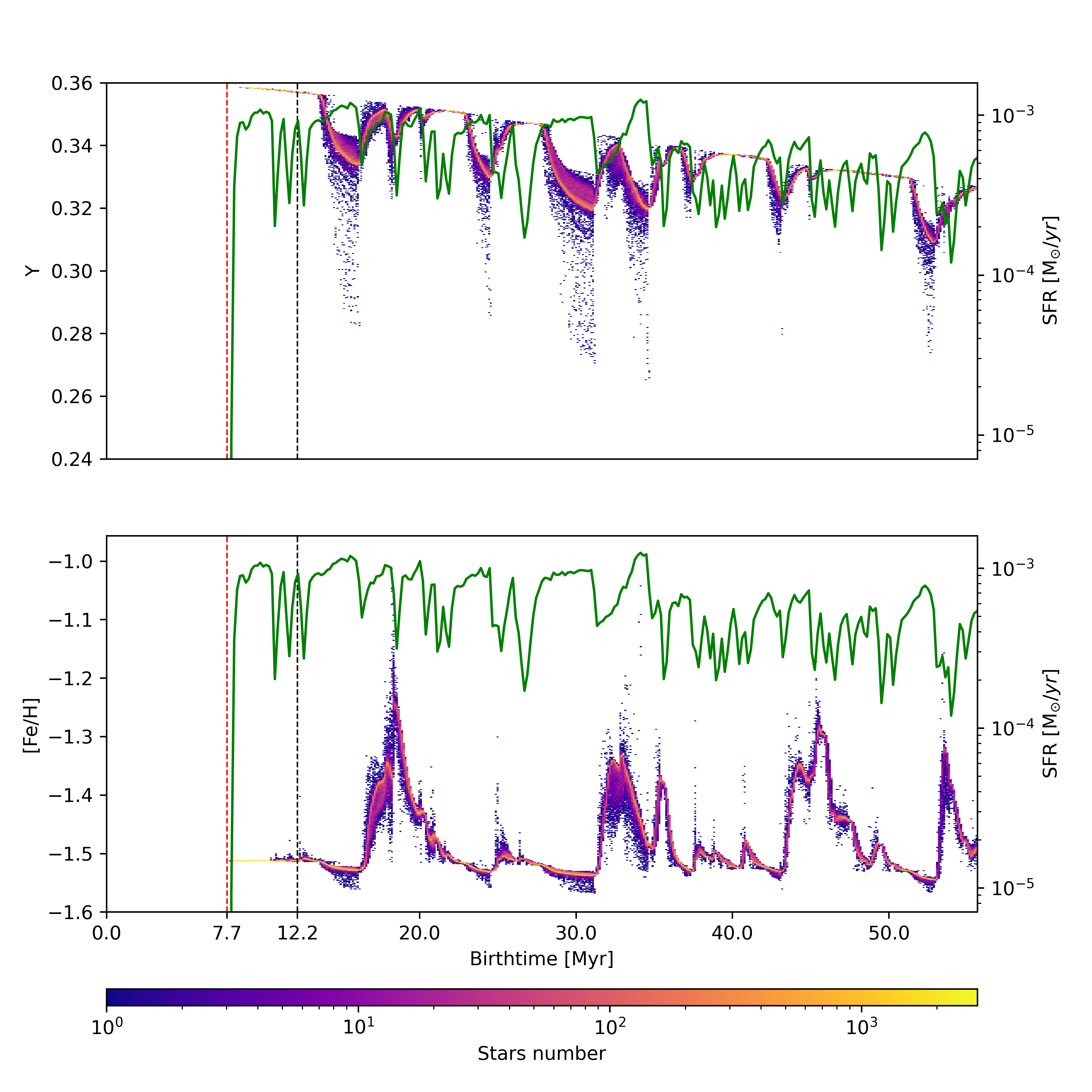}
  \caption{The same as Figure \ref{fig:Y_birth_M5I24}, but for model {\tt M6I24}.}
  \label{fig:Y_birth_M6I24}
\end{figure*} 
 
\subsubsection{Model M5I24}

The model M5I24 has an initial cluster mass $M_{\rm FG}=10^5 \Msun$ and pristine gas density $\rho_{\rm pg} = 10^{-24} \gcm$.
Figure \ref{fig:gas_map_M5I24} shows four snapshots at different evolutionary times of gas density, temperature and pressure for this model, obtained by selecting a slice of the $x-y$ plane passing through the centre of the box. The pristine gas in this model enters the box $3 \Myr$ after the beginning of the simulation and $4.7 \Myr$ before the AGB ejecta start to be released. Despite that, star formation starts only several Myr after, at around $10 \Myr$ as found in \citet{yaghoobi2022} due to the effect of the ram pressure on a shallow gravitational potential well. In the first snapshot of Figure \ref{fig:gas_map_M5I24}, a SG system has already formed with an almost spherical shape as shown also in Figure \ref{fig:part_map_M5I24}. The direction of the infalling gas is weakly perturbed, as visible from the velocity field and, at variance with other models (see \citealt{calura2019,yaghoobi2022}), an accretion column is not formed downstream of the system  due to the cluster's limited ability to accumulate pristine gas behind it. Even though some gas is moving towards the centre of the cluster, its velocity is negligible, as shown by \citet{yaghoobi2022}. At $34 \Myr$, the SG component is more flattened than before in the direction of the infall. The gas in the outer regions is much less dense, hotter and the pressure has increased as a response to a recent SN explosion. The infall has been confined far from the centre as the gas perturbed by the explosion still moves outwards. The $46 \Myr$ snapshot represents the system between two SN explosions as in the first snapshot and resembles the maps shown by \citet{yaghoobi2022}, with the bow-shock profile typical of a massive perturber in motion through a gas distribution (\citealt{calura2019} and references therein). At this time, the infall has filled again the whole box and downstream of the system gas is moving towards the centre but an accretion column is not formed as seen in the simulation without SNe. In the last snapshot at $55 \Myr$ the infall is still filling the void created by a SN explosion that has significantly lowered the density and increased the temperature of both the central region and its surroundings. The SG system in the last three snapshots is flattened in the direction of the infalling gas, as shown in Figure \ref{fig:part_map_M5I24}. 
As found in previous works, the SG is more centrally concentrated than the FG as the AGB ejecta cool down and sink towards the centre, where they then form new stars. 

The distribution of $Y$, shown in Figure \ref{fig:prof_M5I24}, has a similar range as \citet{yaghoobi2022} while its shape is different and forged by the SN feedback. Two peaks are present as a result of the quenching of star formation caused by the explosions which prevent a smooth formation of stars with decreasing helium mass fraction. This is highlighted in Figure \ref{fig:Y_birth_M5I24}, where the stellar helium mass fraction is plotted as a function of the birthtime of SG particles. Overall, $Y$ decreases with time as stars with lower mass produce less helium during the AGB phase and because of the dilution of AGB ejecta with infalling gas. The star formation rate is modulated by SNe Ia and, in particular at around $40\Myr$, the star formation is halted for several Myr and the stars formed after have a much lower $Y$, creating the second peak in the histogram. 
In addition, the SG is dominated by extremely helium-enhanced stars and, at variance with the model with $M_{\rm FG}=10^7\Msun$ of \citet{lacchin2021}, a second peak of less helium-enriched stars is formed as in model M5I24 SNe are rare and are not preventing the infall of gas like in the $10^7\Msun$ cluster models. 
Concerning iron, Figure \ref{fig:prof_M5I24} displays that SG stars are more enriched than the FG. In this model, the fewer explosions limit the iron spread in comparison to the high-mass cluster simulation of \citet{lacchin2020}, leading to a dispersion in SG of ${\rm\sigma^{SG}_{[Fe/H]}=0.005\ dex}$ with a mean value of $-1.46\ {\rm dex}$. To compare with present-day clusters, we assume that most FG stars have been lost during the evolution and constitute only 30\% of the mass at present time, as suggested by \citet{milone2017}. Under this assumption, the iron spread of the whole cluster is estimated to be ${\rm\sigma_{[Fe/H]}=0.04\ dex}$, comparable with the bulk of globular clusters.

\subsubsection{Model M6I24}

Here, we present an analysis of a cluster characterized by a FG with a mass ten times greater (10$^6\Msun$)  than that analyzed in the previous section and the same pristine gas density ($\rho_{\rm pg} = 10^{-24} \gcm$). In contrast to the previous model, the pristine gas begins to enter the box at $t_{\rm inf} = 12.2 \Myr$, with a $ 4.5 \Myr$ delay after the onset of AGB ejecta at $t_{\rm AGB} = 7.7 \Myr$.

In this scenario, the cluster's gravitational potential well is deeper than in the previous simulation. As a result, both ram pressure and the energy injection from SNe Ia are less efficient in sweeping the gas away from the cluster. This effect contributes to a more conspicuous star formation. 

In Figure \ref{fig:gas_map_M6I24}, the gas evolution of the M6I24 model is shown. In the first snapshot, an accretion column of dense and cold gas can be distinguished downstream to the system thanks to the velocity field, while stars are formed predominantly in the very central region. At $34 \Myr$, the majority of the box is filled by highly pressurized, dense and hot gas, a hint of a recent SN explosion. In addition, no infalling gas is present as it has been confined out of the box. The accretion column typical of the SN-free simulations is still present but strongly affected by the explosion and no stars have recently formed along it.  The last two snapshots depict very similar maps: the infall is reaching the centre of the system after a SN explosion which has heated and cleared the surroundings, the accretion column is no longer present and star formation is ongoing only in the central region. At variance with the M5I24 simulation, the stellar component is here more spherical, as a consequence of the later arrival of the infalling gas, as displayed in Figure \ref{fig:part_map_M6I24}. 
In this model, star formation is mildly affected by the presence of Type Ia SNe, which translates into a more extended helium distribution, reported in Figure \ref{fig:prof_M6I24}, with respect both to the high-mass cluster models in \citet{lacchin2021}, due to the more efficient dilution of AGB ejecta, and to the M5I24 model, due to the lower impact of Type Ia SNe on the star formation. Even though in this model the infalling gas is able to reach the centre of the system and dilute the AGB ejecta, stars with low helium enhancement (Y<0.30) are negligible as shown in Figure \ref{fig:prof_M6I24}, at variance with the simulations without Type Ia SNe, meaning that the dilution is happening, but not as efficiently as in \citet{yaghoobi2022}. 
Finally, as in the previous model, the decreasing trend in the stellar $Y$ with time is attributed to the decreasing AGB yield over time plus the dilution of AGB ejecta with helium-poor pristine gas, which is clearly seen in Figure \ref{fig:Y_birth_M6I24}. In a model without Type Ia SNe this drop is monotonic while here it grows again when SNe Ia are exploding. This happens because SNe confine the pristine gas far from the centre so that stars formed soon after a SN explosion are composed of AGB ejecta and, in very small amounts, of Type Ia SNe ones. Then, when the infall restarts to dilute the He-rich gas, $Y$ goes down again. Concerning the iron composition of SG stars, the cluster's higher mass enables the retention of more iron within its gravitational well with respect to the M5I24 model. In addition, a higher cluster mass implies a higher number of SN explosions, enhancing the amount of ejected iron later used to form new stars. Therefore, as shown in Figure \ref{fig:prof_M6I24}, the M6I24 model has a much broader dispersion of ${\rm \sigma^{SG}_{[Fe/H]}=0.06\ dex}$ than the M5I24 model, but also a much lower spread than the high-mass clusters model of \citet{lacchin2021}, which has 10 times more SNe explosions in the same timespan.
Assuming that the cluster has lost most of the FG, i.e. after 12 Gyr the FG only represents the 30\% of the whole cluster, we derive a total iron spread of ${\rm \sigma_{[Fe/H]}=0.07\ dex}$ which is in agreement with typically observed ones \citep{bailin2019}.

\begin{figure*}
        \centering

        \includegraphics[width=0.346\textwidth,trim={0 1cm 0 0.0cm},clip]{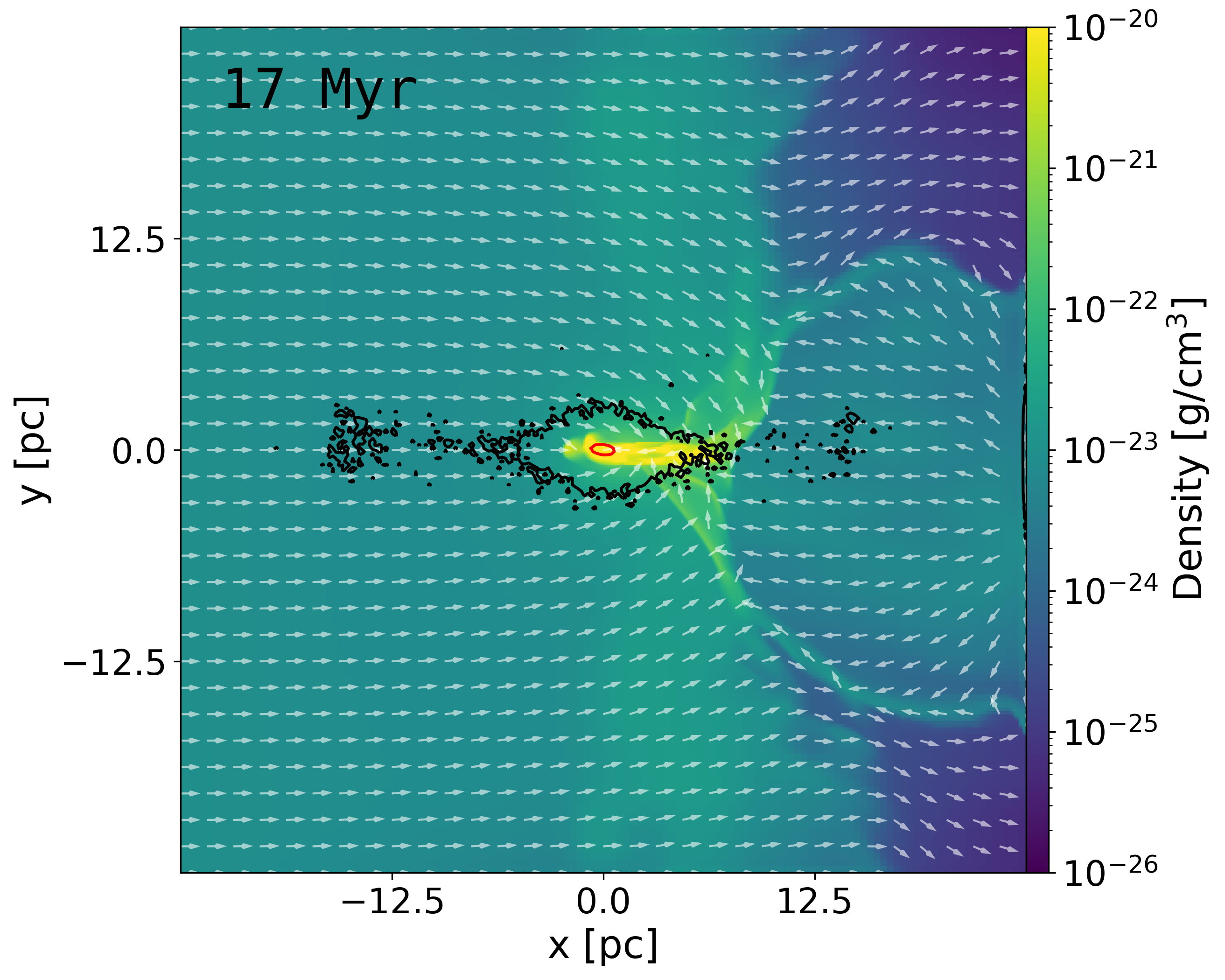}    
        \includegraphics[width=0.31\textwidth,trim={1.6cm 1cm 0 0.0cm},clip]{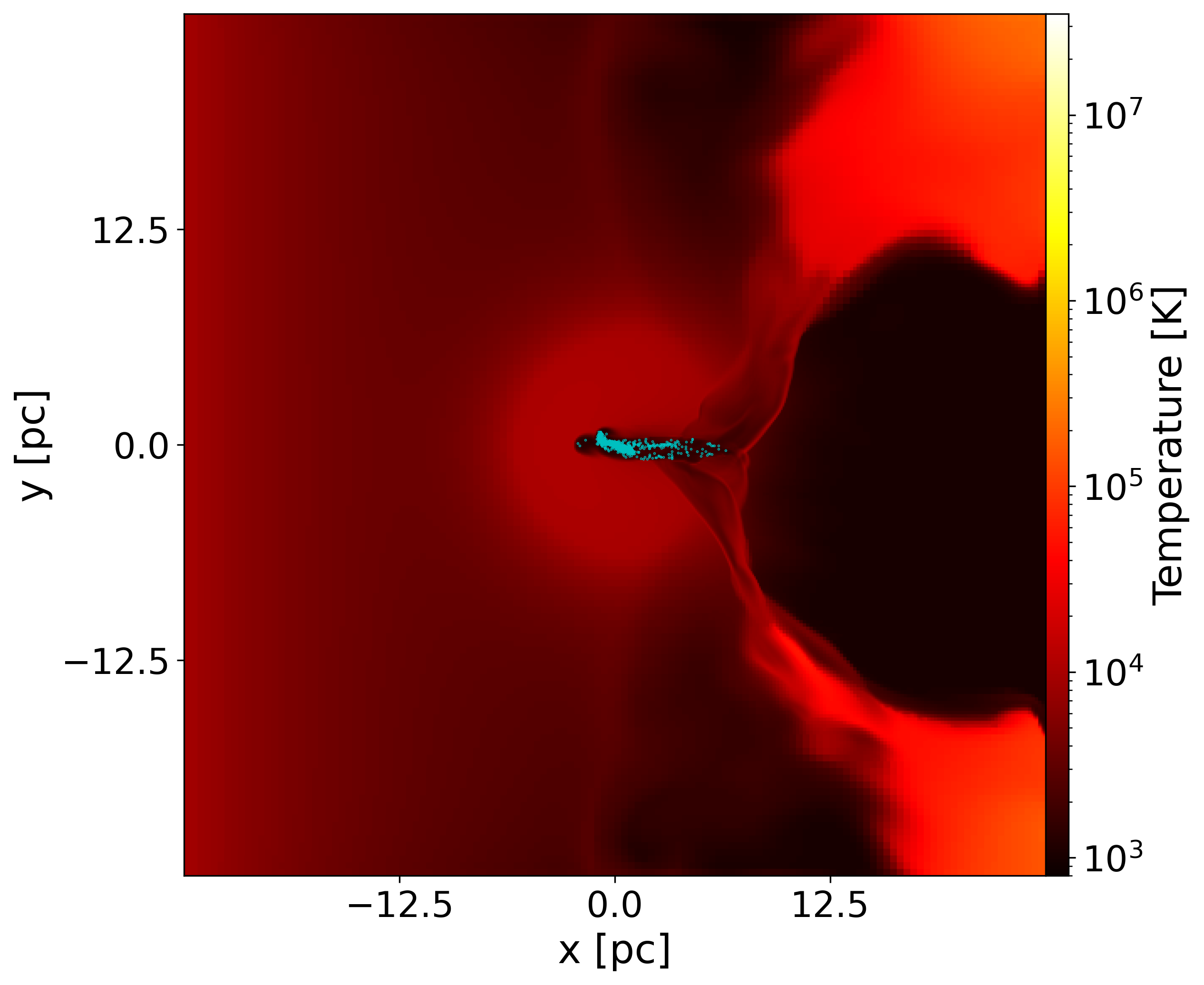}   
         \includegraphics[width=0.31\textwidth,trim={1.6cm 1cm 0 0.0cm},clip]{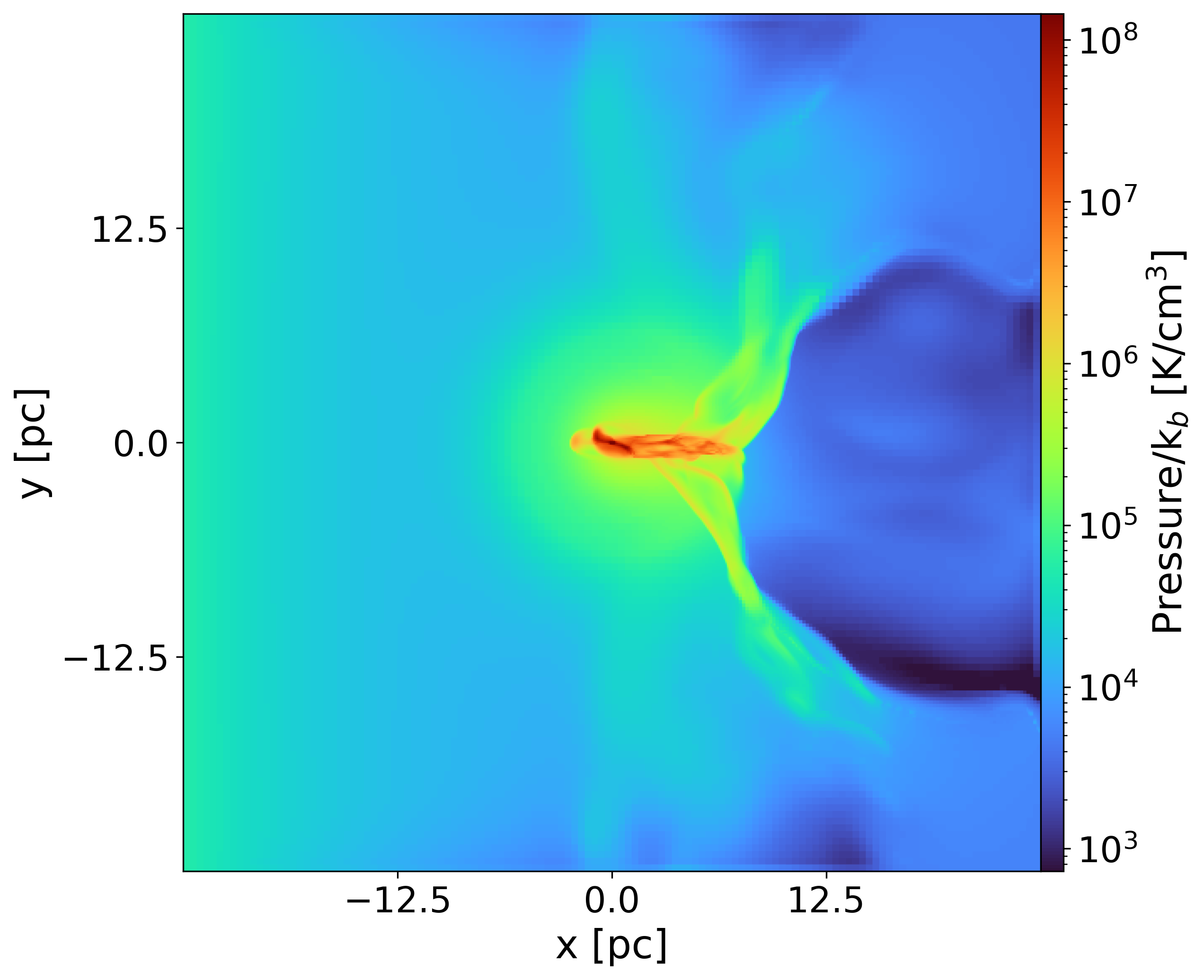}    
         \\

        \includegraphics[width=0.346\textwidth,trim={0 1cm 0 0.0cm},clip]{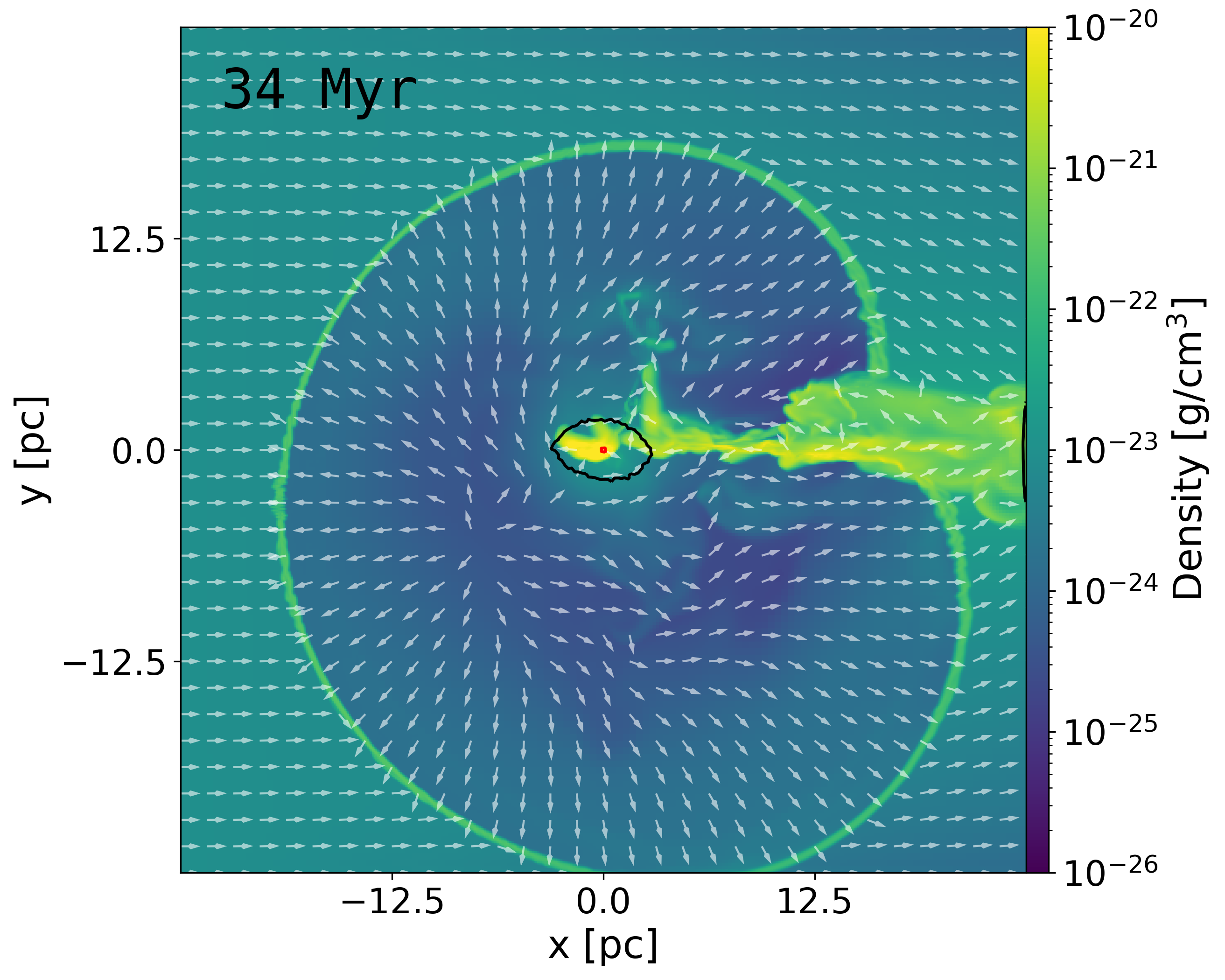}    
        \includegraphics[width=0.31\textwidth,trim={1.6cm 1cm 0 0.0cm},clip]{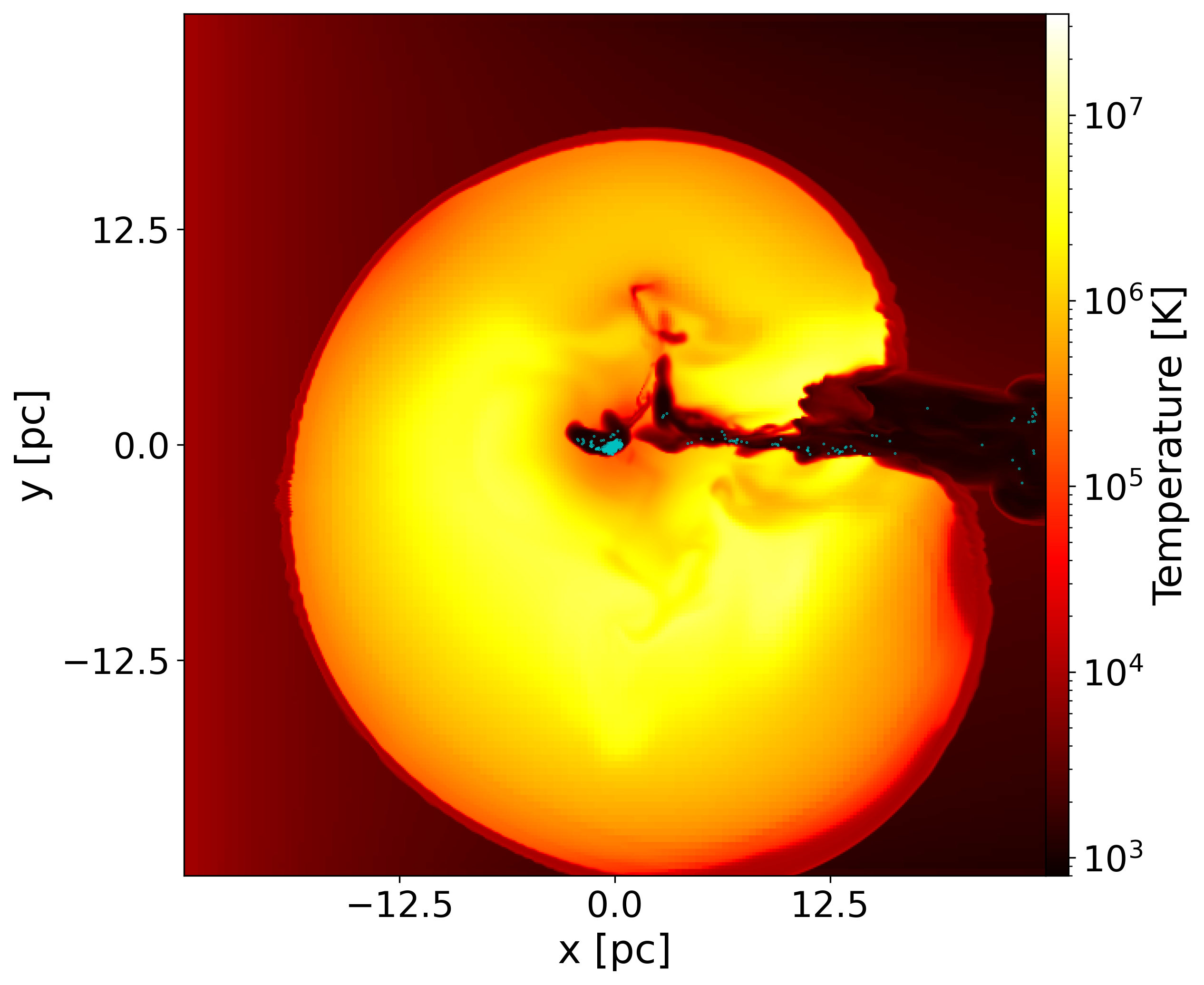}   
         \includegraphics[width=0.31\textwidth,trim={1.6cm 1cm 0 0.0cm},clip]{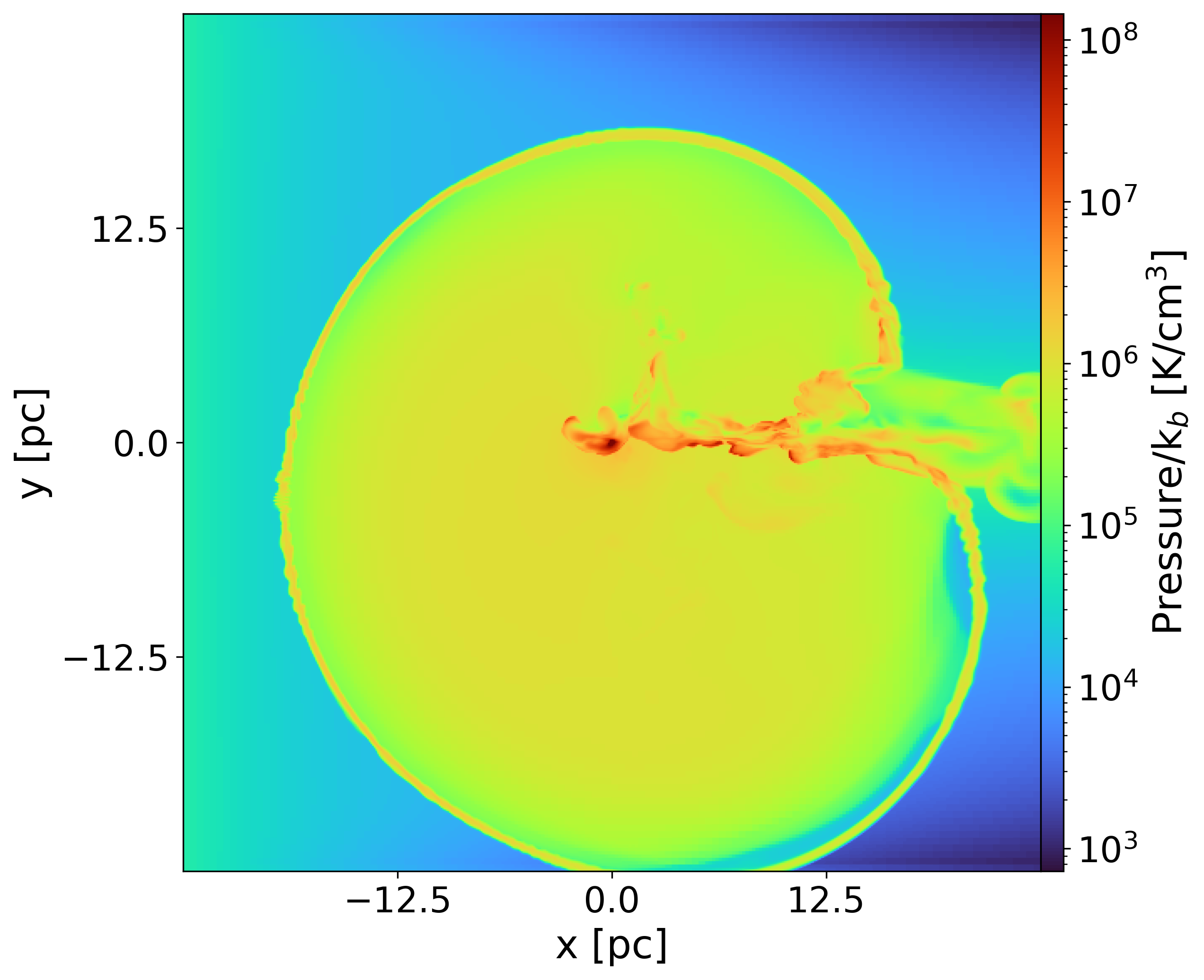}
         \\

        \includegraphics[width=0.346\textwidth,trim={0 1cm 0 0.0cm},clip]{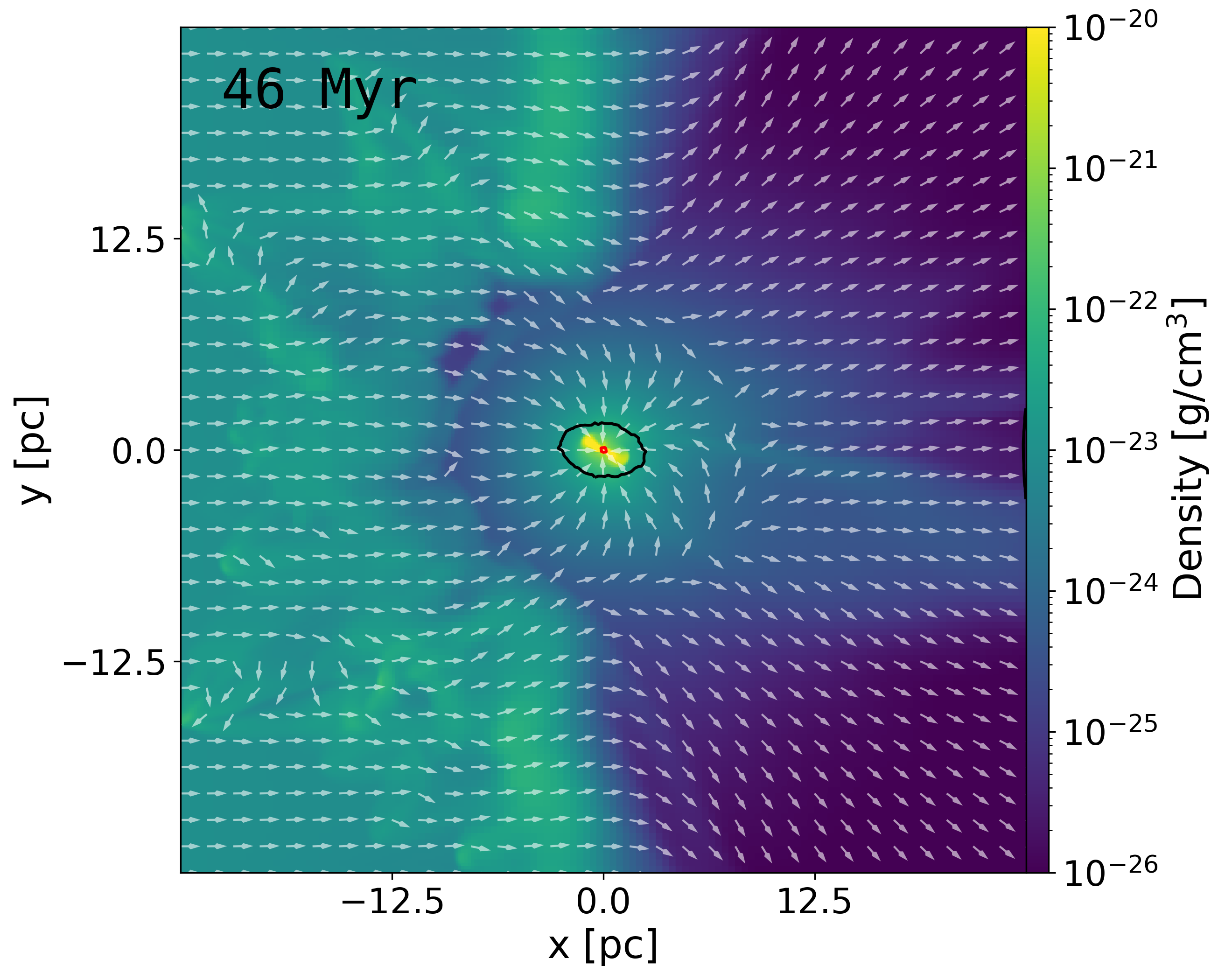}
        \includegraphics[width=0.31\textwidth,trim={1.6cm 1cm 0 0.0cm},clip]{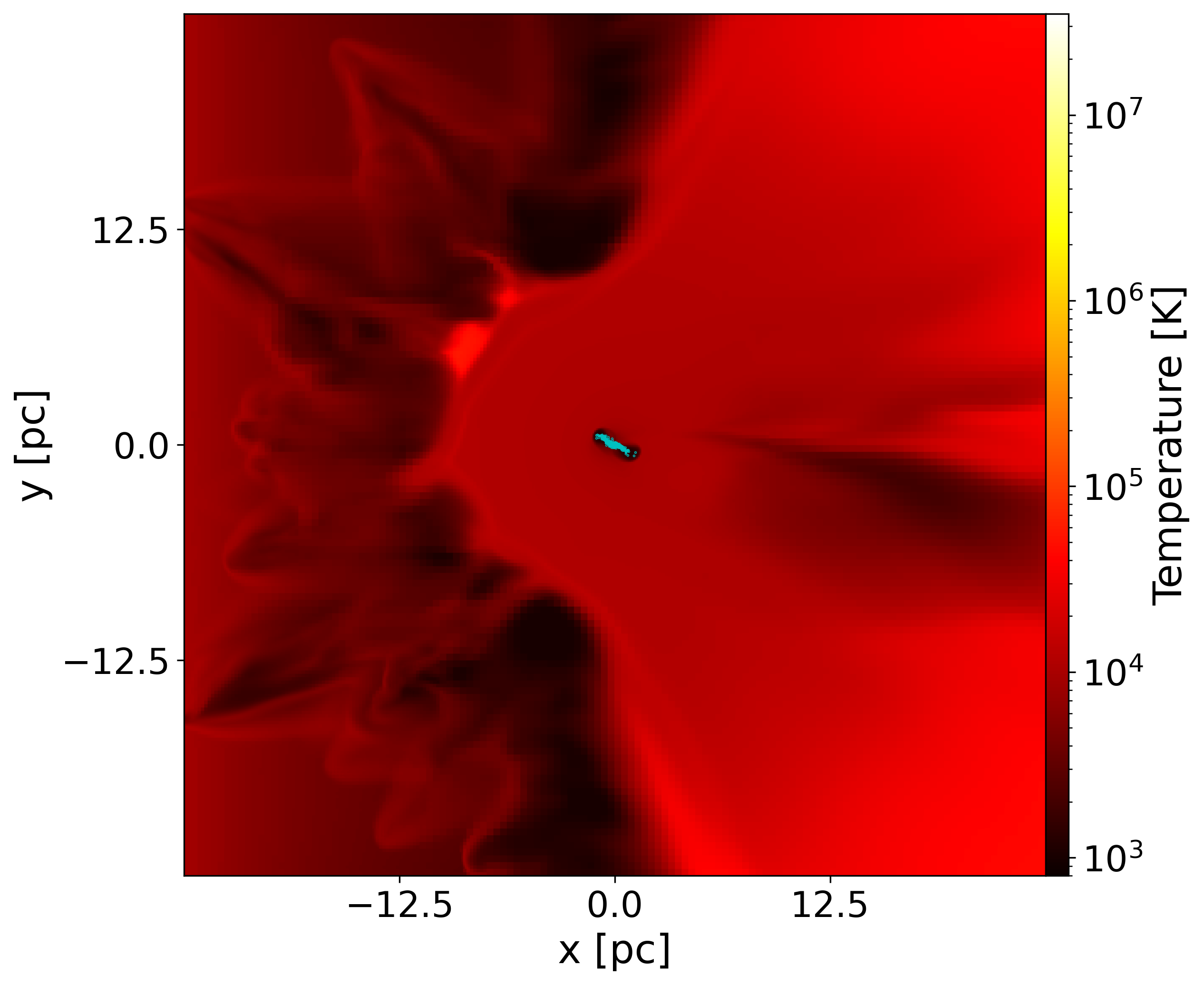}   
         \includegraphics[width=0.31\textwidth,trim={1.6cm 1cm 0 0.0cm},clip]{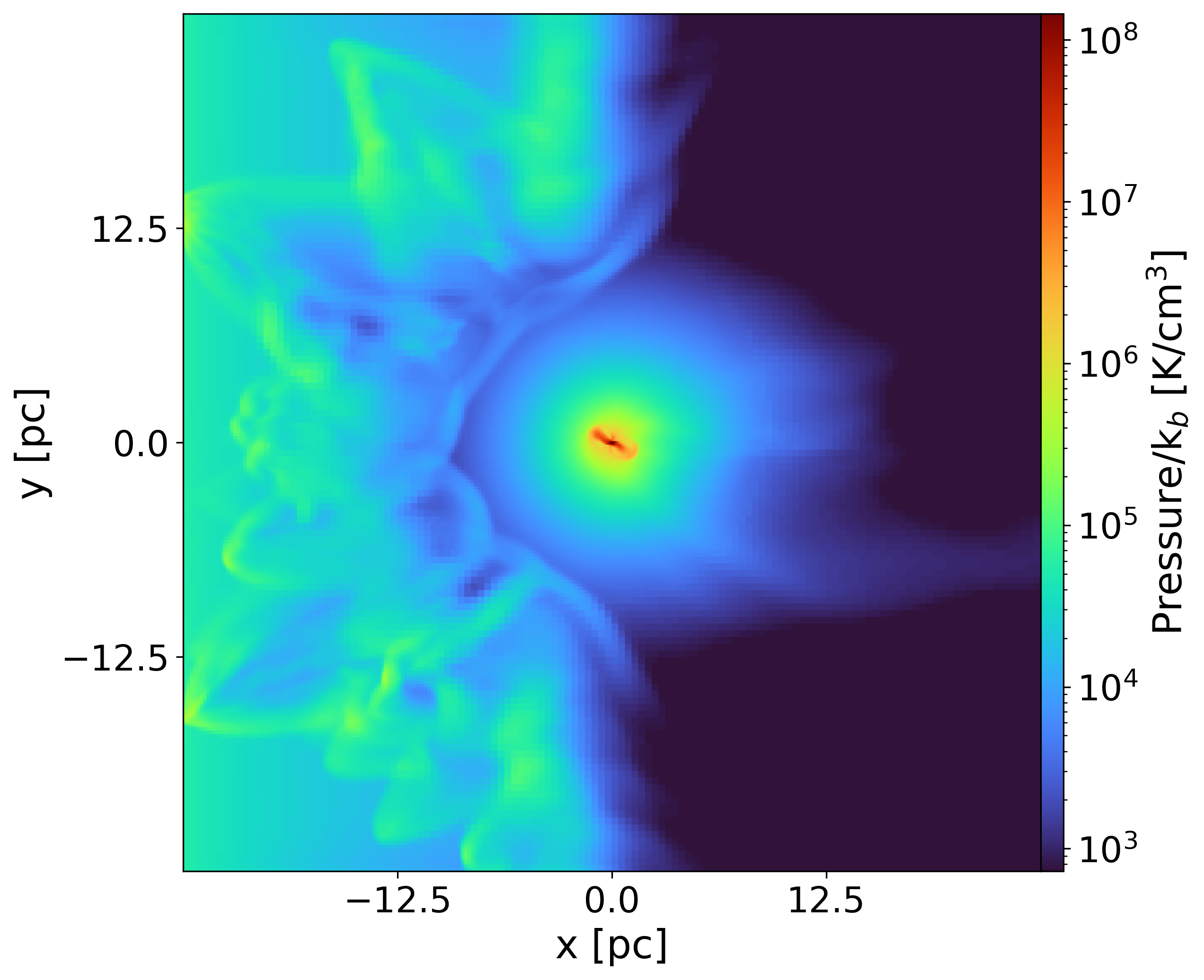}    
         \\

        \includegraphics[width=0.346\textwidth,trim={0 0cm 0 0.0cm},clip]{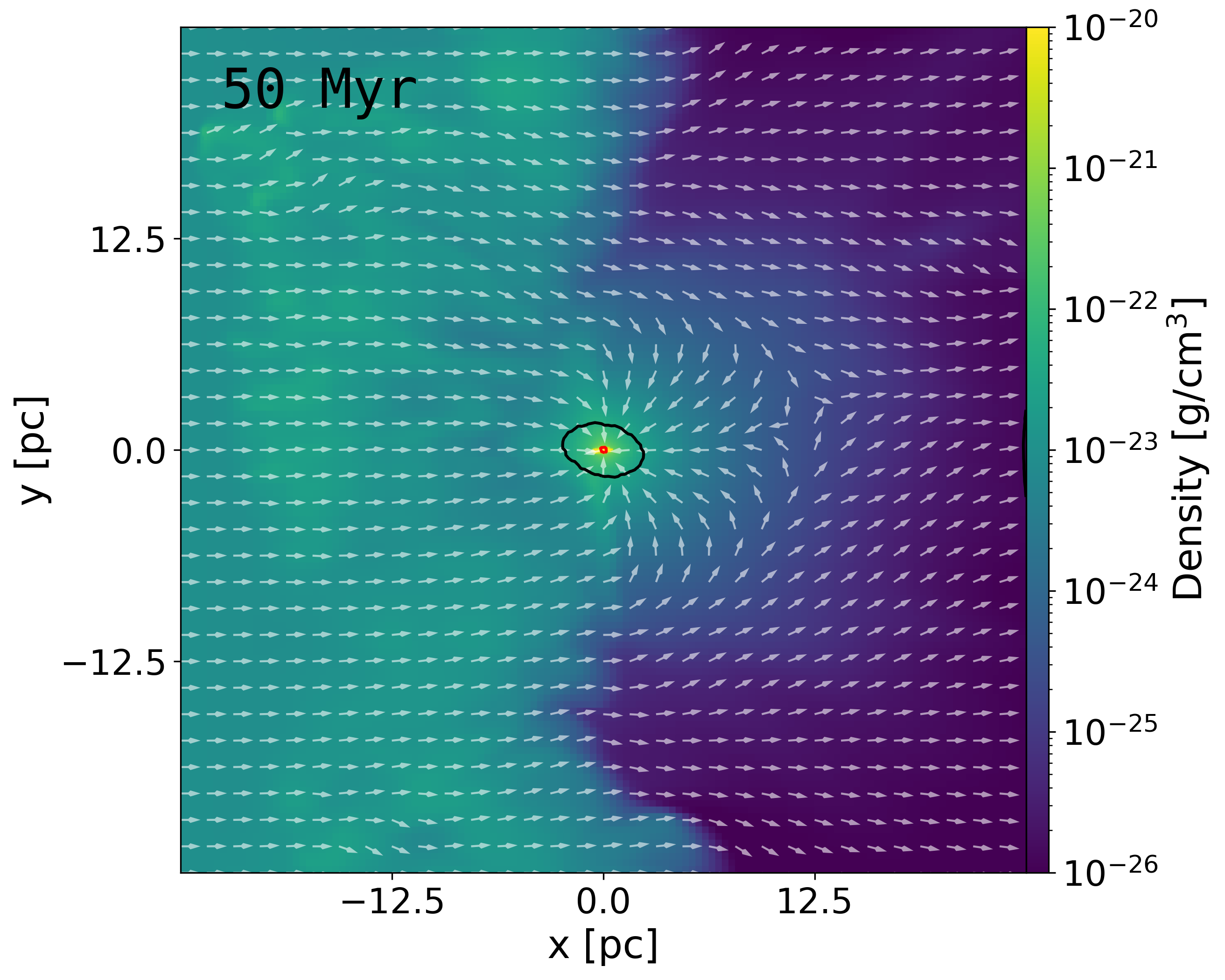}    
        \includegraphics[width=0.31\textwidth,trim={1.6cm 0 0 0.0cm},clip]{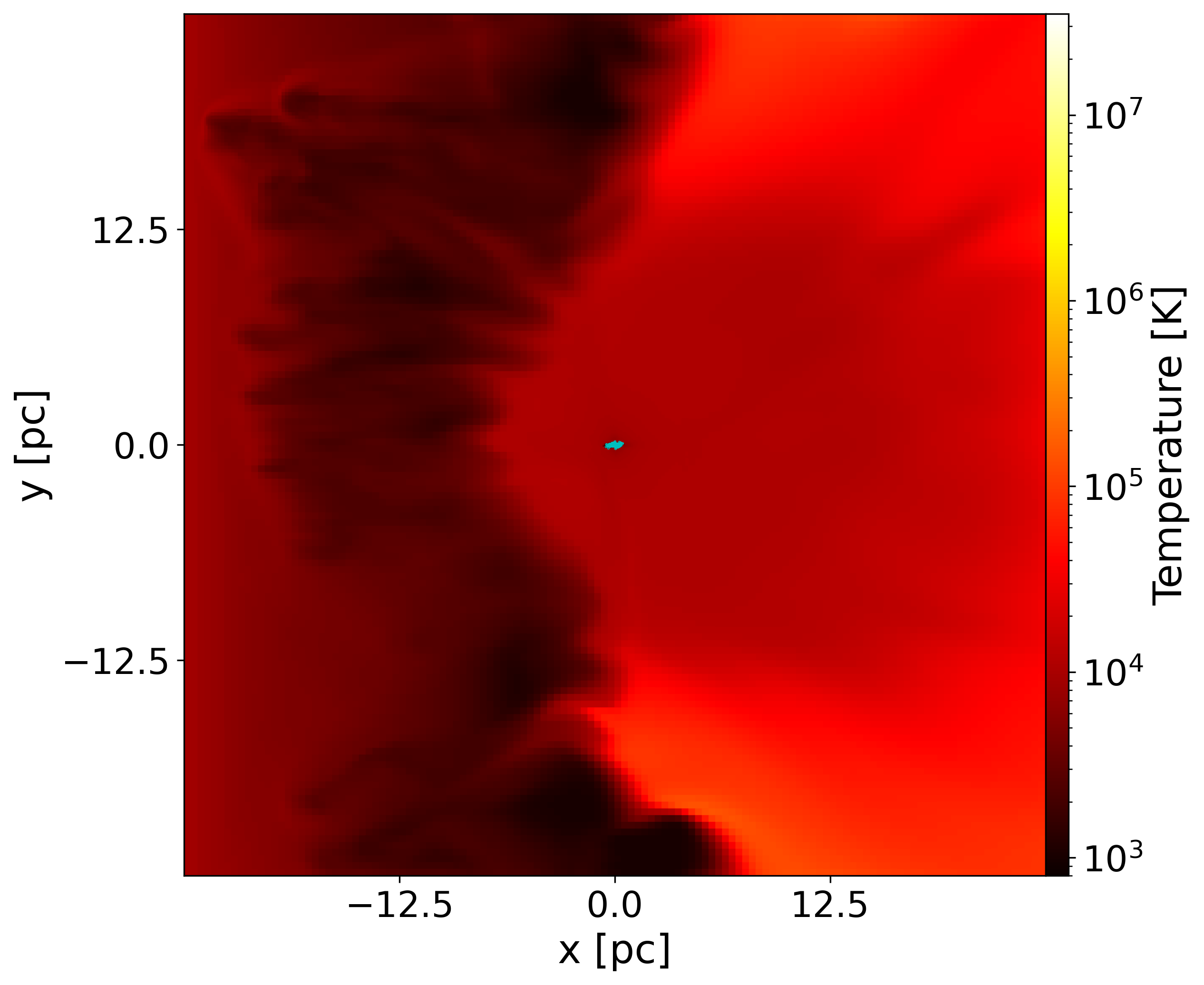}   
         \includegraphics[width=0.31\textwidth,trim={1.6cm 0 0 0.0cm},clip]{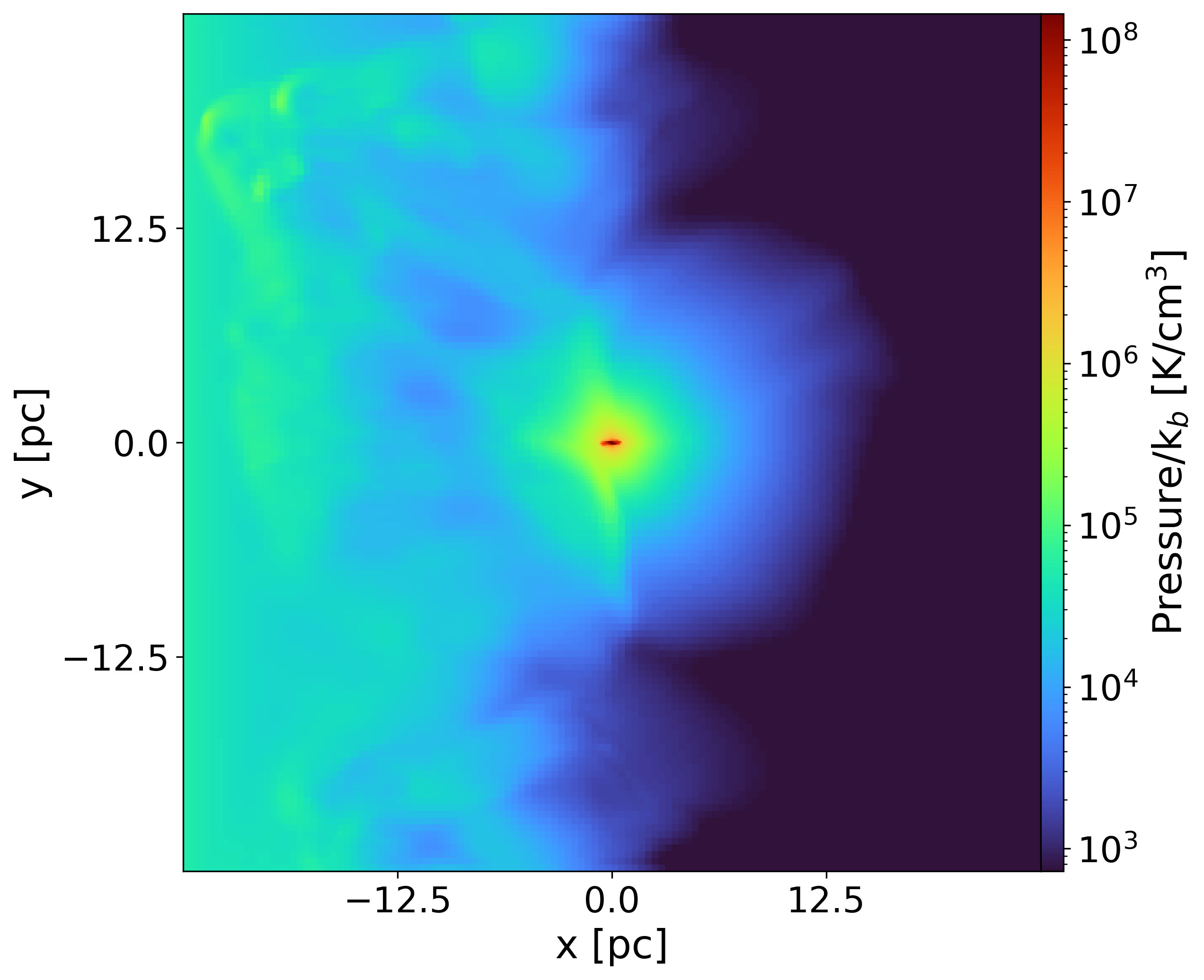}   


        \caption{The same as Figure \ref{fig:gas_map_M5I24}, but for model {\tt M6I23}.}
  \label{fig:gas_map_M6I23}
\end{figure*}

\begin{figure*}
        \centering

        \includegraphics[width=0.268\textwidth,trim={0 0cm 2.935cm 0.0cm},clip]{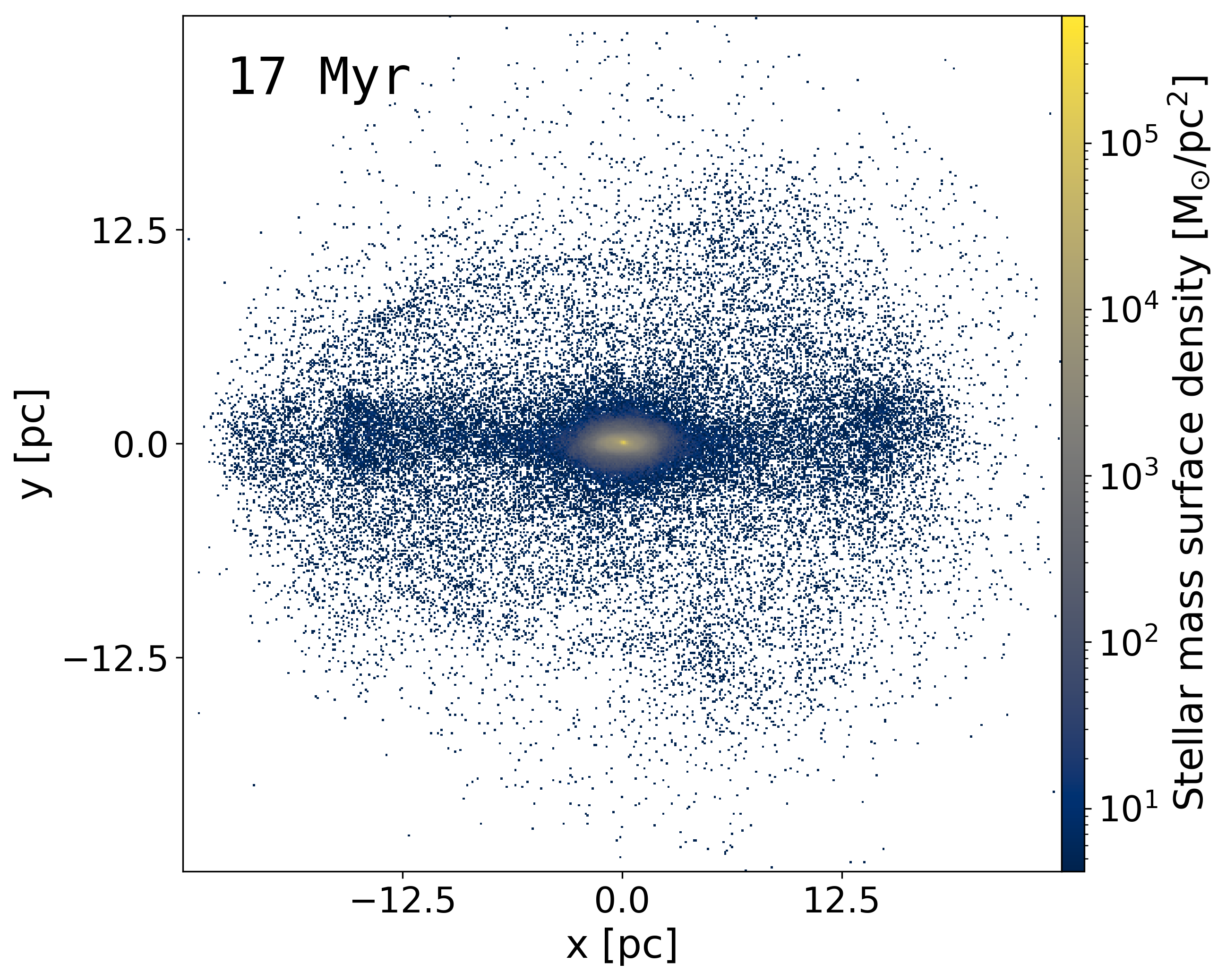}    
        \includegraphics[width=0.225\textwidth,trim={3.1cm 0cm 2.935cm 0.0cm},clip]{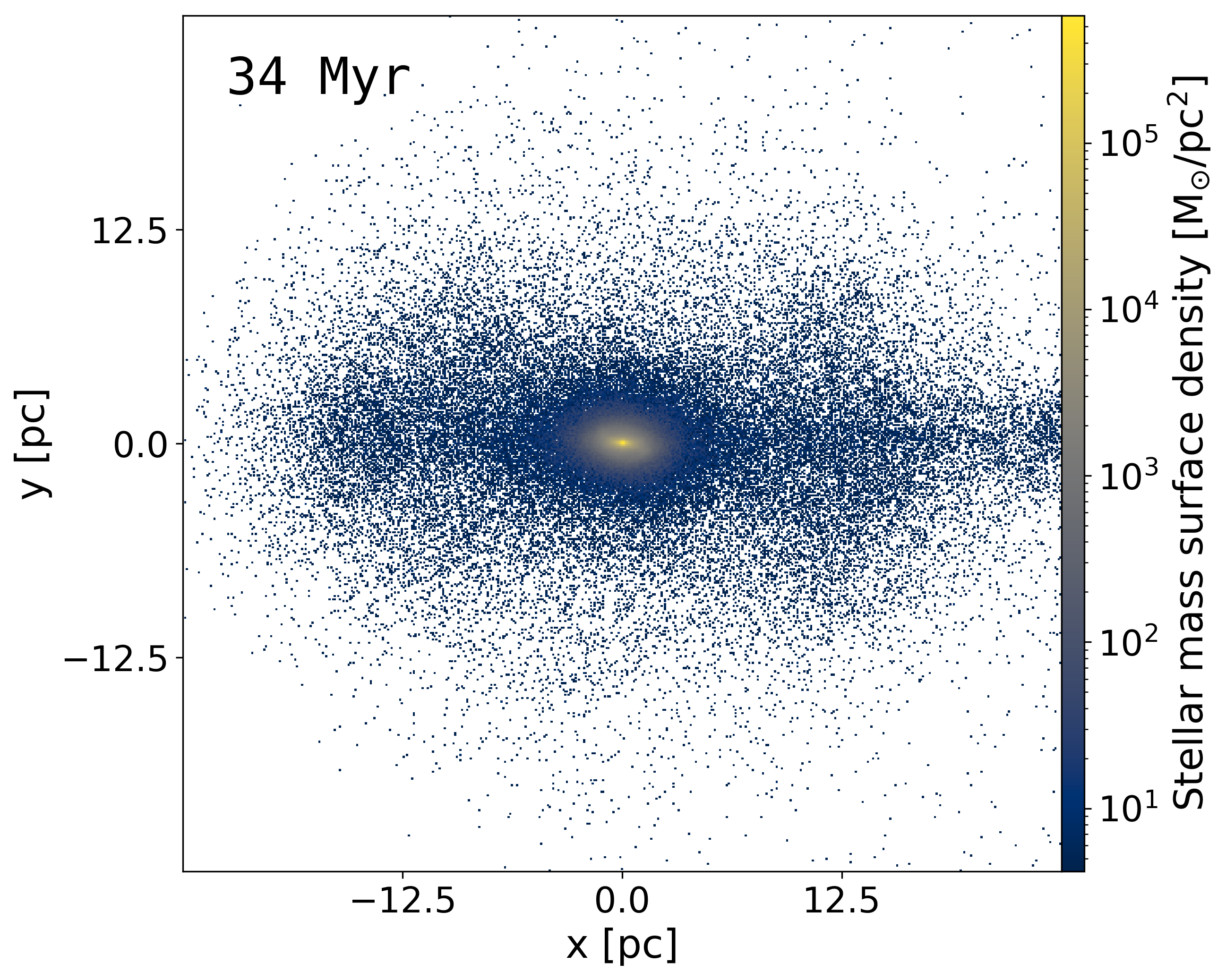}  
          \includegraphics[width=0.225\textwidth,trim={3.1cm 0cm 2.935cm 0.0cm},clip]{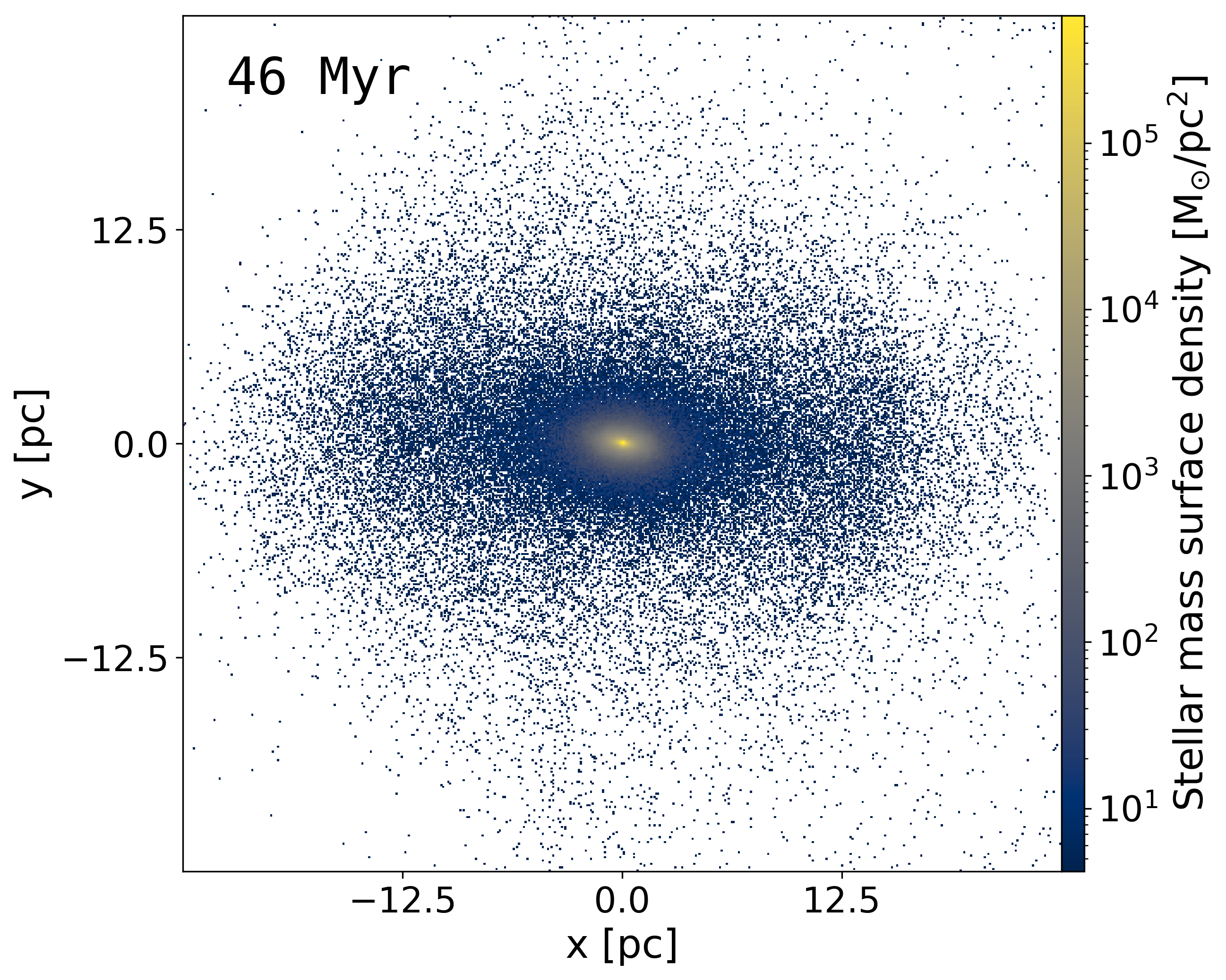}    
        \includegraphics[width=0.267\textwidth,trim={3.1cm 0cm 0cm 0.0cm},clip]{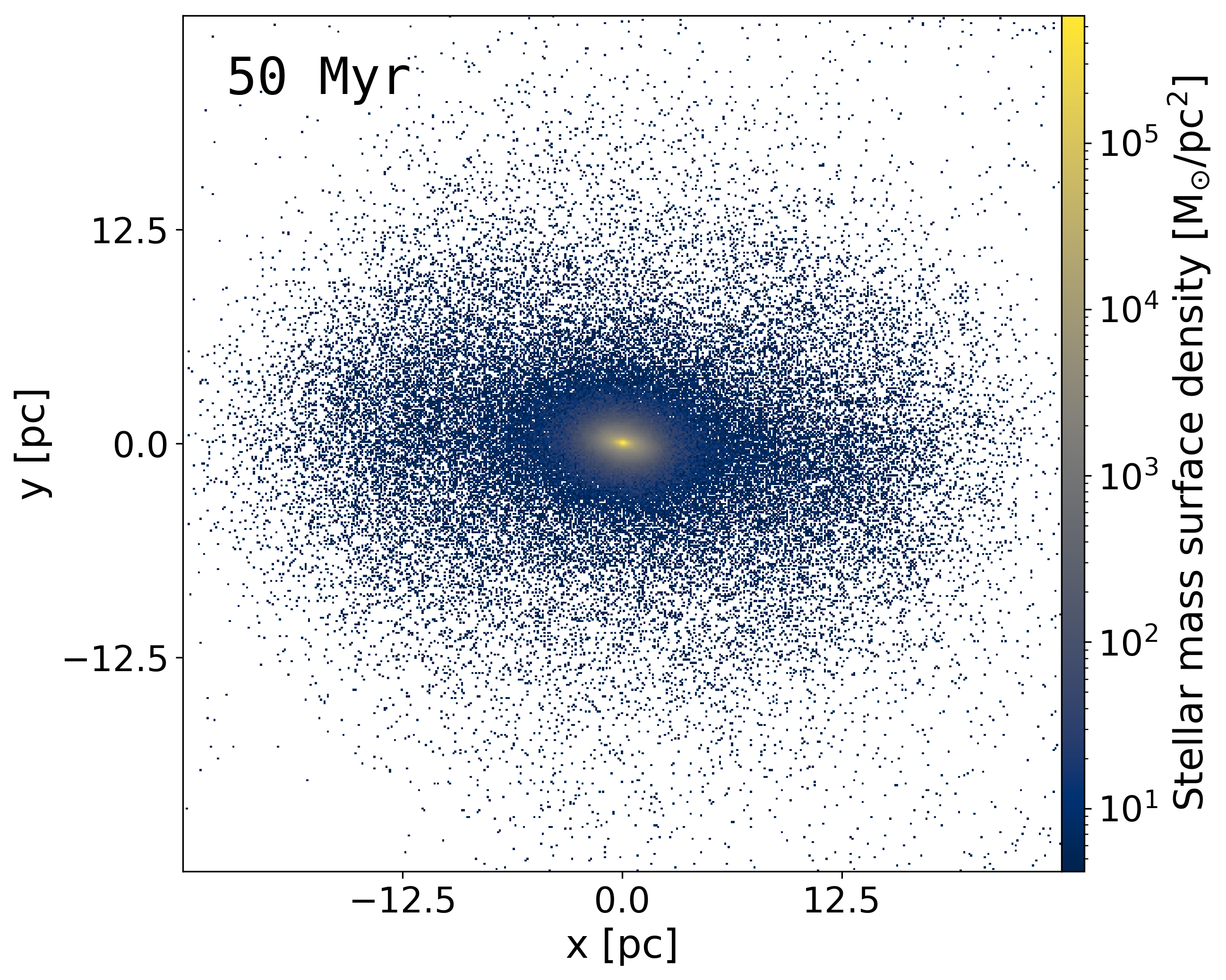}

        \caption{The same as Figure \ref{fig:part_map_M5I24}, but for model {\tt M6I23}.}
  \label{fig:part_map_M6I23}
\end{figure*} 

\begin{figure*}
  \includegraphics[width=0.95\linewidth]{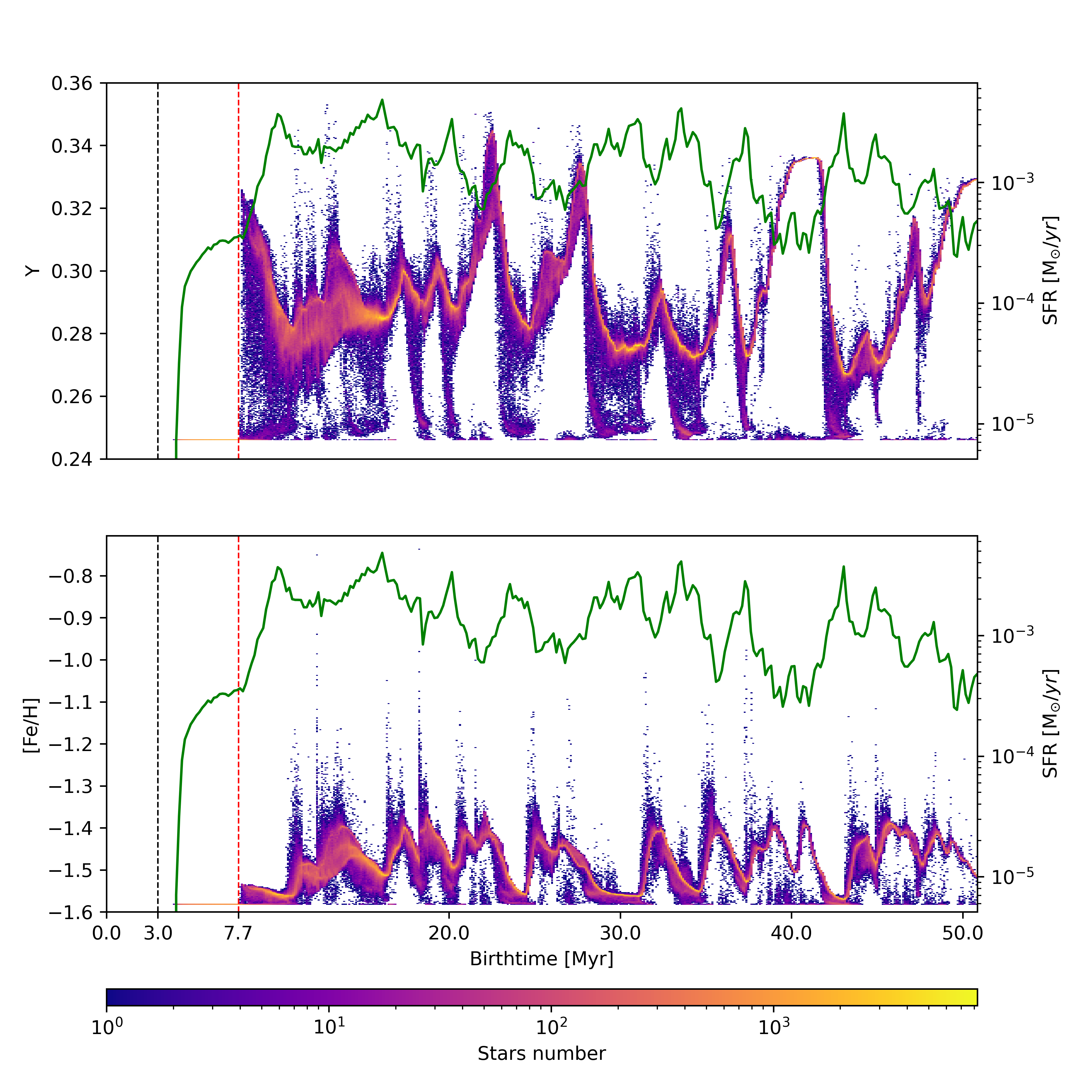}
  \caption{The same as Figure \ref{fig:Y_birth_M5I24}, but for model {\tt M6I23}.}
  \label{fig:Y_birth_M6I23}
\end{figure*} 

\begin{figure*}
        \centering
        \includegraphics[width=0.44\textwidth,trim={0cm 0cm 0cm 0.0cm},clip]{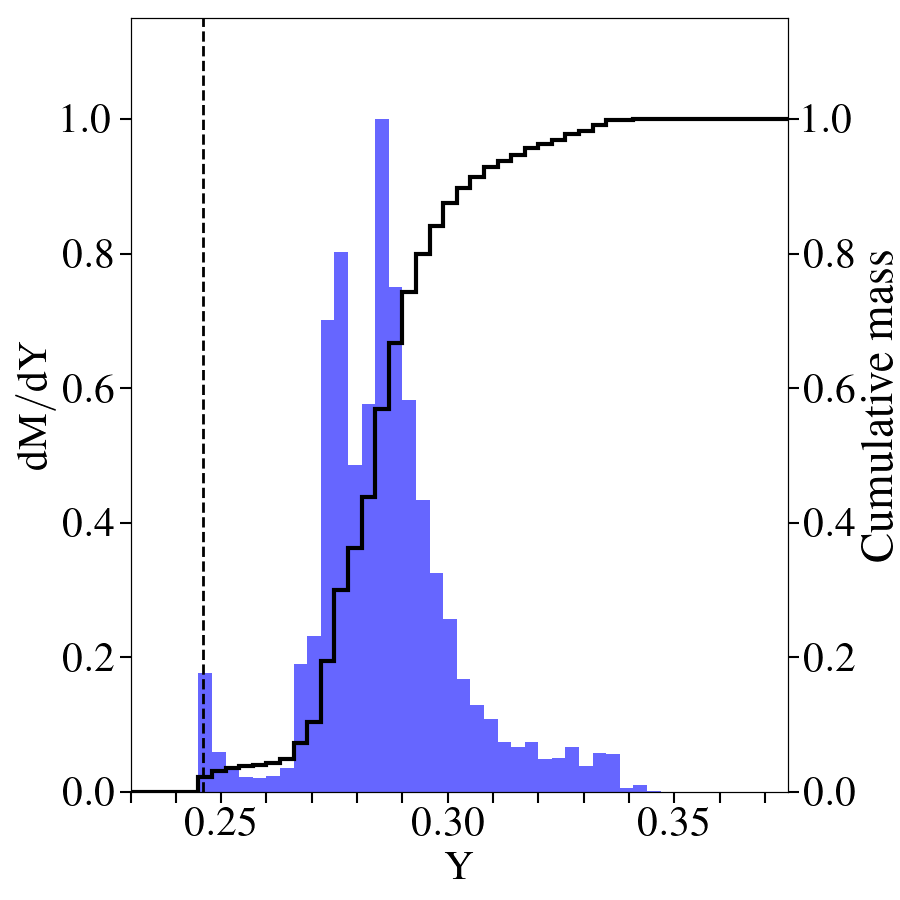} 
        \includegraphics[width=0.44\textwidth,trim={0cm 0cm 0 0.0cm},clip]{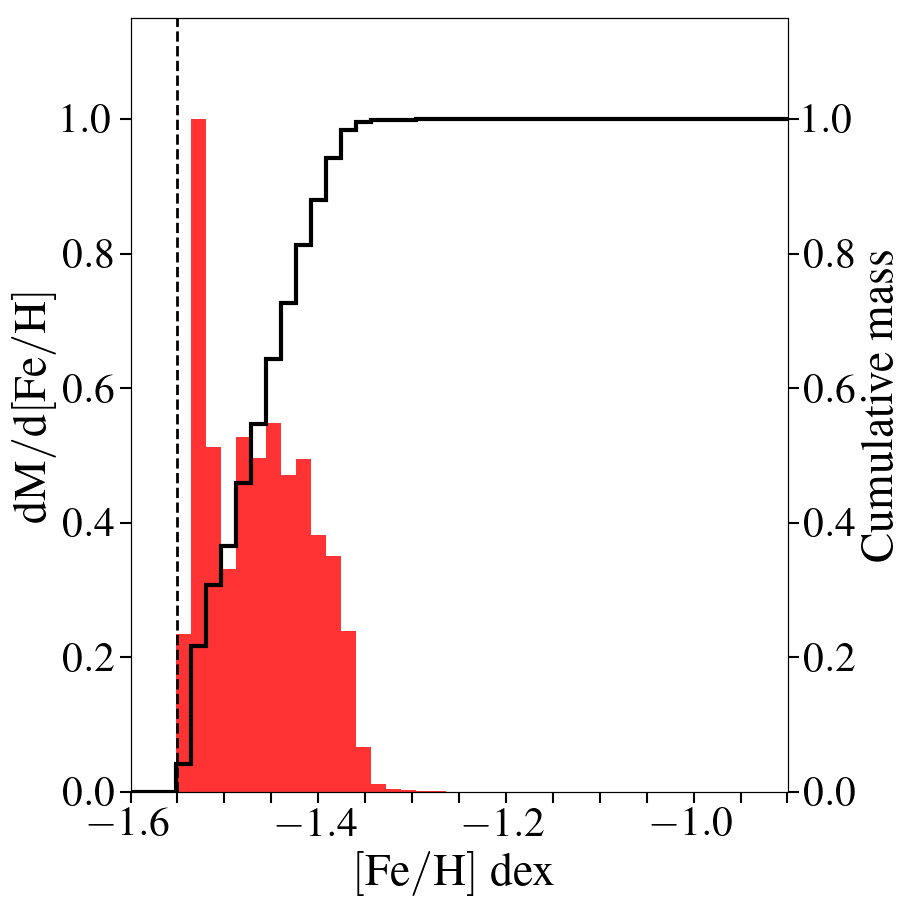} 
        \caption{The same as Figure \ref{fig:prof_M5I24}, but for model {\tt M6I23} at $50\Myr$.}
  \label{fig:prof_M6I23}
\end{figure*}

\subsection{High-density model}

In this section, we present our last model, a cluster characterized by a FG with a mass of 10$^6\Msun$ and with a pristine gas density ten times higher than previous models ($\rho_{\rm pg}= 10^{-23}\gcm$). The infall of pristine gas starts at t$_{\rm inf}$ = 3 Myr and the onset of AGB ejecta is at t$_{\rm AGB}$ = 7.7 Myr.

In Figures \ref{fig:gas_map_M6I23} and \ref{fig:part_map_M6I23}, we present density, pressure, and temperature maps, along with the spatial distribution of SG stars from our M6I23 simulation.
At the first snapshot (t = 17 Myr), a substantial volume within the box is occupied by infalling pristine gas, entering from the left side and moving to the right. On the right side, a notable accumulation of gas with low pressure and temperature is present, forming an accretion tail directed towards the central regions. In the central region, dense and cold gas has accumulated, distributed on an elongated shape along the $x$-axis, corresponding to the direction of the infalling gas, as highlighted by the contours and from the stellar density distribution in Figure \ref{fig:part_map_M6I23}.
The stellar surface density maps reveal significant star formation with stars forming throughout the computational volume with a density peak at the very center ($\Sigma_{\star}>10^4 \Msun/\pc^2$). 
At 34 Myr, similar to the previous models, we observe recent activity from a SN Ia. In this case, in particular, the supernova bubble can be observed with a thin shell at high densities and pressure. The shell surrounds a nearly homogeneous cavity of low-density and highly pressurized hot gas. In the central zones, as in the 17 Myr snapshot, there is a high-pressure and cool gas filament, supplied by an elongated accretion tail along the $x$-axis, but narrower compared to the earlier phase. Along it, stars are recently formed due to its high density.
The stellar mass surface density map in Figure \ref{fig:part_map_M6I23} reveals ongoing star formation even outside the cluster centre, especially downstream of the system where the accretion column of cold and dense pristine gas is present.
At the following snapshot (t = 46 Myr), the system is reaccreting infalling gas, after a momentaneous quenching of the infall due to a SN explosion. The right side of the box displays low density and pressure, accompanied by temperatures around $\sim 10^5 \K$. Meanwhile, the left side consists of infalling pristine gas and numerous instabilities reminiscent of the interaction of the disrupted shell with the pristine gas. As observed in previous snapshots, the central region of the box exhibits the highest density but it appears more compact.
In the last snapshots, at 50 Myr, the system shows minimal changes. The accretion column has been disrupted by the recent SN explosion and the stellar structure has not changed significantly.

As for the other models, the infall of pristine gas is continuously limited by the expanding SN bubbles which stall well outside the cluster. Therefore, after a SN explosion and before the pristine gas is reaccreted, stars in the central region are mainly formed out of helium-rich AGB ejecta. This is clearly shown in Figure \ref{fig:Y_birth_M6I23} where dips in the star formation rate (SFR), due to SN explosions, correspond to enhancements in $Y$. Therefore, the distribution of the stellar $Y$ is shifted to higher values with respect to the case without Type Ia SNe (see \citealt{yaghoobi2022}). Such behaviour is highlighted in Figure \ref{fig:prof_M6I23}, where helium-poor ($Y<0.27$) stars are significantly less than in the model without Type Ia SNe of \citet{yaghoobi2022}. The maximum $Y$ is also much larger, reaching values of 0.35 during the phases in which SNe are limiting the dilution of the AGB ejecta with pristine gas (Figure \ref{fig:Y_birth_M6I23}). The lack of helium-poor SG stars and the presence of extremely helium-enhanced stars in comparison to the models without Type Ia SNe is stressing again the limited dilution experienced by models including SN explosions. We will show in Section \ref{sec:he_star} that this has strong implications for the feasibility of this model.

 The mass distribution of [Fe/H] has a spread of ${\rm\sigma^{SG}_{[Fe/H]}=0.07\ dex}$ centered at $-1.46\ {\rm dex}$, a value much smaller than what has been found for high mass clusters as a result of the lower number of Type Ia SNe explosions experienced in this model. In addition, due to the smaller half-mass radius of the clusters presented here with respect to the ones of \citet{lacchin2021}, SNe ejecta can reach the outskirts and avoid being recycled to form SG stars. If we assume that most of the FG stars have been lost during the cluster's life, accounting for 30\% of the present-day cluster mass, we derive a total cluster spread in [Fe/H] of ${\rm\sigma_{[Fe/H]}=0.07\ dex}$ comparable with the observed values for the bulk of globular clusters.

\section{Discussion}
\label{sec:discussion}

\subsection{On the Validity of the AGB Scenario}

The results presented in this work suggest a tension between the predicted chemical signatures of SG stars and those observed in Galactic GCs, particularly in terms of helium enrichment and dilution efficiency. This raises the question of whether the AGB scenario remains a viable framework for explaining the formation of multiple populations. While our study does not attempt to rule out alternative models, we believe that the AGB scenario still provides a physically motivated and observationally supported avenue, warranting continued refinement and testing.

AGB stars are a \textit{natural component} in self-enrichment models of GC formation. Their slow winds (on the order of $\sim 10$ km s$^{-1}$), rich in helium and light elements (e.g. Na, Al), offer an ideal mechanism for intra-cluster chemical evolution, especially in dense environments where supernova feedback may have already cleared the primordial gas. Unlike massive stars or fast winds, AGB ejecta are more likely to remain gravitationally bound and accumulate in the cluster potential well, providing a plausible fuel source for secondary star formation \citep[e.g.,][]{dercole2008,ventura2009,dantona2016,calura2019,dantona2023}. The chemical patterns produced by AGB nucleosynthesis---namely enhanced He, N, Na and depleted C, O---closely match the abundance trends observed in many SG stars \citep{gratton2019}.

In addition to chemical arguments, AGB-based models have demonstrated significant success in reproducing \textit{spatial and kinematic properties} of SG stars. In particular, simulations based on the AGB framework have correctly predicted the central concentration of SG stars \citep[e.g.,][]{dercole2008,calura2019,yaghoobi2022}, consistent with deep photometric surveys \citep{cordero2014,simioni2016}. Moreover, recent observational studies provide kinematic support for the scenario: SG stars have been found to exhibit \textit{faster rotation} than FG stars in several clusters (e.g., NGC 2808, M13, NGC 5904, \citealt{bellini2015,cordero2017,cordoni2020}), a pattern consistent with SG stars forming from dissipative gas flows in the inner regions of the cluster \citep{bellini2018,lacchin2022,libralato2023,dalessandro2024,cordoni2025}. 
These rotation signals are difficult to explain in models where multiple populations arise purely from external accretion or stellar mergers and instead lend support to an in-situ origin, possibly involving AGB ejecta. 

While it is clear that the AGB scenario still faces important challenges---particularly regarding the mass budget, dilution, timing constraints and other challenges 
\citep{bastian2018,milone2022}---its ability to qualitatively and quantitatively match many key features of GC multiple populations motivates further exploration.
Rather than discarding the scenario entirely in light of the discrepancies introduced by SN~Ia feedback, we argue that targeted refinements---such as more accurate treatment of gas accretion, SN delay-time distributions, or inclusion of additional feedback sources---are a more productive path forward.

\subsection{Type Ia supernova effects of the star formation}

\begin{figure}
        \centering

        \includegraphics[width=0.50\textwidth,trim={0 0.7cm 0 0.0cm},clip]{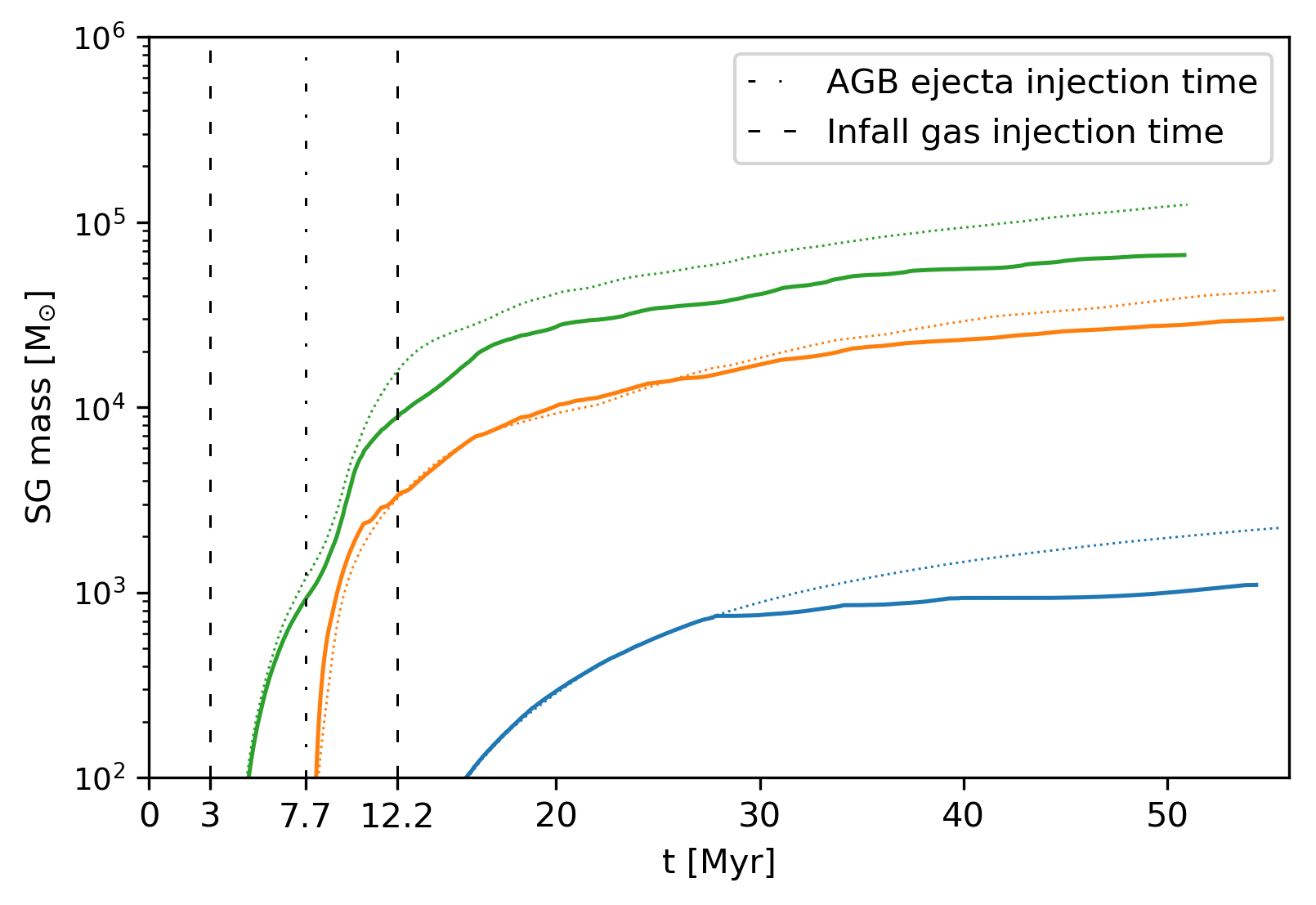}    
         \includegraphics[width=0.50\textwidth,trim={0 0cm 0 0.0cm},clip]{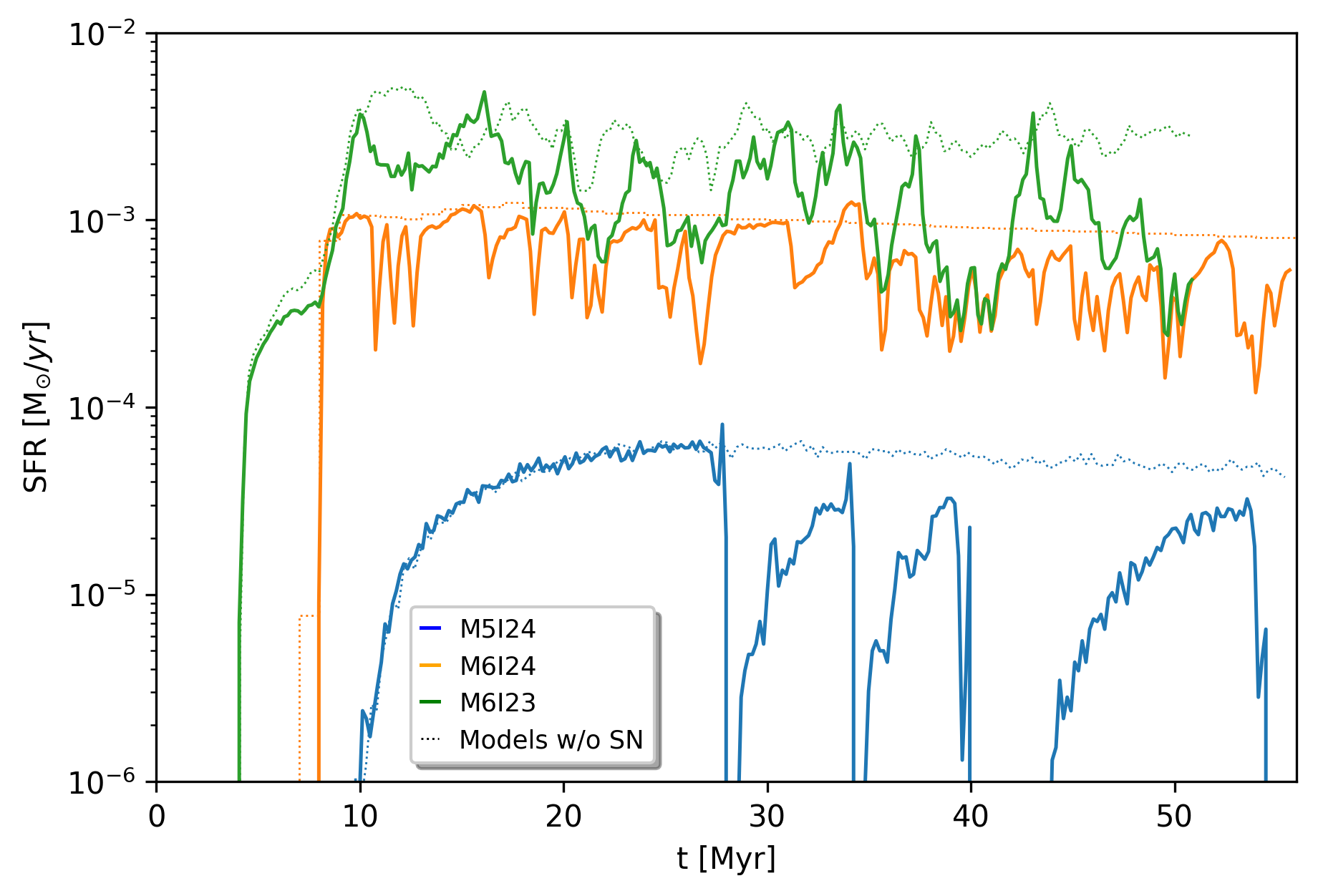}   

        \caption{Upper panel: evolution of the SG stellar mass with time. Lower panel: evolution of the SFR with time. Solid lines represent models with Type Ia SNe, whereas dotted lines of the same color represent the same models but without Type Ia SNe. The blue lines represent the low-mass low-density model while the orange and green lines represent the high-mass low- and high-density models, respectively. In the upper panel, the dashed lines show the times at which the infall enters the box, which depends on the model as reported in Table \ref{tab:simu}. The dash-dotted line displays the beginning of the injection of AGB ejecta, which is the same for all models.}
  \label{fig:SFR}
\end{figure}

In a previous paper \citep{lacchin2021}, we studied the impact of Type Ia SNe on star formation and on the chemical composition of a very massive cluster in the AGB framework to determine whether the results obtained in 1D by \citet{dercole2008} were the same when performing 3D hydrodynamic models. As in \citet{calura2019}, we performed simulations with different ISM densities, to explore the role played by the environment. We found that in the low-density model the SN bubbles extend much further than in the high-density one. Consequently, the ISM gas hardly ever reaches the centre of the system leading to unobserved helium spreads, while star formation is significantly reduced by the explosions. On the other side, for higher ISM densities, the SFR is not affected and a significant dilution of the AGB ejecta leads to helium spread in line with the observations. In this work, we extend the study to lower mass clusters of $10^5-10^6\Msun$, with a half-mass radius of $4\pc$, as expected from young massive clusters \citep{Krumholz2019}. In Table \ref{tab:result}, we report the masses of the SG with and without Type Ia SNe for clusters from $10^5\Msun$ to $10^7\Msun$ encompassing the results obtained in this work with the ones of \citet{calura2019}, \citet{lacchin2021} and \citet{yaghoobi2022}. In the low-density and low-mass model (M5I24), Figure \ref{fig:SFR} shows that star formation is halted by Type Ia SN explosions, but due to the sparsity of the explosions in time, new stars are formed once the gas has cooled down again. At the end of the simulation, the total mass is $1.1\times 10^3\Msun$, which corresponds to 50\% of the models without Type Ia SNe. Instead, for models with $10^6\Msun$ the star formation rate is only slightly affected by SN explosions, but it is not reduced with time. At the end of the simulation, the model M6I24 has formed $3\times 10^4\Msun$ of SG stars, which corresponds to the 61\% of the same model without Type Ia SNe, while for M6I23 the final mass of SG stars is $6.7\times 10^4\Msun$, the 49\% of the case without SNe.

Among the models with $10^6\Msun$, M6I24 shows a smaller difference between the final mass with and without SN feedback than M6I23, at variance with the case at $10^7\Msun$.
In the models presented here, the cluster is much more compact that in \citet{lacchin2021}, with a half-mass radius almost an order of magnitude smaller. This means that, in the high-density models, the radius at which the SN bubble stalls, which depends on the density of the gas, was within the cluster in the $10^7\Msun$ model (see Figure 9 in \citealt{lacchin2021}) while here is much larger than the cluster radius (see Figure \ref{fig:gas_map_M6I23}). The star formation is therefore reduced much more in the simulations performed in this work, as the whole cluster is affected by the SN explosions. In clusters with higher mass, star formation is triggered in the high-density regions at the interface between two or more expanding SN bubbles within the cluster. In addition, in M5I23, the infall of ISM gas is limited by SN explosions, reducing the amount of pristine gas in the central regions, similarly to what was happening in the low-density model with $M_{\rm FG}=10^7\Msun$. 
At around $40\Myr$ the presence of SNe exploding at short time separation is preventing the infalling gas from reaching the central part of the system and, therefore, new stars are born only from AGB ejecta, as visible in Figure \ref{fig:Y_birth_M6I23}. The similar SFR experienced by models M6I23 and M6I4 at  $40\Myr$ and later at around $48\Myr$ confirms this behaviour. 

\begin{table}
\caption{Models description. {\it Columns}: (1) name of the model, (2) FG mass, (3) SG mass (SG mass in the model without Type Ia SNe) (4)  the fraction of SG stars, $f_{SG}$. The SG masses in parenthesis are the ones obtained for the same model but without Type Ia SNe. For models M7I23 and M7I24 the values with Type Ia SNe are from \citet{lacchin2021}, while the ones without are from \citet{calura2019}.  } 
\begin{tabular}{cccc} 
\hline
\hline
 Model &${\rm M_{FG} [M_\odot]}$& ${\rm M_{SG}[\Msun]}$  &$f_{SG}$\\
\hline
${\rm M5I24}$                  & $10^5 $   & $  1.1\times 10^{3}\ (2.2\times10^{3}) $  &1.1\%\\
${\rm M6I23}$                  & $10^6 $   & $  6.7\times 10^{4}\ (1.3\times10^{5})$  &6.7\%\\
${\rm M6I24}$                  & $10^6 $   & $  3.0\times 10^{4}\ (4.9\times10^{4})$  &3\%\\
${\rm M7I23}$                  & $10^7 $   & $  2.3\times 10^{6}\ (5.0\times10^{6})$  &23\%\\
${\rm M7I24}$                  & $10^7 $   & $  1.2\times 10^{5}\ (7.0\times10^{5})$  &1.2\% \\

\hline
\hline
\end{tabular}

\label{tab:result}

\end{table}

\subsection{On the Timing and Spatial Distribution of Type Ia Supernovae}

In our models, Type Ia supernovae (SNe Ia) are assumed to follow the delay-time distribution (DTD) derived by \citet{greggio2005} for the single-degenerate (SD) scenario. As described in Section~\ref{sec_ia}, this results in the onset of SNe Ia roughly contemporaneously with the beginning of AGB ejecta release, i.e. $\sim$39 Myr after the formation of the first generation (FG) of stars. The integrated number of explosions in our simulations---about 10 for a cluster with $10^5$ M$_\odot$ and 100 for one with $10^6$ M$_\odot$---reflects this early burst of activity. While this approach has been adopted in previous works, we acknowledge that it may not accurately reflect the physical conditions of low-metallicity globular clusters. 

In particular, the early emergence of SNe Ia in our models tends to suppress the standard pathway of SG formation, in which AGB ejecta are diluted by infalling pristine gas. Since SNe Ia begin exploding as soon as the AGB winds are released, they inhibit the cooling and settling of ambient material before a typical SG can form. This overlap is especially problematic given that the SD channel relies on mass accretion onto white dwarfs---a process that is highly sensitive to metallicity through the formation of optically thick winds \citep{kobayashi1998}. These winds are required to maintain accretion within the narrow range necessary for a white dwarf to reach the Chandrasekhar mass and explode. In low-metallicity environments like those typical of GCs ($Z \lesssim 0.001$), such winds are strongly suppressed, and hence the SD channel is expected to be significantly delayed or even inefficient.

\citet{kobayashi2020} studied the dependence of DTD on metallicity, finding that 
for $Z = 0.004$---already higher than most GC metallicities--- 
the DTD begins at log(age) $\sim$ 8.25 (i.e., $\gtrsim 180$ Myr), i. e. signficantly after the AGB pollution phase. This would naturally leave room for the canonical SG formation process to unfold without interference from SNe Ia in the early stages. These considerations suggest that our assumption of a prompt SD channel may overestimate the dynamical impact of SNe Ia in the early evolution of GCs.

An alternative scenario involves the double-degenerate (DD) channel, in which two white dwarfs merge. This channel is thought to dominate at early times in low-metallicity systems and may begin earlier than the SD route.

Moreover, it may be that both SD and DD progenitors are more centrally concentrated than what assumed here, 
due to mass segregation and dynamical interactions \citep{voss2012}. If this is the case, their explosions would inject energy into a smaller volume, potentially affecting only the central region while leaving the outer layers---and possibly even the cooling inflow---relatively undisturbed. 

\citet{dercole2008} used 1D hydrodynamical simulations to show that centrally located SNe Ia can quickly quench SG star formation by expelling gas from the cluster core. While insightful, such 1D models necessarily neglect anisotropies, clumpiness, and angular momentum effects that are important in 3D.
It would therefore be valuable to revisit this configuration using high-resolution 3D simulations to assess whether centrally concentrated SNe Ia retain the same quenching efficiency under more realistic geometric and dynamical conditions.

Moreover, the detailed impact of DD SNe Ia on cluster gas dynamics has not been fully explored, and future simulations should account for their spatial and time distribution and energetics in a realistic way.

Finally, we highlight an important asymmetry in our current results: while SNe Ia in our simulations are generally effective at halting the infall of pristine gas, they are not always successful at quenching star formation from the accumulated material. This might suggest that the assumed energy deposition is too diffuse, or alternatively, that additional concentrated feedback mechanisms (e.g., X-ray binaries, pulsar winds; see Section~\ref{sec_sources}) may be needed to suppress residual star formation.


\subsection{He enrichment of second-generation stars}
\label{sec:he_star}
\begin{figure}
        \centering

        \includegraphics[width=0.50\textwidth,trim={0 0cm 0 0.0cm},clip]{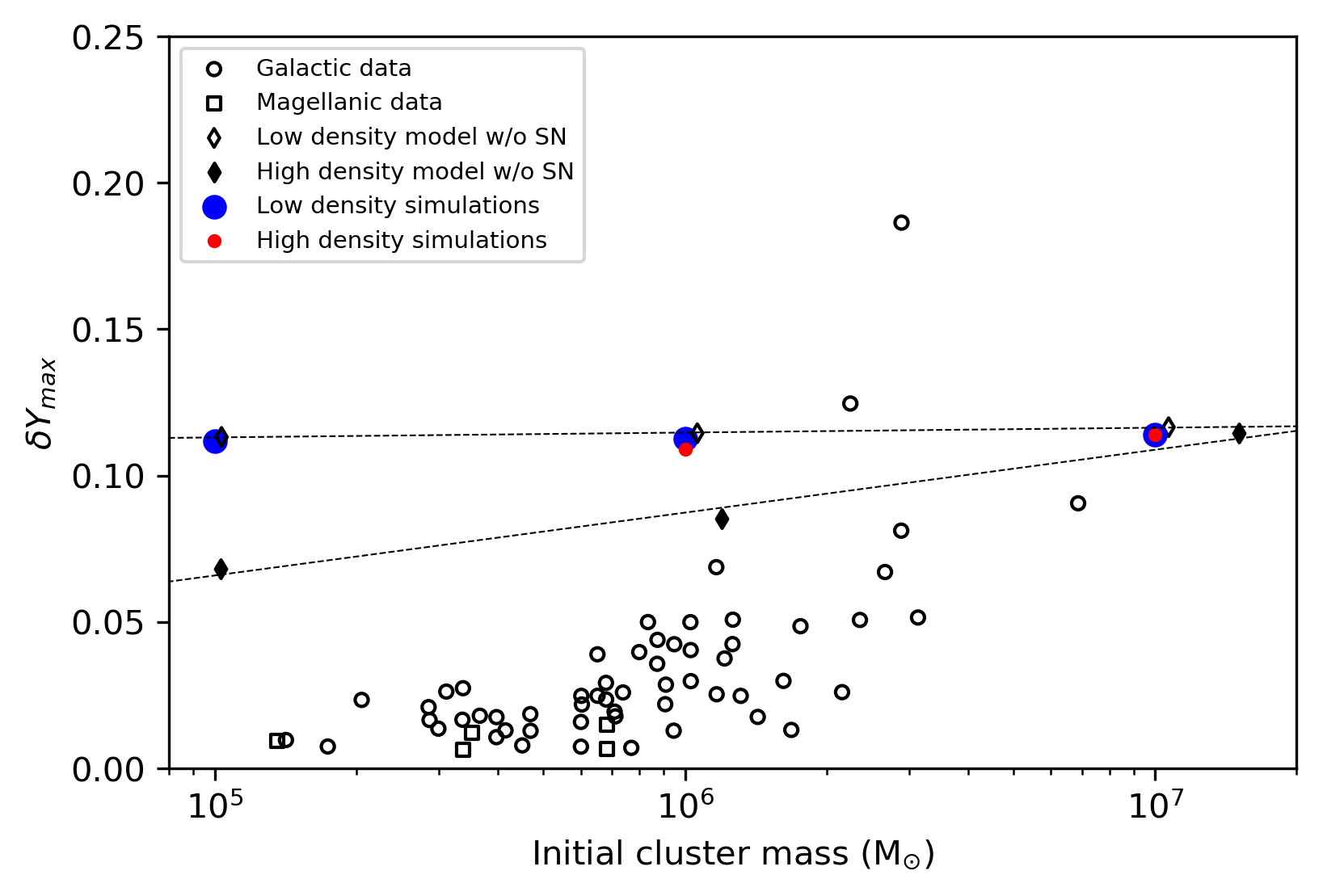}    

        \caption{Maximum enhancement of the helium mass fraction in SG stars as a function of initial cluster mass. Diamonds represent models without Type Ia SNe from \citet{calura2019} and \citet{yaghoobi2022}: filled diamonds represent the high-density models, while empty ones the low-density models. The data for the models with Type Ia SNe with the highest initial cluster masses are taken from \citet{lacchin2021}. Open circles and squares are observational data obtained by \citet{milone2020} for the Milky Way and Magellanic Clouds clusters (initial masses of each cluster are from \citealt{baumgardt2018} and \citealt{goudfrooij2014}.) }
  \label{fig:dYmax_vs_Mini}
\end{figure}

In the last decade, the Hubble Space Telescope photometry has allowed to derive helium composition in several globular cluster stars, both in the Milky Way and in Magellanic Clouds \citep{milone2015,marino2018,lagioia2018,lagioia2019,milone2020} revealing a correlation between the maximum He spread $\delta Y_{max}$ and both the present and initial cluster mass \citep{milone2018,zennaro2019,milone2020}. Therefore, massive clusters have been found to display larger He enhancements a trend that has been found in \citet{yaghoobi2022} for the high-density models. 
Figure \ref{fig:dYmax_vs_Mini} shows the maximum He spread among both FG and SG stars $\delta Y_{max}=Y_{max}-Y_0$, where $Y_{max}$ is the maximum He mass fraction in the cluster and $Y_0$ the He mass fraction of the FG stars, as a function of initial cluster mass for both the models with (this work and \citealt{lacchin2021}) and without (\citealt{yaghoobi2022} and \citealt{calura2019}) SNe Ia. 
In the low-density models, the $\delta Y_{max}$ is independent of the cluster mass, as it has been found even in the simulations without Type Ia SNe as most of the SG stars are formed out of undiluted AGB ejecta. Once we include SNe Ia, the infall of ISM gas is reduced even more leading to the formation of a higher number of stars from undiluted material. As the AGB yields do not depend on the cluster mass, the first stars formed have a helium mass fraction around $Y\sim 0.36 $ in all models.
In high-density models without Type Ia SNe, instead, a correlation with cluster mass was found as the majority of the material composing SG stars was coming from pristine gas. 
On the other hand, when Type Ia SNe are added, even the high-density models show no dependence on the cluster mass, as in the low-density ones. SN explosions together with reducing the SFR, are preventing the infall of pristine gas, and, therefore, stars that are formed after the SN explosion and before the pristine gas has time to reach again the central regions would be mainly composed of AGB ejecta, significantly increasing the $Y$ in SG stars. Such enhancement can be clearly seen in Figure \ref{fig:Y_birth_M6I23}, where peaks of $Y$ correspond to decreasing SFR due to Type Ia SN explosions.

\subsection{Iron enrichment}

The total iron spread does not vary much in clusters with the same mass and different ISM density. The iron rich stars are formed soon after a decrease of the star formation rate before the AGB ejecta and the pristine gas restart to pollute the system as shown in Figure \ref{fig:Y_birth_M6I24} and \ref{fig:Y_birth_M6I23}. The peaks in the [Fe/H] are decreasing fast over time, as AGB material dilutes SN ejecta.

Concerning the iron content in SG stars only, the dispersions are much smaller than the ones obtained in the $10^7\Msun$ clusters due to the fewer SN explosions. Observationally, more attention has been given to first-generation stars where an iron spread has been detected in several clusters \citep{legnardi2022,lardo2022,lardo2023}. On the other side, the SG spread is much more difficult to derive: \citet{legnardi2022} compared the two populations for a few clusters finding that the SG is more homogeneous in iron than the FG. In all our clusters the spread of both the SG only and the whole cluster is small, in agreement with the current observations \citep{bailin2019}. Nevertheless, in every model presented here, the FG is assumed to be homogeneous in the iron content, therefore it is likely that having a spread in the FG iron composition would lead to a broader SG one, perhaps larger than the typical ones.

In all the models, SG stars are found to be more iron-rich than FG stars. When considering the iron dispersion of the whole cluster assuming that most of the FG stars have been lost to match the present-day fraction of SG stars, no significant variations have been found with respect to the dispersion of the SG only.

\subsection{Additional Sources of Energy Influencing Star Formation Quenching}
\label{sec_sources}
While Type Ia supernovae are a key source of feedback in our models, other astrophysical processes may significantly
contribute to the quenching of SG star formation in GCs. 
Below we discuss a number of other possible mechanisms, supported by recent literature, 
that could provide additional energy input---either preventing the accretion of pristine gas or disrupting star-forming regions.

\subsubsection{Photoionizing radiation}

Massive stars and post-AGB objects emit substantial ultraviolet radiation capable of ionizing the surrounding medium. 
Such ionizing radiation emitted by evolved stars has been proposed as a fundamental
ingredient to explain the very small content of ionized gas of GCs at the present-day \citep{chantereau2020}.
As for young clusters, recent simulations by \citet{yaghoobi2024} suggest that photoionizing radiation alone can suppress cooling 
flows in low-mass clusters, especially in systems where the potential well is shallow.
These authors find that in most configurations---except for the most massive clusters 
(with stellar mass $>10^6$ M$\odot$, see also \citealt{yaghoobi2022})--the gas remains too hot and diffuse to trigger SG star formation.
They also suggest that in the local Universe, the ionizing luminosity from a modest population 
of post-main-sequence stars ($Q_{\rm H} \sim 10^{46}$--$10^{47}$ s$^{-1}$ $M_{\odot}^{-1}$) in low-mass young clusters 
may maintain the interstellar medium (ISM) at temperatures $T \sim 10^4$--$10^5$ K stalling further collapse and preventing star formation. 
That explains why local young clusters do not show evidence of ongoing star formation and why they are free of cold gas.
However, 
due to a more significant population of massive clusters and larger reservoir of dense gas,   
the same may not be true in high-redshift galaxies \citep{yaghoobi2024}. 

\subsubsection{X-ray binaries}

X-ray binaries---especially high-mass systems formed early in cluster evolution, after the collapse of massive stars, 
generally named high-mass X-ray binaries \citep{fornasini2023} ---
can inject energy into the ambient gas through both hard X-ray radiation and mechanical feedback from jets or winds.   
\citet{power2009} have argued that their collective feedback, particularly if concentrated in the central region,
could heat or expel gas in systems with shallow gravitational wells, giving a significant 
contribution to the high-redshift X-ray background.   
The total energy output from a single, bright low-mass X-ray binary over a few Myr can
reach $\sim 10^{49}$--$10^{50}$ erg,  in some cases 
comparable to the cumulative energy emitted by a massive star via stellar winds \citep{calura2015,calura2025}. 

After their formation, at 50-100 Myr,  their luminosity has decreased progressively of several orders of magnitude
\cite{power2009};
still, they could play and important role at preventing further accretion in evolved systems \cite{dercole2008}.

\subsubsection{Pulsar winds}

Millisecond pulsars are ubiquitous in GCs and may give a non-negligible contribution to long-term stellar feedback. 
Through magnetized winds and relativistic particles, they can inflate bubbles and inject non-thermal pressure into the ISM.

With typical X-ray luminosities  of $10^{33}-10^{34}$ erg/s, their cumulative energy release is lower than
the one of other sources such as X-ray binaries and their radial distribution indicates a typical higher concentration
in the cluster centre \citep{ye2019}. 

However, considering  the ubiquitous presence and continuous nature of these sources,
they may contribute in a non-negligible fashion to regulating central gas accretion in GCs.

\subsubsection{Planetary nebulae}

In their late planetary nebula phases, red giant stars 
emit shells of ionized gas that may affect the interstellar and intracluster gas. 
Although less energetic than the sources above, the fast winds of planetary nebulae (PNe) 
can provide local pressure support. Their terminal velocities ($v \sim 100-1000$ km s$^{-1}$, \citealt{grewing1989})  
and short durations ($t \sim 10^4$ yr) \cite{banedes2015} yield energy releases of $\sim 10^{33}$--$10^{37}$ erg per event 
\citep{schoenberner2014,tan2024}. 
In principle, if occurring in large numbers, these events can perturb local gas structure and
reduce the effectiveness of inflowing streams, thus with potential effects on
the late star formation history of a star cluster.
Their progenitors are intermediate mass stars, therefore the youngest PNe can
occurr on timescales ranging from  $\sim 40$ Myr to several hundred Myr and represent a diffuse source. 
Based on present estimates, their cumulative effect is likely subdominant compared
to supernovae or X-ray binaries \cite{dercole2008}, but they can play some role to explain
the loss of the intracluster medium in old (from $\sim$ a few Gyr) GCs.

Despite the growing recognition of the importance of these additional energy sources,
their implementation in present-day simulations of globular cluster formation remains partial and fragmented.
Most studies to date have explored these mechanisms in isolation, often using simplified prescriptions
for their energy injection and without accounting for their spatial clustering or temporal evolution.
Furthermore, the possible interplay among different feedback channels---such as the combined effect of SNe Ia, X-ray binaries, and pulsar winds---
has not been systematically investigated. Future high-resolution simulations should aim to incorporate these processes
simultaneously, with physically motivated models for their spatial distribution, duty cycles, and feedback efficiency.
Only through such comprehensive approaches can we fully understand the range of pathways leading to the formation
(or suppression) of second-generation stars in globular clusters.

\section{Conclusions}
\label{sec:conclusions}
By means of 3D hydrodynamic simulations, we explored the role of Type Ia SNe on the formation of second-generation stars in globular clusters of $10^5-10^6\Msun$ and surrounded by a uniform interstellar medium of $10^{-23}-10^{-24}\gcm$. We focused on the ability of Type Ia SNe to quench the star formation and on the chemical composition of newborn SG stars. In all models, we assume that the SG are formed out of the ejecta of first-generation AGB stars plus ISM gas which is supposed to dilute the AGB ejecta to recall the anticorrelations observed in globular clusters. As shown by \citet{lacchin2021}, dilution can be significantly reduced by the presence of Type Ia SNe in massive systems, leading to clusters composed of extremely helium-rich stars, at variance with observations. 
Even though the number of exploding SN Ia is small, given the lower mass of the clusters studied here, the shallower potential well of the FG reduces the amount of pristine gas that can be accreted and dilute the AGB ejecta.

Our results can be summarized as follows:

\begin{enumerate}
    \item In the $10^5\Msun$ cluster, the star formation rate is strongly affected by the presence of Type Ia SNe, with quenching events every time a SN explodes. However, given the few explosions, star formation restarts once the gas cools down again. On the other side, for clusters with a mass of $10^6\Msun$ the star formation rate is mildly affected by Type Ia SNe, without a proper quenching of SF but with SFR decreases when SNe Ia explode.
    
    \item The iron spread in all our models is much smaller than the one derived in \citet{lacchin2021} due to the lower number of SNe Ia explosions. In this case, the iron content in SG, as well as for the whole cluster, is more in line with what is observed.

    \item As in the more massive cluster simulations of \citet{lacchin2021}, Type Ia SNe limit the dilution of the AGB ejecta with pristine gas falling into the system. Therefore, even in the models presented here, the helium mass fraction $Y$ of SG stars is on average much higher than the one found in simulations without Type Ia SN feedback by \citet{yaghoobi2022}. Such behaviour has dramatic effects on the highest helium mass fraction in stars, which significantly increases especially in the high-density models. Therefore, the maximum enhancement of helium mass fraction $ \delta Y_{max}$ is above 0.1 even for low mass clusters, where, instead, without Type Ia SN a correlation was found with initial cluster mass. With SNe Ia explosions this trend is no more visible and at all cluster masses $\delta Y_{max}$ is comparable with the difference between the most massive AGB ejecta $Y\sim 0.36$ and the primordial one of $0.246$.

\end{enumerate}

Combining the results we obtain here with the ones obtained by \citet{lacchin2021}, we can conclude Type Ia SN feedback alone does not immediately quench the star formation as previously thought, and, on top of that, it leads to unobserved chemical compositions due to the significant reduction of the dilution which is instead necessary to reproduce the anticorrelations detected in globular clusters. Recently, \citet{yaghoobi2024} (but see also \citealt{chantereau2020}) have found that photoionizing radiation can deeply affect star formation in low-mass clusters like the ones we study in this work. In their paper, they found that only in the M6I23 model SG stars can form, while for all other models ionizing radiation heats and expands the gas preventing it from forming a cooling flow and reaching high enough densities to be eligible for SF. 
Nevertheless, not even assuming photoionizing radiation clusters can quench or significantly reduce the SG formation within a hundred Myr. Other mechanisms should be invoked to explain how the SG formation is halted, such as X-ray binaries, planetary nebulae \citep{dercole2008} or pulsar-winds \citep{krause2016}. Such processes act at the same time, so modeling them and studying their interplay is essential to reach clearer insights on the formation of SG stars, but also on the globular cluster journey to become gas-free systems. 


\bigskip
\textit{ \small Acknowledgements:}
 {\small EL acknowledges financial support from the European Research Council for the ERC Consolidator
grant DEMOBLACK, under contract no. 770017. FC acknowledges support from grant PRIN MIUR 2017- 20173ML3WW 001, from the INAF main-stream (1.05.01.86.31) and from PRIN INAF 1.05.01.85.01. We acknowledge EuroHPC Joint Undertaking for awarding us access to Discoverer at SofiaTech, Bulgaria. The research activities described in this paper have been co-funded by the European Union - NextGeneration EU within PRIN 2022 project n.20229YBSAN - Globular clusters in cosmological simulations and in lensed fields: from their birth to the present epoch. }

%
\bibliographystyle{aa} 
\bibliography{aanda}

\end{document}